\documentclass[aps,twocolumn,prl,superscriptaddress,amsmath,amssymb]{revtex4-2} 
\usepackage{dcolumn}
\usepackage{amsmath}
\usepackage{mathrsfs}
\usepackage{txfonts}
\usepackage{bm}
\usepackage[T1]{fontenc}
\usepackage{xspace}
\usepackage{ulem}
\usepackage{comment}
\usepackage{braket}
\usepackage{lipsum}
\usepackage{bbold}
\usepackage{mathtools}
\allowdisplaybreaks[4]

\ifx\pdfoutput\undefined
\usepackage[dvipdfmx]{graphicx}
\usepackage[dvipdfmx]{hyperref}
\usepackage[dvipdfmx]{color}
\usepackage[dvipdfmx]{xcolor}
\else
\usepackage{graphicx}
\usepackage{hyperref}
\usepackage{color}
\usepackage{xcolor}
\usepackage[version=3]{mhchem}
\fi

\hypersetup{
        colorlinks=true,
        citecolor=blue,
        urlcolor=blue,
        linkcolor=blue
}

\begin{document}
\let\emph\textit

\title{
    Chiral Valence-Bond Solid and Chiral Triplons
}
\author{Shinnosuke Koyama}
\author{Kazumasa Hattori}
\affiliation{
Department of Physics, Tokyo Metropolitan University,
1-1, Minami-osawa, Hachioji, Tokyo 192-0397, Japan
}

\date{\today}
\begin{abstract}
We propose a chiral valence-bond solid  (VBS) in which spin-singlet bond order itself spontaneously acquires handedness without magnetic order. Using a spin-lattice model on the diamond lattice, we demonstrate two enantiomorphic VBS states. In the presence of anisotropic spin interactions, this emergent handedness is inherited by chiral triplons with spin-split bands and spin textures that reverse between the two enantiomers. These textures generate a temperature-gradient-induced net spin response whose sign is set by the VBS handedness. The imaginary off-diagonal components of the dynamical spin structure factor encode the triplon polarization and change sign between the two enantiomers, enabling direct detection of chiral triplons and the selected handedness by polarized inelastic neutron scattering. Our results establish a direct connection between spontaneous chirality and magnetic excitations, opening avenues for exploring the interplay between quantum magnetism and chiral physics.

\end{abstract}
\maketitle

Chirality refers to the property of an object or state
that cannot be superimposed on its mirror image
\cite{kelvin1894,barron2012}.
It appears across diverse areas of science,
including molecular enantiomers~\cite{pasteur1848,cahn1966},
biomolecular helices~\cite{watson1953},
parity-violating weak interactions~\cite{wu1957},
and crystals~\cite{flack2003,bousquet2025}.
This distinction between left and right gives rise to
characteristic physical responses
\cite{berova2007,gohler2011_CISS,bloom2024_CISS}.
From the viewpoint of symmetry breaking,
an achiral-to-chiral transition that
preserves time-reversal symmetry
can be characterized by
a time-reversal-even pseudoscalar
that is odd under improper spatial operations~\cite{barron2004}.
Within the multipole description,
this pseudoscalar corresponds to
the electric toroidal~(ET) monopole $G_0$
\cite{hayami2018_prb,kishine2022,kusunose2024_apl},
whose sign distinguishes the two enantiomers
\cite{oiwa2022,inda2024}.
Microscopically, such pseudoscalars can be realized by
lattice-displacement modes~\cite{song2016,romao2024,gomez2024,matsubara2026_arxiv}
or electronic degrees of freedom
\cite{kusunose2020_jpsj,hoshino2023,kusunose2024_apl,inda2024,oiwa2025,miki2025,ishitobi2026}.

Quantum magnets offer an unusual route to chirality.
In a valence-bond solid (VBS), spins form an ordered pattern of singlet dimers
that can break lattice symmetries without magnetic order~\cite{sachdev2008,lhuillier2001}.
This suggests that VBS order itself may become chiral:
a three-dimensional singlet-bond pattern can acquire handedness
while the state remains nonmagnetic and time-reversal symmetric.
Unlike magnetic chirality in
helically ordered moments~\cite{cheong2022} or scalar spin chirality~\cite{wen1989,taguchi2001},
this chirality resides in the spatial pattern of spin-singlet correlations.

Handedness can also be encoded in elementary excitations.
In chiral crystals, chiral phonons exhibit circular lattice motion
and carry mechanical angular momentum whose sign reverses between enantiomers
\cite{kishine2020,tsunetsugu2023,ishito2023,ueda2023,oishi2024,zhang2025_nature}.
This reversal links phonon dynamics to the handedness of the underlying crystal.
In a VBS phase, the elementary spin excitations are triplons~\cite{sachdev1990}.
Can triplons likewise inherit the handedness of the underlying bond order,
allowing their spin dynamics to provide a spectroscopic signature of the selected enantiomer?

\begin{figure}[b]
\includegraphics[width=0.9\columnwidth]{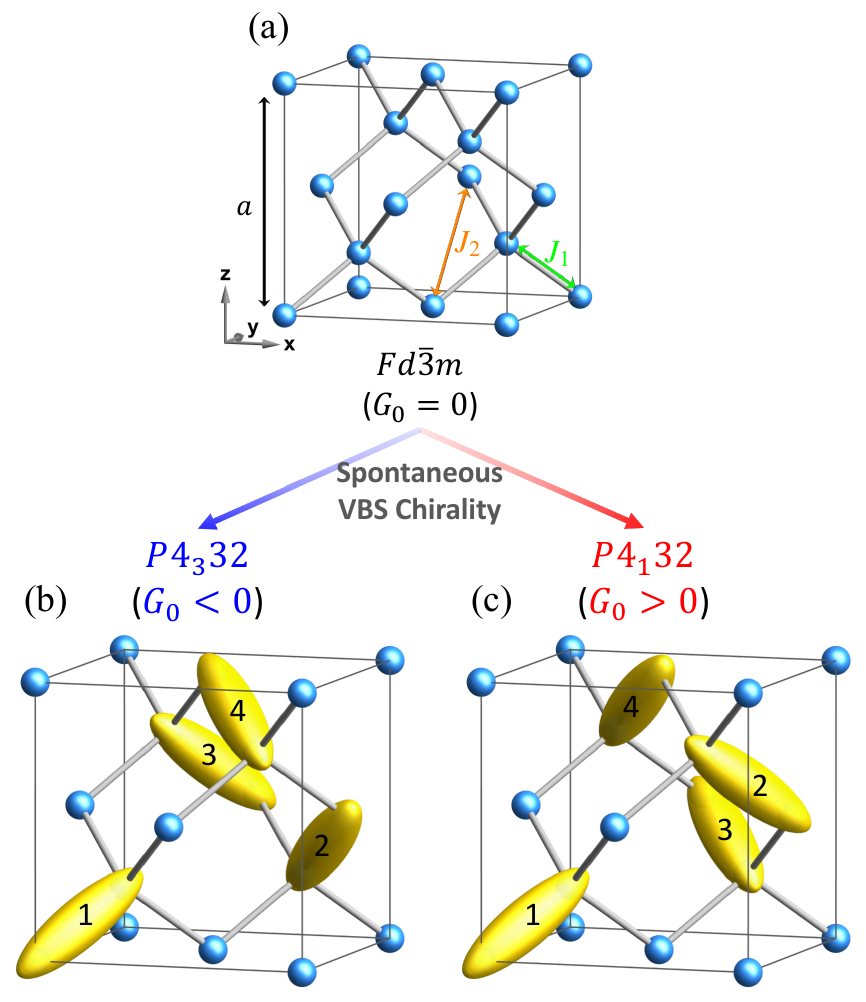}
\caption{
(a) Diamond-lattice cubic cell with
NN $J_1$ (green) and NNN $J_2$ (orange) interactions.
(b,c) Mirror-related chiral VBS patterns.
The yellow ellipsoids and numbers denote
the spin-singlet dimers and their sublattice labels, respectively.
The signs of $G_0$ follow the convention in
Eq.~\eqref{eq:G0_VBS}.
}
\label{fig:diamond_chiralVBS}
\end{figure}

In this Letter, we introduce \emph{chiral VBS} order,
in which spin-singlet bond order itself carries handedness.
We demonstrate its cubic realization on the diamond lattice
[Fig.~\ref{fig:diamond_chiralVBS}]
and the construction of $G_0$
from bond-order components.
We further show that the elementary excitations
inherit this handedness and exhibit spin-split bands.
We call these modes \emph{chiral triplons}.
Their antisymmetric dynamical spin correlations
reverse sign between the two enantiomers,
enabling polarized inelastic neutron scattering
to identify the selected VBS handedness directly
in the magnetic excitation spectrum.


We first illustrate chiral VBS order
using the centrosymmetric diamond lattice
with space group $Fd\bar{3}m$ [Fig.~\ref{fig:diamond_chiralVBS}(a)].
When spin-singlet dimers form an ordered pattern,
the system can spontaneously select either of
the two mirror-related configurations
shown in Figs.~\ref{fig:diamond_chiralVBS}(b) and~\ref{fig:diamond_chiralVBS}(c).
These states belong to the enantiomorphic space groups $P4_332$
and $P4_132$, respectively,
and break inversion and all mirror symmetries.
Their opposite handedness is represented
by opposite signs of the VBS ET monopole $G_0$,
defined below~[Eq.~\eqref{eq:G0_VBS}].

To realize this chiral VBS microscopically,
we consider a spin-lattice Hamiltonian
on the diamond lattice with
nearest-neighbor~(NN) and next-nearest-neighbor~(NNN) 
Heisenberg interactions~[Fig.~\ref{fig:diamond_chiralVBS}(a)].
For the lattice degrees of freedom,
we assume that the two-dimensional
${\rm X}_4$, ${\rm Y}_4$, and ${\rm Z}_4$
modes at the X, Y, and Z points
of the face-centered cubic~(fcc) Brillouin zone~(BZ) are soft.
These points form a star of $\bm{k}$ with
$\bm{k}_{\rm X}=2\pi(1,0,0)/a$,
$\bm{k}_{\rm Y}=2\pi(0,1,0)/a$, and
$\bm{k}_{\rm Z}=2\pi(0,0,1)/a$,
where $a$ is the cubic-cell lattice constant.
We treat the corresponding six normal coordinates
as independent classical variables.
Their triple-$q$ superpositions generate
the bond modulations underlying
the chiral VBS patterns in
Figs.~\ref{fig:diamond_chiralVBS}(b)
and~\ref{fig:diamond_chiralVBS}(c)~\cite{matsubara2026_arxiv,SM}.
The cubic cell shown in Fig.~\ref{fig:diamond_chiralVBS}(a)
is the smallest unit cell accommodating
all six modes.
Collecting the normal coordinates as
$\bm{Q} = (Q_{\rm X_4}^1, Q_{\rm X_4}^2,
Q_{\rm Y_4}^1, Q_{\rm Y_4}^2,
Q_{\rm Z_4}^1, Q_{\rm Z_4}^2)$
and choosing this basis to diagonalize the dynamical matrix,
we write the spin-lattice Hamiltonian as
\begin{align}
    \mathcal{H}(\bm{Q})
    =&
    \sum_{\braket{ij}}
    J_{1,ij}(\bm{Q}) \bm{S}_i \cdot \bm{S}_j
    +
    \sum_{\braket{\braket{ij}}}
    J_{2,ij}(\bm{Q}) \bm{S}_i \cdot \bm{S}_j\notag\\
    &+
    \frac{N_{\rm cell} K_{\rm X_4}}{2} \bm{Q}^2,
    \label{eq:H(Q)}
\end{align}
where $\braket{ij}$ and $\braket{\braket{ij}}$
denote the NN and NNN pairs, respectively, and
$J_{1,ij}(\bm{Q})$ and $J_{2,ij}(\bm{Q})$
are the corresponding Heisenberg interactions.
See  the Supplemental Material~(SM)~\cite{SM}.
The operator $\bm{S}_i$ describes an $S=1/2$ spin at site
$i$.
The last term is the elastic energy, where
$N_{\rm cell}$ is the number of cubic unit cells and
$K_{\rm X_4}$ is the force constant of the ${\rm X}_4$ mode.
Cubic symmetry gives
$K_{\rm Y_4}=K_{\rm Z_4}=K_{\rm X_4}$.

We determine the $\bm{Q}$ dependence of
the exchange interactions by expanding them
to first order in the small displacements and imposing
the parent $Fd\bar{3}m$ symmetry.
The resulting symmetry constraints fix the bond dependence
of the otherwise arbitrary linear coefficients.
Thus, the exchange interactions can be written as 
\begin{align}
    \label{eq:J1ij}
    J_{n,ij}(\bm{Q})
    &= J_n +
    \lambda_n \sum_{\Lambda\in\{{\rm X_4,Y_4,Z_4}\}}\sum_{\mu=1}^2
    A_{n,ij}^{\Lambda\mu} Q_{\Lambda}^\mu,\quad (n=1,2),
\end{align}
where $A_{n,ij}^{\Lambda\mu}$ are
symmetry-determined expansion coefficients.
The zeroth-order terms are
$J_1=J_{1,ij}(\bm{0})$ and $J_2=J_{2,ij}(\bm{0})$~[Fig.~\ref{fig:diamond_chiralVBS}(a)].
We consider the case $J_1> 8J_2 > 0$,
where the incommensurate spiral orders
realized for $J_2>J_1/8$~\cite{bergman2007} are absent.
Assuming that the Heisenberg interactions
depend on the bond length as
$r^{-\eta}$, we parametrize both $\lambda_1$ and $\lambda_2$
by a single exponent $\eta$.
This gives
$\lambda_1 = J_1\eta / (\sqrt{6}\tau_1)$ and
$\lambda_2 = J_2\eta/ (2\tau_2)$,
where
$\tau_1 = \sqrt{3}a/4$ and $\tau_2 = a/\sqrt{2}$
are the NN and NNN bond lengths, respectively~\cite{SM}.

Substituting Eq.~\eqref{eq:J1ij}
into Eq.~\eqref{eq:H(Q)}
yields the NN bond-order parameters
coupled to $Q_{\Lambda}^\mu$.
We define the $\Lambda$ components of
the NN bond-order parameter by
\begin{align}
    \label{eq:Phi}
    \Phi_{\Lambda}^\mu =
    -\frac{1}{\sqrt{16}}\sum_{\braket{ij}}^{16} A_{1,ij}^{\Lambda\mu} \braket{\bm{S}_i \cdot \bm{S}_j},
\end{align}
where the sum runs over the
$16$ NN bonds in a cubic unit cell.
Thus, the VBS order parameter $\Phi_{\Lambda}^\mu$
projects the spatial pattern of NN bond strengths
onto the $\Lambda$ component.
Through the spin-lattice coupling,
a finite $\Phi_{\Lambda}^\mu$
induces the corresponding
lattice displacement $Q_\Lambda^\mu$.

Following the symmetry construction
for lattice-displacement modes
\cite{matsubara2026_arxiv},
we define the ET monopole characterizing
the chiral transitions to the enantiomorphic
$P4_332$ and $P4_132$ VBS states
in terms of the NN bond-order components as
\begin{align}
    \label{eq:G0_VBS}
    G_0 \propto \Phi_{\rm X_4}^1 \Phi_{\rm Y_4}^1 \Phi_{\rm Z_4}^1
    -
    \Phi_{\rm X_4}^2 \Phi_{\rm Y_4}^2 \Phi_{\rm Z_4}^2.
\end{align}
When the $\mu=1$ or $\mu=2$ bond-order components
form an equal-amplitude triple-$q$ structure,
the space-group symmetry is reduced to
$P4_132$ or $P4_332$, respectively.

We treat $\bm{Q}$ as
variational parameters and apply
two-site cluster mean-field theory~(CMFT)~\cite{barwolf2025,koyama2025}
to $\mathcal{H}(\bm{Q})$ in Eq.~\eqref{eq:H(Q)}.
The coverings of the eight-site cubic unit cell
by NN two-site clusters fall into
three symmetry-inequivalent classes.
We perform the CMFT calculation
for each class
and adopt the lowest-energy
solution as the ground state.
For $J_2=0.10J_1$,
close to the spiral instability
of the diamond-lattice
$J_1$-$J_2$ model~\cite{bergman2007},
we also consider competing
single-$q$ variational states~\cite{SM}.

\begin{figure}[t]
\includegraphics[width=1\columnwidth]{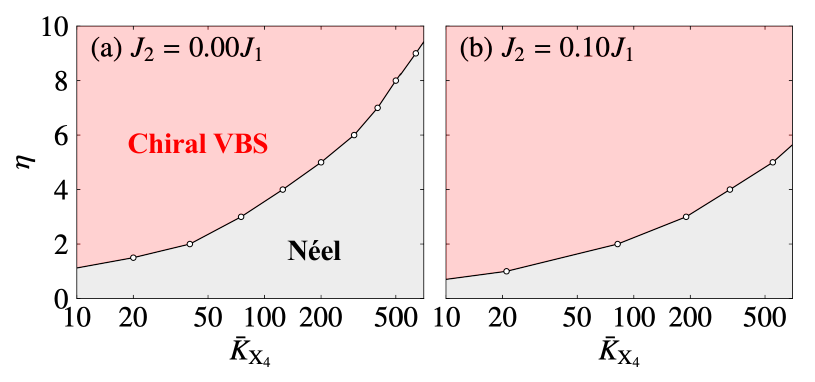}
\caption{
CMFT phase diagrams for (a) $J_{2} = 0.00 J_{1}$
and (b) $J_{2} = 0.10 J_{1}$.
Red and gray regions denote the triple-$q$ chiral VBS
and N\'eel phases, respectively.
}
\label{fig:phase diagram}
\end{figure}

Figures~\ref{fig:phase diagram}(a) and~\ref{fig:phase diagram}(b)
show the ground-state phase diagrams as functions of
$\bar{K}_{\rm X_4}=K_{\rm X_4}a^2/J_1$ and $\eta$.
The chiral VBS region with finite $G_0$ in Eq.~\eqref{eq:G0_VBS}
expands for $J_2/J_1 =0.10$,
indicating that the NNN interaction suppresses
the N\'eel order and stabilizes the
chiral VBS phase.
For
$\bm{\Phi}=(\Phi_{\rm X_4}^1, \Phi_{\rm X_4}^2,
\Phi_{\rm Y_4}^1, \Phi_{\rm Y_4}^2,
\Phi_{\rm Z_4}^1, \Phi_{\rm Z_4}^2)
$,
the corresponding solutions are of triple-$q$ type
$\bm{\Phi}_{P4_332}=\Phi(0,1,0,1,0,1)$
and
$\bm{\Phi}_{P4_132}=\Phi(1,0,1,0,1,0)$,
with $\Phi > 0$,
as illustrated in
Figs.~\ref{fig:diamond_chiralVBS}(b)
and~\ref{fig:diamond_chiralVBS}(c), respectively.
Thus, the chiral VBS phase spontaneously develops
a finite $G_0$ and selects one of the two enantiomers.
The chiral VBS bond patterns
obtained here are closely related to
the $R$ state of the
3:1 magnetization plateau in pyrochlore antiferromagnets
\cite{bergman2006_prl,bergman2006_prb1,bergman2006_prb2,sikora2009,sikora2011}
through a dual relation between
the pyrochlore and diamond lattices.
In contrast, the present nonmagnetic VBS
enables us to explore the physics governed purely
by the ET monopole.

We next introduce the Dzyaloshinskii-Moriya interaction~(DMI)
\cite{dzyaloshinskii1958,moriya1960},
arising from spin-orbit coupling~(SOC),
to study elementary excitations in the chiral VBS state.
The DMI was omitted in the phase-diagram calculation
to emphasize that the chiral VBS order itself
is driven by the Heisenberg spin-lattice
model in Eq.~\eqref{eq:H(Q)}.
Once the VBS-induced displacement becomes finite,
however, inversion symmetry
at all NN-bond centers is broken,
making a DMI linear in $Q_{\Lambda}^{\mu}$
symmetry-allowed~\cite{moriya1960,sergienko2006}.
This DMI
does not drive the chiral VBS order
but reveals the spin-polarization structure
of elementary excitations.
We repeat the CMFT calculation
for the Hamiltonian including the DMI
and use the resulting local ground and excited states
to analyze these excitations.

The DMI part of the spin-lattice Hamiltonian
is written as
\begin{align}
    \mathcal{H}_{\rm DM}(\bm{Q})
    =
    \sum_{\braket{ij}}
    \bm{D}_{ij}(\bm{Q})\cdot
    (\bm{S}_i \times \bm{S}_j),
\end{align}
where
$\bm{D}_{ij}(\bm{Q})$
denotes the DMI vector on the NN bond
and satisfies
$\bm{D}_{ij}(\bm{Q}) = -\bm{D}_{ji}(\bm{Q})$.
To first order in $\bm{Q}$,
it is given by
\begin{align}
    \bm{D}_{ij}(\bm{Q})
    &=
    \sum_{\Lambda\in\{\mathrm{X}_4,\mathrm{Y}_4,\mathrm{Z}_4\}}
    \sum_{\mu=1}^{2}
    \left(
    g_1\bm{B}_{1,ij}^{\Lambda\mu}
    +
    g_2\bm{B}_{2,ij}^{\Lambda\mu}
    \right)
    Q_{\Lambda}^{\mu},
\end{align}
where
$\bm{B}_{n,ij}^{\Lambda\mu}$
are vector coefficients allowed
by diamond-lattice symmetry~\cite{SM}.
We denote the lattice displacements
associated with the $P4_332$ and $P4_132$ VBS states by
$\bm{Q}_{P4_332}$ and $\bm{Q}_{P4_132}$, respectively.
These finite displacements
yield $\bm{D}_{ij}(\bm{Q})\neq \bm{0}$,
whereas $\bm{D}_{ij}(\bm{0})=\bm{0}$.
The total Hamiltonian is
$\mathcal{H}_{\rm tot}(\bm{Q}) =
\mathcal{H}(\bm{Q}) + \mathcal{H}_{\rm DM}(\bm{Q})$.
For a representative parameter set  $J_2=0.10J_1,\eta=6, \bar{K}_{\rm X_4} = 600$,
and
$g_1 = g_2 = 0.2 \lambda_1$, we have confirmed that
the chiral VBS phase remains stable,
although the DMI induces singlet--triplet mixing
in the local ground states.

We analyze the elementary excitations
in the chiral VBS state
using linear flavor-wave theory~(LFWT)
\cite{joshi1999,shiina2003,nasu2021,koyama2025,SM}.
We denote the local mean-field ground state
and three local excited states of a two-site cluster by
$\ket{l;0}$ and $\ket{l;\rho}$~($\rho=1,2,3$), respectively.
The cluster label is $d=(R,l)$,
where $R$ is the cubic-unit-cell label
and $l=1,2,3,4$ labels the dimer sublattice
shown in Figs.~\ref{fig:diamond_chiralVBS}(b) and~\ref{fig:diamond_chiralVBS}(c).
Within LFWT, the spin operator
$S_{d,c}^\alpha$ at site $c=1,2$ in cluster $d$
is expressed as
$S_{d,c}^\alpha \simeq
\sum_{\rho=1}^3\braket{l;\rho|S^\alpha_{d,c}|l;0} a_{d,\rho}^\dagger + \mathrm{H.c.}$,
where
$a_{d,\rho}^\dagger$ is a boson creation operator
for the $\rho$th local excitation of cluster $d$.
We refer to the resulting collective excitations as triplons.
With the saddle-point displacement
$\bm{Q}_0=\bm{Q}_{P4_332}$ or $\bm{Q}_{P4_132}$ fixed,
LFWT yields the quadratic bosonic
Hamiltonian
$\mathcal{H}_{\rm tot}(\bm{Q}_0) \simeq \mathcal{H}_{\rm FW}(\bm{Q}_0)$.
The triplon dispersions are obtained
by diagonalizing $\mathcal{H}_{\rm FW}$
using the Bogoliubov transformation~\cite{colpa}
and the $n$th energy at the momentum $\bm{k}$ is denoted as $\varepsilon_{\bm{k},n}$.
We also denote the Bogoliubov vacuum and the $n$th one-triplon state at $\bm{k}$ by
$|0\rangle\rangle$ and  $|\bm{k},n\rangle\rangle$, respectively.

\begin{figure}[t]
\includegraphics[width=1.0\columnwidth]{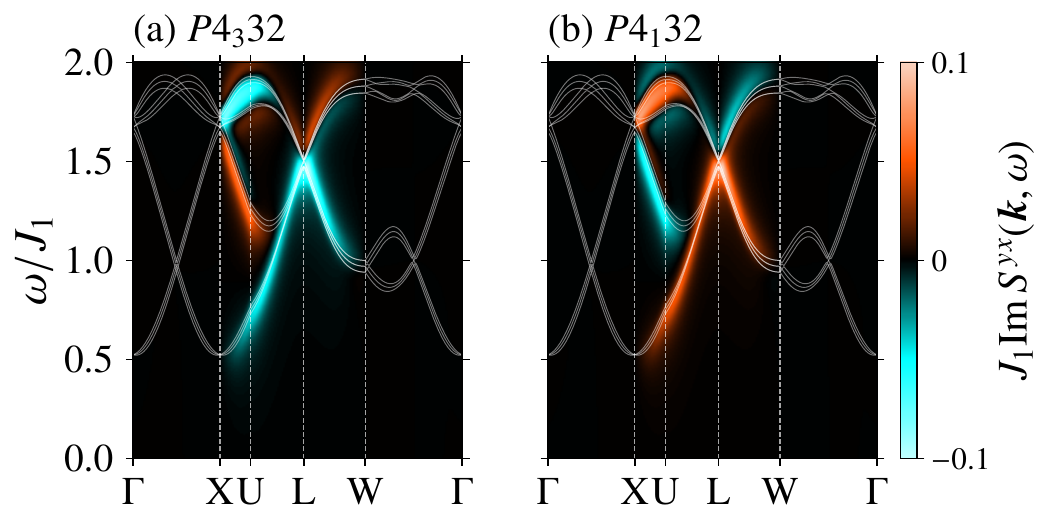}
\caption{
Color maps of the 
dynamical spin structure factor
$\mathrm{Im} S^{yx}(\bm{k},\omega)$
for (a) $P4_332$ and (b) $P4_132$.
The spectra are broadened using
a Lorentzian of width $0.025 J_1$.
}
\label{fig:ImSyx}
\end{figure}

To characterize the internal polarization of triplons,
we introduce the transition spin amplitude
$
F_{lc,n}^{\alpha}(\bm{k})
=
\langle\langle 0|
S_{\bm{k},lc}^{\alpha}
|\bm{k},n\rangle\rangle,
$
where $S_{\bm{k},lc}^{\alpha}$ is the Fourier component
of $S_{d,c}^{\alpha}$.
The corresponding real-time spin oscillation is
$
\delta\bm{S}_{lc,n}(\bm{k},t)=
{\rm Re}
[
\bm{F}_{lc,n}(\bm{k})
e^{-i\varepsilon_{\bm{k},n}t/\hbar}
].
$
We characterize its local polarization by
\begin{align}
    \label{eq:chi_lc}
    \bm{\chi}_{lc,n}(\bm{k})
    =
    -\frac{1}{2}{\rm Im}
    [
    \bm{F}_{lc,n}(\bm{k})
    \times
    \bm{F}_{lc,n}^{*}(\bm{k})
    ].
\end{align}
Writing $\bm{F}_{lc,n}=\bm{A}_{lc,n}+i\bm{B}_{lc,n}$
with real $\bm{A}_{lc,n}$ and $\bm{B}_{lc,n}$
gives
$\bm{\chi}_{lc,n}=\bm{A}_{lc,n}\times\bm{B}_{lc,n}$,
which specifies the axis and sense of rotation in internal spin space~\cite{SM}.
A nonzero $\bm{\chi}_{lc,n}$ indicates that the triplon is
circularly or elliptically polarized.
This is analogous to circularly polarized phonons,
for which complex phonon eigenvectors describe local circular motion and give rise to phonon angular momentum~\cite{zhang2014,zhang2015,zhang2022}.
See also the discussion accompanying Fig.~\ref{fig:ImSyx}.

\begin{figure}[t]
\includegraphics[width=1.0\columnwidth]{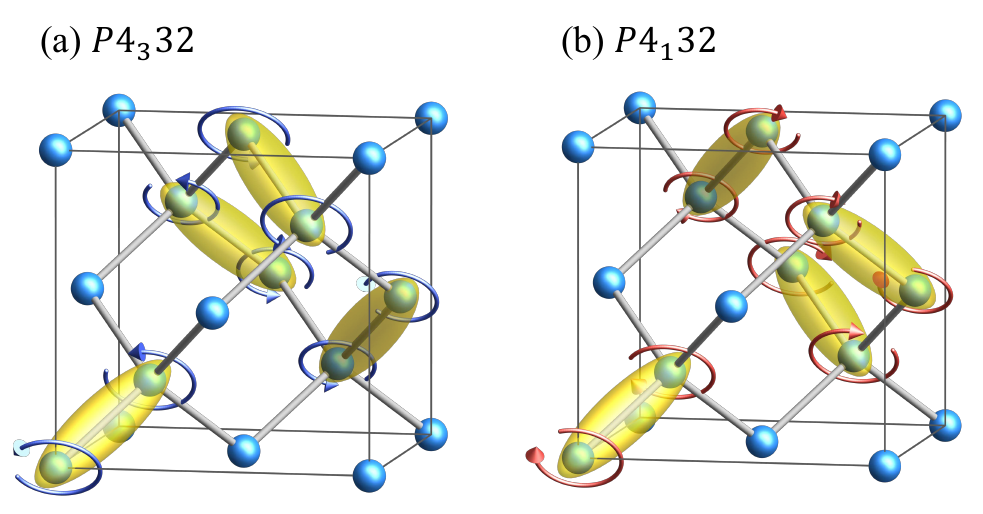}
\caption{
Spin motion of the lowest triplon band for
(a) $P4_332$ and (b) $P4_132$
at the U point in the fcc BZ, $\bm{k}_{\rm U} = \pi(2,0.5,0.5)/a$.
The red and blue arrows represent 
$\chi^z_{lc,n}(\bm{k}_{\rm U})>0$ and $\chi^z_{lc,n}(\bm{k}_{\rm U})<0$, respectively.
}
\label{fig:spin_motion_U}
\end{figure}

Figures~\ref{fig:spin_motion_U}(a) and~\ref{fig:spin_motion_U}(b)
show the spin motion of the
lowest triplon band at the U point in the fcc BZ,
${\bm k}_{\rm U} =\pi(2,0.5,0.5)/a$.
Since $k_{{\rm U},y} = k_{{\rm U},z}$,
the mirror operation
$\sigma_{yz}: (x,y,z) \mapsto (x,z,y)$
belongs to the little group of $\bm{k}_{\rm U}$
in the parent $Fd\bar{3}m$ phase
and maps the two enantiomers onto each other
[Figs.~\ref{fig:diamond_chiralVBS}(b) and \ref{fig:diamond_chiralVBS}(c)]. 
Accordingly, spin rotations on dimers
related by $\sigma_{yz}$ have opposite senses in the two enantiomers.

The quantity $\bm{\chi}_{lc,n}(\bm{k})$ characterizes
the site-resolved internal spin polarization and
is closely related to the spin carried by a triplon.
To establish this relation,
we project the spin at each site of dimer $d$
onto the local excited states:
\begin{align}
    \bm{\mathcal{S}}_{d,c}
    \equiv
    P_{d}^{\rm ex}\bm{S}_{d,c} P_d^{\rm ex}
    =
    \sum_{\rho,\rho'=1}^3
    \braket{l;\rho|\bm{S}_{d,c}|l;\rho'} a_{d,\rho}^\dagger a_{d,\rho'},
\end{align}
where
$P_d^{\rm ex}$ is the projection operator
onto the local excited-state
subspace $\rho=1,2,3$.
Its site-resolved expectation value in a one-triplon state
is
$\bm{s}_{lc,n}(\bm{k})
=
\braket{\braket{
\bm{k},n|
\sum_R\bm{\mathcal{S}}_{R,lc}
|\bm{k},n
}}$,
where $\bm{\mathcal{S}}_{R,lc} = \bm{\mathcal{S}}_{d,c}$
for $d=(R,l)$.
The projected total spin $\bm{\mathcal{S}}_d = \bm{\mathcal{S}}_{d,1} + \bm{\mathcal{S}}_{d,2}$
is the natural local spin operator
in the triplon sector~\cite{thomasen2021,koyama2025,esaki2025}.
When the local mean-field Hamiltonian~[Eq.~(S31) in SM~\cite{SM}]
is time-reversal-symmetric and
its eigenstates can be chosen to form a
time-reversal-invariant basis,
each Cartesian component
$\mathcal{S}_{d,c}^{\alpha}$~($\alpha=x,y,z$)
is represented by a purely imaginary matrix.
Consequently,
in the boson-number-conserving case,
$\bm{s}_{lc,n}(\bm{k})=\bm{0}$
when the expansion coefficients of the one-triplon state
in this basis are real,
whereas a nonzero $\bm{s}_{lc,n}(\bm{k})$
requires relative phases among these coefficients.
In particular,
in the pure-singlet--triplet limit
with the boson number conserved,
the exact relation
$\bm{s}_{lc,n}(\bm{k}) = 4\bm{\chi}_{lc,n}(\bm{k})$ holds,
establishing their microscopic correspondence.
In general cases, this quantitative relation is modified,
but the two quantities remain similar measures
of the same internal spin polarization.

\begin{figure}[t]
\includegraphics[width=1.0\columnwidth]{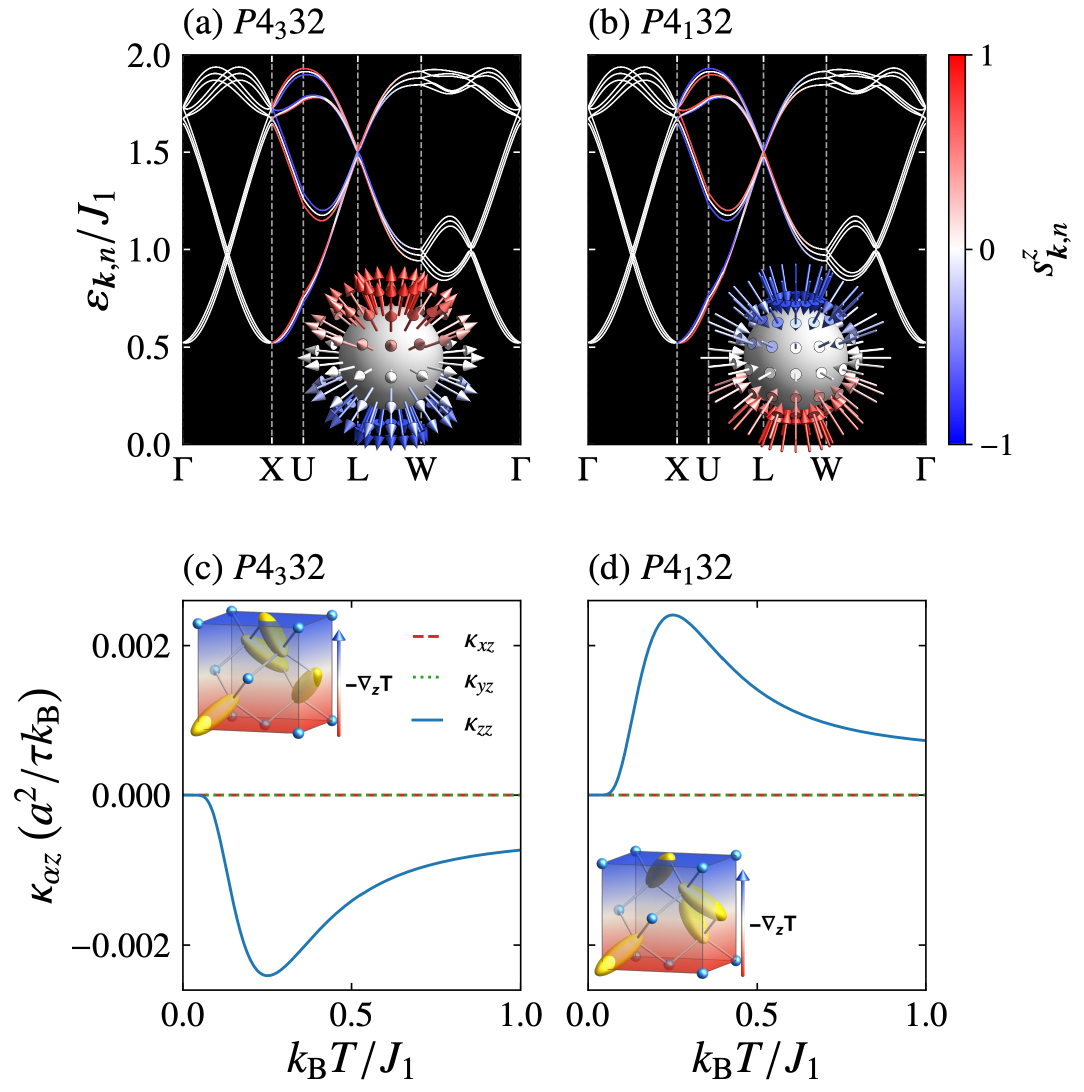}
\caption{
Triplon dispersions colored by $s_{\bm{k},n}^z$
for (a) $P4_332$ and (b) $P4_132$.
At degeneracies, $s_{\bm{k},n}^z$ is averaged over the
degenerate subspace.
The insets show $\bm{s}_{\bm{k},4}$ near
the $\mathrm{\Gamma}$ point.
(c,d) Temperature dependence of
$\kappa_{\alpha z}$~$(\alpha=x,y,z)$
for the corresponding enantiomers,
where $\tau$ is the relaxation time.
}
\label{fig:Sz and TIM}
\end{figure}

Figures~\ref{fig:Sz and TIM}(a) and \ref{fig:Sz and TIM}(b)
show the triplon dispersions for
the chiral VBS states with space groups
$P4_332$ and $P4_132$, respectively.
The color scale encodes $s^z_{\bm{k},n}$
for each triplon mode,
where
$s^\alpha_{\bm{k},n}=
\braket{\braket{
\bm{k},n|\mathcal{S}^\alpha_{\rm tot}
|\bm{k},n}}$
and
$\mathcal{S}^\alpha_{\rm tot}=
\sum_d \mathcal{S}^\alpha_d$.
The sign of $s^z_{\bm{k},n}$ reverses
between the two enantiomers
and vanishes
along $k_z = 0$ paths~($\mathrm{\Gamma}$--X and W--$\mathrm{\Gamma}$).
Together with the $s^x_{\bm{k},n}$
and $s^y_{\bm{k},n}$ maps in the SM~\cite{SM},
these results reveal
hedgehog-like spin textures
in the triplon bands.
The insets directly visualize such spin textures,
suggesting an effective coupling
$G_0 \bm{k} \cdot \bm{\mathcal{S}}_{\rm tot}$.
This is the triplon analogue
of chiral SOC in electronic~\cite{hirayama2015,sakano2020} and phonon~\cite{tsunetsugu2026} states.
We refer to such a mode as a chiral triplon.

These spin textures enable
a thermal Edelstein response,
in which a temperature gradient $\nabla T$
generates nonequilibrium spin angular momentum
\cite{hamada2018,li2020,yao2025,neumann2026_arxiv}.
We define it by
$\hbar \braket{\mathcal{S}^\alpha_{\rm tot}}_{\nabla T} / V
= \kappa_{\alpha\beta} (-\nabla_\beta T)$,
where
$V$ is the volume and
$\braket{\cdots}_{\nabla T}$
denotes the thermal average
under $\nabla T$.
We evaluate $\kappa_{\alpha\beta}$ using the Kubo formula~\cite{kubo1957,luttinger1964,li2020};
its explicit expression is given in
the SM~\cite{SM}.
For $G_0\ne 0$, 
symmetry allows
$\kappa_{\alpha\beta} \propto G_0 \delta_{\alpha\beta}$~\cite{hayami2018_prb}.
Consistently, Figs.~\ref{fig:Sz and TIM}(c)
and~\ref{fig:Sz and TIM}(d) show
$\kappa_{xz}=\kappa_{yz}=0$ and a finite
$\kappa_{zz}$ with opposite signs
for the two enantiomers.
We have also confirmed numerically that
$\kappa_{xx}=\kappa_{yy}=\kappa_{zz}$.

We finally discuss
the experimental detection
of chiral triplons through
polarized inelastic neutron scattering~\cite{roessli2002,lorenzo2007,nambu2020},
which can probe the antisymmetric off-diagonal part
of the dynamical spin structure factor $S^{\alpha\beta}(\bm{k},\omega)$
\cite{knetter2004,lohofer2015,martinez2026_arxiv}.
As a tensor in the spin indices,
$S^{\alpha\beta}(\bm{k},\omega)$
transforms as the direct product of two axial
vectors under the point group O$_h$:
${\rm T}_{1g} \otimes {\rm T}_{1g} = {\rm A}_{1g} \oplus {\rm E}_g \oplus {\rm T}_{1g}
\oplus {\rm T}_{2g}$.
The T$_{1g}$ component is
the antisymmetric off-diagonal part,
which can be represented by
the imaginary off-diagonal components as
$\bm{S}_{{\rm T}_{1g}}(\bm{k},\omega)
=[
\mathrm{Im}S^{zy}(\bm{k},\omega),
\mathrm{Im}S^{xz}(\bm{k},\omega),
\mathrm{Im}S^{yx}(\bm{k},\omega)
]$.
At lowest order in $\bm{k}$,
symmetry allows $\bm{S}_{{\rm T}_{1g}}(\bm{k},\omega) \propto G_0 \bm{k}$.
Figures~\ref{fig:ImSyx}(a) and~\ref{fig:ImSyx}(b)
show $\mathrm{Im} S^{yx}(\bm{k},\omega)$
in the chiral VBS phase with $P4_332$ and $P4_132$, respectively.
The spectral sign structure reverses
between the two enantiomers,
reflecting the sign reversal of $G_0$.
To relate $\bm{S}_{{\rm T}_{1g}}$
to the internal polarization of chiral triplons,
we define the transition spin amplitude
summed over the cubic unit cell,
$\bm{F}_{n}(\bm{k}) = \sum_{l,c} \bm{F}_{lc,n}(\bm{k})$
and the corresponding polarization,
$
\bm{\chi}_{n}(\bm{k})
=
-\frac{1}{2}{\rm Im}
[\bm{F}_{n}(\bm{k})
\times
\bm{F}_{n}^{*}(\bm{k})]
$.
Unlike the site-resolved polarization
$\bm{\chi}_{lc,n}(\bm{k})$
defined in Eq.~\eqref{eq:chi_lc},
$\bm{\chi}_{n}(\bm{k})$
characterizes the polarization of the total
spin amplitude $\bm{F}_{n}(\bm{k})$.
It is directly related to the
antisymmetric dynamical spin structure factor through the exact relation
$
8\bm{S}_{{\rm T}_{1g}}(\bm{k},\omega)=
\sum_{n}\bm{\chi}_n(\bm{k}) \delta(\omega - \varepsilon_{\bm{k},n})
$~\cite{SM}.
Thus,
a finite $\bm{S}_{{\rm T}_{1g}}(\bm{k},\omega)$
directly signals a nonzero $\bm{\chi}_{n}(\bm{k})$
and provides a smoking-gun signature
of chiral triplons.
The enantiomer-dependent spectral sign structure
also allows the selected handedness to be identified.

We emphasize that chiral VBS states as demonstrated in the diamond model in this Letter
can occur in any dimer coverings breaking all improper spatial symmetries 
and need not arise from a cubic-to-cubic chiral transition.
By analogy with ET monopoles constructed from
lattice-displacement modes~\cite{matsubara2026_arxiv},
corresponding ET monopoles
can be formed from symmetry-adapted combinations
of bond-order patterns.
Related helical arrangements of
spin-singlet dimers have been discussed
within a coupled spin-orbital-lattice order~\cite{matteo2005}.
The mechanism for chiral triplons is likewise
not specific to triple-$q$ VBS order.
In a dimerized quantum magnet,
either chiral VBS order or a chiral lattice distortion,
in the presence of DMI and/or other SOC terms,
can produce internally spin-polarized triplons.
Spin-Peierls systems provide a natural setting
for exploring these phenomena,
as spin-lattice coupling drives spin-singlet bond order.

Our findings motivate reexamining known materials
for previously unrecognized chiral properties
of their ordered states and magnetic excitations.
The organic ionic Mott insulator TTF--BA provides an example of
a VBS transition into a chiral structure~\cite{garcia2005}. 
GaNb$_4$S$_8$ and GaTa$_4$Se$_8$
are further candidates for chiral VBS order:
their low-temperature phases are chiral~\cite{geirhos2021,langmann2024}
with VBS order~\cite{waki2010,ishikawa2020,yang2022}.
Magnetic excitation gaps have been reported or inferred
in all three~\cite{sunami2020,waki2010,ishikawa2020},
making them potential platforms for chiral triplons
or related spin-orbital excitations.
In $\alpha'$-NaV$_2$O$_5$~\cite{sawa2002,ohwada2005,seo1998,mostovoy1999},
GaNb$_4$Se$_8$~\cite{yang2024}, and
the organic Mott insulator (TMTTF)$_2$PF$_6$~\cite{kitou2017},
VBS order develops on a chiral background,
rather than generating the chirality itself.
In particular,
triplet excitations have been observed
in $\alpha'$-NaV$_2$O$_5$~\cite{luther1998,yoshihama1998,grenier2001,room2004}
and (TMTTF)$_2$PF$_6$~\cite{dumm2000,leylekian2009,pouget2017}.
These materials are therefore interesting candidates
for chiral triplons in the presence of
anisotropic spin interactions~\cite{furukawa2009,yasin2012,luther1998,room2004,eremin2006,ishikawa2020}.

In summary, we have demonstrated a triple-$q$ chiral VBS state on the diamond lattice,
in which VBS order induces an ET monopole and spontaneously selects one of two enantiomorphic space groups while preserving cubic rotational symmetry.
In the presence of DMI, this emergent handedness is inherited by triplon modes through their circular or elliptical spin polarization and imprinted on their antisymmetric dynamical spin correlations.
The reversal of these correlations between the two enantiomers makes the selected VBS chirality directly observable in the magnetic excitation spectrum through polarized inelastic neutron scattering experiments.
Our results establish VBS order as a source of chirality
and link its handedness to magnetic excitation spectra,
offering opportunities for further study
in quantum magnetism and chiral physics.

The authors thank
T. Sato,
D. Sasamoto,
and
J. Nasu
for helpful discussions.
This work was supported by Grant-in-Aid
for Scientific Research from JSPS,
KAKENHI Grant
Nos. JP23H04866 and JP23H04869.

\bibliography{./refs}

\begin{thebibliography}{98}%
\makeatletter
\providecommand \@ifxundefined [1]{%
 \@ifx{#1\undefined}
}%
\providecommand \@ifnum [1]{%
 \ifnum #1\expandafter \@firstoftwo
 \else \expandafter \@secondoftwo
 \fi
}%
\providecommand \@ifx [1]{%
 \ifx #1\expandafter \@firstoftwo
 \else \expandafter \@secondoftwo
 \fi
}%
\providecommand \natexlab [1]{#1}%
\providecommand \enquote  [1]{``#1''}%
\providecommand \bibnamefont  [1]{#1}%
\providecommand \bibfnamefont [1]{#1}%
\providecommand \citenamefont [1]{#1}%
\providecommand \href@noop [0]{\@secondoftwo}%
\providecommand \href [0]{\begingroup \@sanitize@url \@href}%
\providecommand \@href[1]{\@@startlink{#1}\@@href}%
\providecommand \@@href[1]{\endgroup#1\@@endlink}%
\providecommand \@sanitize@url [0]{\catcode `\\12\catcode `\$12\catcode `\&12\catcode `\#12\catcode `\^12\catcode `\_12\catcode `\%12\relax}%
\providecommand \@@startlink[1]{}%
\providecommand \@@endlink[0]{}%
\providecommand \url  [0]{\begingroup\@sanitize@url \@url }%
\providecommand \@url [1]{\endgroup\@href {#1}{\urlprefix }}%
\providecommand \urlprefix  [0]{URL }%
\providecommand \Eprint [0]{\href }%
\providecommand \doibase [0]{https://doi.org/}%
\providecommand \selectlanguage [0]{\@gobble}%
\providecommand \bibinfo  [0]{\@secondoftwo}%
\providecommand \bibfield  [0]{\@secondoftwo}%
\providecommand \translation [1]{[#1]}%
\providecommand \BibitemOpen [0]{}%
\providecommand \bibitemStop [0]{}%
\providecommand \bibitemNoStop [0]{.\EOS\space}%
\providecommand \EOS [0]{\spacefactor3000\relax}%
\providecommand \BibitemShut  [1]{\csname bibitem#1\endcsname}%
\let\auto@bib@innerbib\@empty
\bibitem [{\citenamefont {Kelvin}(1894)}]{kelvin1894}%
  \BibitemOpen
  \bibfield  {author} {\bibinfo {author} {\bibfnamefont {W.~T.}\ \bibnamefont {Kelvin}},\ }\href@noop {} {\emph {\bibinfo {title} {{The Molecular Tactics of a Crystal}}}}\ (\bibinfo  {publisher} {Clarendon Press},\ \bibinfo {address} {Oxford},\ \bibinfo {year} {1894})\BibitemShut {NoStop}%
\bibitem [{\citenamefont {Barron}(2012)}]{barron2012}%
  \BibitemOpen
  \bibfield  {author} {\bibinfo {author} {\bibfnamefont {L.~D.}\ \bibnamefont {Barron}},\ }\bibfield  {title} {\bibinfo {title} {{From Cosmic Chirality to Protein Structure: Lord Kelvin's Legacy}},\ }\href {https://doi.org/https://doi.org/10.1002/chir.22017} {\bibfield  {journal} {\bibinfo  {journal} {Chirality}\ }\textbf {\bibinfo {volume} {24}},\ \bibinfo {pages} {879} (\bibinfo {year} {2012})}\BibitemShut {NoStop}%
\bibitem [{\citenamefont {Pasteur}(1848)}]{pasteur1848}%
  \BibitemOpen
  \bibfield  {author} {\bibinfo {author} {\bibfnamefont {L.}~\bibnamefont {Pasteur}},\ }\bibfield  {title} {\bibinfo {title} {{Recherches sur les relations qui peuvent exister entre la forme cristalline, la composition chimique et le sens de la polarisation rotatoire}},\ }\href@noop {} {\bibfield  {journal} {\bibinfo  {journal} {Ann. Chim. Phys.}\ }\textbf {\bibinfo {volume} {24}},\ \bibinfo {pages} {442} (\bibinfo {year} {1848})}\BibitemShut {NoStop}%
\bibitem [{\citenamefont {Cahn}\ \emph {et~al.}(1966)\citenamefont {Cahn}, \citenamefont {Ingold},\ and\ \citenamefont {Prelog}}]{cahn1966}%
  \BibitemOpen
  \bibfield  {author} {\bibinfo {author} {\bibfnamefont {R.~S.}\ \bibnamefont {Cahn}}, \bibinfo {author} {\bibfnamefont {C.}~\bibnamefont {Ingold}},\ and\ \bibinfo {author} {\bibfnamefont {V.}~\bibnamefont {Prelog}},\ }\bibfield  {title} {\bibinfo {title} {{Specification of Molecular Chirality}},\ }\href {https://doi.org/https://doi.org/10.1002/anie.196603851} {\bibfield  {journal} {\bibinfo  {journal} {{Angew. Chem. Int. Ed. Engl.}}\ }\textbf {\bibinfo {volume} {5}},\ \bibinfo {pages} {385} (\bibinfo {year} {1966})}\BibitemShut {NoStop}%
\bibitem [{\citenamefont {Watson}\ and\ \citenamefont {Crick}(1953)}]{watson1953}%
  \BibitemOpen
  \bibfield  {author} {\bibinfo {author} {\bibfnamefont {J.}~\bibnamefont {Watson}}\ and\ \bibinfo {author} {\bibfnamefont {F.}~\bibnamefont {Crick}},\ }\bibfield  {title} {\bibinfo {title} {{Molecular Structure of Nucleic Acids: A Structure for Deoxyribose Nucleic Acid}},\ }\href {https://doi.org/10.1038/171737a0} {\bibfield  {journal} {\bibinfo  {journal} {Nature}\ }\textbf {\bibinfo {volume} {171}},\ \bibinfo {pages} {737} (\bibinfo {year} {1953})}\BibitemShut {NoStop}%
\bibitem [{\citenamefont {Wu}\ \emph {et~al.}(1957)\citenamefont {Wu}, \citenamefont {Ambler}, \citenamefont {Hayward}, \citenamefont {Hoppes},\ and\ \citenamefont {Hudson}}]{wu1957}%
  \BibitemOpen
  \bibfield  {author} {\bibinfo {author} {\bibfnamefont {C.~S.}\ \bibnamefont {Wu}}, \bibinfo {author} {\bibfnamefont {E.}~\bibnamefont {Ambler}}, \bibinfo {author} {\bibfnamefont {R.~W.}\ \bibnamefont {Hayward}}, \bibinfo {author} {\bibfnamefont {D.~D.}\ \bibnamefont {Hoppes}},\ and\ \bibinfo {author} {\bibfnamefont {R.~P.}\ \bibnamefont {Hudson}},\ }\bibfield  {title} {\bibinfo {title} {{Experimental Test of Parity Conservation in Beta Decay}},\ }\href {https://doi.org/10.1103/PhysRev.105.1413} {\bibfield  {journal} {\bibinfo  {journal} {Phys. Rev.}\ }\textbf {\bibinfo {volume} {105}},\ \bibinfo {pages} {1413} (\bibinfo {year} {1957})}\BibitemShut {NoStop}%
\bibitem [{\citenamefont {Flack}(2003)}]{flack2003}%
  \BibitemOpen
  \bibfield  {author} {\bibinfo {author} {\bibfnamefont {H.~D.}\ \bibnamefont {Flack}},\ }\bibfield  {title} {\bibinfo {title} {{Chiral and Achiral Crystal Structures}},\ }\href {https://doi.org/https://doi.org/10.1002/hlca.200390109} {\bibfield  {journal} {\bibinfo  {journal} {Helv. Chim. Acta}\ }\textbf {\bibinfo {volume} {86}},\ \bibinfo {pages} {905} (\bibinfo {year} {2003})}\BibitemShut {NoStop}%
\bibitem [{\citenamefont {Bousquet}\ \emph {et~al.}(2025)\citenamefont {Bousquet}, \citenamefont {Fava}, \citenamefont {Romestan}, \citenamefont {G^^c3^^b3mez-Ortiz}, \citenamefont {McCabe},\ and\ \citenamefont {Romero}}]{bousquet2025}%
  \BibitemOpen
  \bibfield  {author} {\bibinfo {author} {\bibfnamefont {E.}~\bibnamefont {Bousquet}}, \bibinfo {author} {\bibfnamefont {M.}~\bibnamefont {Fava}}, \bibinfo {author} {\bibfnamefont {Z.}~\bibnamefont {Romestan}}, \bibinfo {author} {\bibfnamefont {F.}~\bibnamefont {G^^c3^^b3mez-Ortiz}}, \bibinfo {author} {\bibfnamefont {E.~E.}\ \bibnamefont {McCabe}},\ and\ \bibinfo {author} {\bibfnamefont {A.~H.}\ \bibnamefont {Romero}},\ }\bibfield  {title} {\bibinfo {title} {{Structural chirality and related properties in periodic inorganic solids: review and perspectives}},\ }\href {https://doi.org/10.1088/1361-648X/adb674} {\bibfield  {journal} {\bibinfo  {journal} {J. Phys.: Condens. Matter}\ }\textbf {\bibinfo {volume} {37}},\ \bibinfo {pages} {163004} (\bibinfo {year} {2025})}\BibitemShut {NoStop}%
\bibitem [{\citenamefont {Berova}\ \emph {et~al.}(2007)\citenamefont {Berova}, \citenamefont {Di~Bari},\ and\ \citenamefont {Pescitelli}}]{berova2007}%
  \BibitemOpen
  \bibfield  {author} {\bibinfo {author} {\bibfnamefont {N.}~\bibnamefont {Berova}}, \bibinfo {author} {\bibfnamefont {L.}~\bibnamefont {Di~Bari}},\ and\ \bibinfo {author} {\bibfnamefont {G.}~\bibnamefont {Pescitelli}},\ }\bibfield  {title} {\bibinfo {title} {{Application of electronic circular dichroism in configurational and conformational analysis of organic compounds}},\ }\href {https://doi.org/10.1039/B515476F} {\bibfield  {journal} {\bibinfo  {journal} {Chem. Soc. Rev.}\ }\textbf {\bibinfo {volume} {36}},\ \bibinfo {pages} {914} (\bibinfo {year} {2007})}\BibitemShut {NoStop}%
\bibitem [{\citenamefont {G{\"o}hler}\ \emph {et~al.}(2011)\citenamefont {G{\"o}hler}, \citenamefont {Hamelbeck}, \citenamefont {Markus}, \citenamefont {Kettner}, \citenamefont {Hanne}, \citenamefont {Vager}, \citenamefont {Naaman},\ and\ \citenamefont {Zacharias}}]{gohler2011_CISS}%
  \BibitemOpen
  \bibfield  {author} {\bibinfo {author} {\bibfnamefont {B.}~\bibnamefont {G{\"o}hler}}, \bibinfo {author} {\bibfnamefont {V.}~\bibnamefont {Hamelbeck}}, \bibinfo {author} {\bibfnamefont {T.~Z.}\ \bibnamefont {Markus}}, \bibinfo {author} {\bibfnamefont {M.}~\bibnamefont {Kettner}}, \bibinfo {author} {\bibfnamefont {G.~F.}\ \bibnamefont {Hanne}}, \bibinfo {author} {\bibfnamefont {Z.}~\bibnamefont {Vager}}, \bibinfo {author} {\bibfnamefont {R.}~\bibnamefont {Naaman}},\ and\ \bibinfo {author} {\bibfnamefont {H.}~\bibnamefont {Zacharias}},\ }\bibfield  {title} {\bibinfo {title} {{Spin Selectivity in Electron Transmission Through Self-Assembled Monolayers of Double-Stranded DNA}},\ }\href {https://doi.org/10.1126/science.1199339} {\bibfield  {journal} {\bibinfo  {journal} {Science}\ }\textbf {\bibinfo {volume} {331}},\ \bibinfo {pages} {894} (\bibinfo {year} {2011})}\BibitemShut {NoStop}%
\bibitem [{\citenamefont {Bloom}\ \emph {et~al.}(2024)\citenamefont {Bloom}, \citenamefont {Paltiel}, \citenamefont {Naaman},\ and\ \citenamefont {Waldeck}}]{bloom2024_CISS}%
  \BibitemOpen
  \bibfield  {author} {\bibinfo {author} {\bibfnamefont {B.~P.}\ \bibnamefont {Bloom}}, \bibinfo {author} {\bibfnamefont {Y.}~\bibnamefont {Paltiel}}, \bibinfo {author} {\bibfnamefont {R.}~\bibnamefont {Naaman}},\ and\ \bibinfo {author} {\bibfnamefont {D.~H.}\ \bibnamefont {Waldeck}},\ }\bibfield  {title} {\bibinfo {title} {{Chiral Induced Spin Selectivity}},\ }\href {https://doi.org/10.1021/acs.chemrev.3c00661} {\bibfield  {journal} {\bibinfo  {journal} {Chem. Rev.}\ }\textbf {\bibinfo {volume} {124}},\ \bibinfo {pages} {1950} (\bibinfo {year} {2024})}\BibitemShut {NoStop}%
\bibitem [{\citenamefont {Barron}(2004)}]{barron2004}%
  \BibitemOpen
  \bibfield  {author} {\bibinfo {author} {\bibfnamefont {L.~D.}\ \bibnamefont {Barron}},\ }\href@noop {} {\emph {\bibinfo {title} {{Molecular Light Scattering and Optical Activity}}}},\ \bibinfo {edition} {2nd}\ ed.\ (\bibinfo  {publisher} {Cambridge University Press},\ \bibinfo {address} {Cambridge},\ \bibinfo {year} {2004})\BibitemShut {NoStop}%
\bibitem [{\citenamefont {Hayami}\ \emph {et~al.}(2018)\citenamefont {Hayami}, \citenamefont {Yatsushiro}, \citenamefont {Yanagi},\ and\ \citenamefont {Kusunose}}]{hayami2018_prb}%
  \BibitemOpen
  \bibfield  {author} {\bibinfo {author} {\bibfnamefont {S.}~\bibnamefont {Hayami}}, \bibinfo {author} {\bibfnamefont {M.}~\bibnamefont {Yatsushiro}}, \bibinfo {author} {\bibfnamefont {Y.}~\bibnamefont {Yanagi}},\ and\ \bibinfo {author} {\bibfnamefont {H.}~\bibnamefont {Kusunose}},\ }\bibfield  {title} {\bibinfo {title} {{Classification of atomic-scale multipoles under crystallographic point groups and application to linear response tensors}},\ }\href {https://doi.org/10.1103/PhysRevB.98.165110} {\bibfield  {journal} {\bibinfo  {journal} {Phys. Rev. B}\ }\textbf {\bibinfo {volume} {98}},\ \bibinfo {pages} {165110} (\bibinfo {year} {2018})}\BibitemShut {NoStop}%
\bibitem [{\citenamefont {Kishine}\ \emph {et~al.}(2022)\citenamefont {Kishine}, \citenamefont {Kusunose},\ and\ \citenamefont {Yamamoto}}]{kishine2022}%
  \BibitemOpen
  \bibfield  {author} {\bibinfo {author} {\bibfnamefont {J.-i.}\ \bibnamefont {Kishine}}, \bibinfo {author} {\bibfnamefont {H.}~\bibnamefont {Kusunose}},\ and\ \bibinfo {author} {\bibfnamefont {H.~M.}\ \bibnamefont {Yamamoto}},\ }\bibfield  {title} {\bibinfo {title} {{On the Definition of Chirality and Enantioselective Fields}},\ }\href {https://doi.org/https://doi.org/10.1002/ijch.202200049} {\bibfield  {journal} {\bibinfo  {journal} {Isr. J. Chem.}\ }\textbf {\bibinfo {volume} {62}},\ \bibinfo {pages} {e202200049} (\bibinfo {year} {2022})}\BibitemShut {NoStop}%
\bibitem [{\citenamefont {Kusunose}\ \emph {et~al.}(2024)\citenamefont {Kusunose}, \citenamefont {Kishine},\ and\ \citenamefont {Yamamoto}}]{kusunose2024_apl}%
  \BibitemOpen
  \bibfield  {author} {\bibinfo {author} {\bibfnamefont {H.}~\bibnamefont {Kusunose}}, \bibinfo {author} {\bibfnamefont {J.-i.}\ \bibnamefont {Kishine}},\ and\ \bibinfo {author} {\bibfnamefont {H.~M.}\ \bibnamefont {Yamamoto}},\ }\bibfield  {title} {\bibinfo {title} {{Emergence of chirality from electron spins, physical fields, and material-field composites}},\ }\href {https://doi.org/10.1063/5.0214919} {\bibfield  {journal} {\bibinfo  {journal} {Appl. Phys. Lett.}\ }\textbf {\bibinfo {volume} {124}},\ \bibinfo {pages} {260501} (\bibinfo {year} {2024})}\BibitemShut {NoStop}%
\bibitem [{\citenamefont {Oiwa}\ and\ \citenamefont {Kusunose}(2022)}]{oiwa2022}%
  \BibitemOpen
  \bibfield  {author} {\bibinfo {author} {\bibfnamefont {R.}~\bibnamefont {Oiwa}}\ and\ \bibinfo {author} {\bibfnamefont {H.}~\bibnamefont {Kusunose}},\ }\bibfield  {title} {\bibinfo {title} {{Rotation, Electric-Field Responses, and Absolute Enantioselection in Chiral Crystals}},\ }\href {https://doi.org/10.1103/PhysRevLett.129.116401} {\bibfield  {journal} {\bibinfo  {journal} {Phys. Rev. Lett.}\ }\textbf {\bibinfo {volume} {129}},\ \bibinfo {pages} {116401} (\bibinfo {year} {2022})}\BibitemShut {NoStop}%
\bibitem [{\citenamefont {Inda}\ \emph {et~al.}(2024)\citenamefont {Inda}, \citenamefont {Oiwa}, \citenamefont {Hayami}, \citenamefont {Yamamoto},\ and\ \citenamefont {Kusunose}}]{inda2024}%
  \BibitemOpen
  \bibfield  {author} {\bibinfo {author} {\bibfnamefont {A.}~\bibnamefont {Inda}}, \bibinfo {author} {\bibfnamefont {R.}~\bibnamefont {Oiwa}}, \bibinfo {author} {\bibfnamefont {S.}~\bibnamefont {Hayami}}, \bibinfo {author} {\bibfnamefont {H.~M.}\ \bibnamefont {Yamamoto}},\ and\ \bibinfo {author} {\bibfnamefont {H.}~\bibnamefont {Kusunose}},\ }\bibfield  {title} {\bibinfo {title} {{Quantification of chirality based on electric toroidal monopole}},\ }\href {https://doi.org/10.1063/5.0204254} {\bibfield  {journal} {\bibinfo  {journal} {J. Chem. Phys.}\ }\textbf {\bibinfo {volume} {160}},\ \bibinfo {pages} {184117} (\bibinfo {year} {2024})}\BibitemShut {NoStop}%
\bibitem [{\citenamefont {Song}\ \emph {et~al.}(2016)\citenamefont {Song}, \citenamefont {Zhao}, \citenamefont {Fang},\ and\ \citenamefont {Dai}}]{song2016}%
  \BibitemOpen
  \bibfield  {author} {\bibinfo {author} {\bibfnamefont {Z.}~\bibnamefont {Song}}, \bibinfo {author} {\bibfnamefont {J.}~\bibnamefont {Zhao}}, \bibinfo {author} {\bibfnamefont {Z.}~\bibnamefont {Fang}},\ and\ \bibinfo {author} {\bibfnamefont {X.}~\bibnamefont {Dai}},\ }\bibfield  {title} {\bibinfo {title} {{Detecting the chiral magnetic effect by lattice dynamics in Weyl semimetals}},\ }\href {https://doi.org/10.1103/PhysRevB.94.214306} {\bibfield  {journal} {\bibinfo  {journal} {Phys. Rev. B}\ }\textbf {\bibinfo {volume} {94}},\ \bibinfo {pages} {214306} (\bibinfo {year} {2016})}\BibitemShut {NoStop}%
\bibitem [{\citenamefont {Romao}\ and\ \citenamefont {Juraschek}(2024)}]{romao2024}%
  \BibitemOpen
  \bibfield  {author} {\bibinfo {author} {\bibfnamefont {C.~P.}\ \bibnamefont {Romao}}\ and\ \bibinfo {author} {\bibfnamefont {D.~M.}\ \bibnamefont {Juraschek}},\ }\bibfield  {title} {\bibinfo {title} {{Phonon-Induced Geometric Chirality}},\ }\href {https://doi.org/10.1021/acsnano.4c05978} {\bibfield  {journal} {\bibinfo  {journal} {ACS Nano}\ }\textbf {\bibinfo {volume} {18}},\ \bibinfo {pages} {29550} (\bibinfo {year} {2024})}\BibitemShut {NoStop}%
\bibitem [{\citenamefont {G\'omez-Ortiz}\ \emph {et~al.}(2024)\citenamefont {G\'omez-Ortiz}, \citenamefont {Fava}, \citenamefont {McCabe}, \citenamefont {Romero},\ and\ \citenamefont {Bousquet}}]{gomez2024}%
  \BibitemOpen
  \bibfield  {author} {\bibinfo {author} {\bibfnamefont {F.}~\bibnamefont {G\'omez-Ortiz}}, \bibinfo {author} {\bibfnamefont {M.}~\bibnamefont {Fava}}, \bibinfo {author} {\bibfnamefont {E.~E.}\ \bibnamefont {McCabe}}, \bibinfo {author} {\bibfnamefont {A.~H.}\ \bibnamefont {Romero}},\ and\ \bibinfo {author} {\bibfnamefont {E.}~\bibnamefont {Bousquet}},\ }\bibfield  {title} {\bibinfo {title} {{Structural chirality measurements and computation of handedness in periodic solids}},\ }\href {https://doi.org/10.1103/PhysRevB.110.174112} {\bibfield  {journal} {\bibinfo  {journal} {Phys. Rev. B}\ }\textbf {\bibinfo {volume} {110}},\ \bibinfo {pages} {174112} (\bibinfo {year} {2024})}\BibitemShut {NoStop}%
\bibitem [{\citenamefont {Matsubara}\ and\ \citenamefont {Hattori}(shed)}]{matsubara2026_arxiv}%
  \BibitemOpen
  \bibfield  {author} {\bibinfo {author} {\bibfnamefont {K.}~\bibnamefont {Matsubara}}\ and\ \bibinfo {author} {\bibfnamefont {K.}~\bibnamefont {Hattori}},\ }\bibfield  {title} {\bibinfo {title} {{Order parameter scaling of chirality in structural phase transitions}},\ }\href {https://arxiv.org/abs/2605.27812} {\bibfield  {journal} {\bibinfo  {journal} {arXiv:2605.27812}\ } (\bibinfo {year} {unpublished})}\BibitemShut {NoStop}%
\bibitem [{\citenamefont {Kusunose}\ \emph {et~al.}(2020)\citenamefont {Kusunose}, \citenamefont {Oiwa},\ and\ \citenamefont {Hayami}}]{kusunose2020_jpsj}%
  \BibitemOpen
  \bibfield  {author} {\bibinfo {author} {\bibfnamefont {H.}~\bibnamefont {Kusunose}}, \bibinfo {author} {\bibfnamefont {R.}~\bibnamefont {Oiwa}},\ and\ \bibinfo {author} {\bibfnamefont {S.}~\bibnamefont {Hayami}},\ }\bibfield  {title} {\bibinfo {title} {{Complete Multipole Basis Set for Single-Centered Electron Systems}},\ }\href {https://doi.org/10.7566/JPSJ.89.104704} {\bibfield  {journal} {\bibinfo  {journal} {J. Phys. Soc. Jpn.}\ }\textbf {\bibinfo {volume} {89}},\ \bibinfo {pages} {104704} (\bibinfo {year} {2020})}\BibitemShut {NoStop}%
\bibitem [{\citenamefont {Hoshino}\ \emph {et~al.}(2023)\citenamefont {Hoshino}, \citenamefont {Suzuki},\ and\ \citenamefont {Ikeda}}]{hoshino2023}%
  \BibitemOpen
  \bibfield  {author} {\bibinfo {author} {\bibfnamefont {S.}~\bibnamefont {Hoshino}}, \bibinfo {author} {\bibfnamefont {M.-T.}\ \bibnamefont {Suzuki}},\ and\ \bibinfo {author} {\bibfnamefont {H.}~\bibnamefont {Ikeda}},\ }\bibfield  {title} {\bibinfo {title} {{Spin-Derived Electric Polarization and Chirality Density Inherent in Localized Electron Orbitals}},\ }\href {https://doi.org/10.1103/PhysRevLett.130.256801} {\bibfield  {journal} {\bibinfo  {journal} {Phys. Rev. Lett.}\ }\textbf {\bibinfo {volume} {130}},\ \bibinfo {pages} {256801} (\bibinfo {year} {2023})}\BibitemShut {NoStop}%
\bibitem [{\citenamefont {Oiwa}\ and\ \citenamefont {Kusunose}(2025)}]{oiwa2025}%
  \BibitemOpen
  \bibfield  {author} {\bibinfo {author} {\bibfnamefont {R.}~\bibnamefont {Oiwa}}\ and\ \bibinfo {author} {\bibfnamefont {H.}~\bibnamefont {Kusunose}},\ }\bibfield  {title} {\bibinfo {title} {{Predominant electronic order parameter for structural chirality: Role of spinless electronic toroidal multipoles in Te and Se}},\ }\href {https://doi.org/10.1103/1zq8-pqh8} {\bibfield  {journal} {\bibinfo  {journal} {Phys. Rev. Res.}\ }\textbf {\bibinfo {volume} {7}},\ \bibinfo {pages} {033250} (\bibinfo {year} {2025})}\BibitemShut {NoStop}%
\bibitem [{\citenamefont {Miki}\ \emph {et~al.}(2025)\citenamefont {Miki}, \citenamefont {Ikeda}, \citenamefont {Suzuki},\ and\ \citenamefont {Hoshino}}]{miki2025}%
  \BibitemOpen
  \bibfield  {author} {\bibinfo {author} {\bibfnamefont {T.}~\bibnamefont {Miki}}, \bibinfo {author} {\bibfnamefont {H.}~\bibnamefont {Ikeda}}, \bibinfo {author} {\bibfnamefont {M.-T.}\ \bibnamefont {Suzuki}},\ and\ \bibinfo {author} {\bibfnamefont {S.}~\bibnamefont {Hoshino}},\ }\bibfield  {title} {\bibinfo {title} {{Quantification of Electronic Asymmetry: Chirality and Axiality in Solids}},\ }\href {https://doi.org/10.1103/PhysRevLett.134.226401} {\bibfield  {journal} {\bibinfo  {journal} {Phys. Rev. Lett.}\ }\textbf {\bibinfo {volume} {134}},\ \bibinfo {pages} {226401} (\bibinfo {year} {2025})}\BibitemShut {NoStop}%
\bibitem [{\citenamefont {Ishitobi}\ and\ \citenamefont {Hattori}(2026)}]{ishitobi2026}%
  \BibitemOpen
  \bibfield  {author} {\bibinfo {author} {\bibfnamefont {T.}~\bibnamefont {Ishitobi}}\ and\ \bibinfo {author} {\bibfnamefont {K.}~\bibnamefont {Hattori}},\ }\bibfield  {title} {\bibinfo {title} {{Purely Electronic Chirality without Structural Chirality}},\ }\href {https://doi.org/10.1103/dc1q-xzbd} {\bibfield  {journal} {\bibinfo  {journal} {Phys. Rev. Lett.}\ }\textbf {\bibinfo {volume} {136}},\ \bibinfo {pages} {056402} (\bibinfo {year} {2026})}\BibitemShut {NoStop}%
\bibitem [{\citenamefont {Sachdev}(2008)}]{sachdev2008}%
  \BibitemOpen
  \bibfield  {author} {\bibinfo {author} {\bibfnamefont {S.}~\bibnamefont {Sachdev}},\ }\bibfield  {title} {\bibinfo {title} {{Quantum magnetism and criticality}},\ }\href {https://doi.org/10.1038/nphys894} {\bibfield  {journal} {\bibinfo  {journal} {Nat. Phys.}\ }\textbf {\bibinfo {volume} {4}},\ \bibinfo {pages} {173} (\bibinfo {year} {2008})}\BibitemShut {NoStop}%
\bibitem [{\citenamefont {Lhuillier}\ and\ \citenamefont {Misguich}(2001)}]{lhuillier2001}%
  \BibitemOpen
  \bibfield  {author} {\bibinfo {author} {\bibfnamefont {C.}~\bibnamefont {Lhuillier}}\ and\ \bibinfo {author} {\bibfnamefont {G.}~\bibnamefont {Misguich}},\ }\bibinfo {title} {{Frustrated Quantum Magnets}},\ in\ \href {https://doi.org/10.1007/3-540-45649-X_6} {\emph {\bibinfo {booktitle} {{High Magnetic Fields: Applications in Condensed Matter Physics and Spectroscopy}}}},\ \bibinfo {editor} {edited by\ \bibinfo {editor} {\bibfnamefont {C.}~\bibnamefont {Berthier}}, \bibinfo {editor} {\bibfnamefont {L.~P.}\ \bibnamefont {L{\'e}vy}},\ and\ \bibinfo {editor} {\bibfnamefont {G.}~\bibnamefont {Martinez}}}\ (\bibinfo  {publisher} {{Springer Berlin Heidelberg}},\ \bibinfo {address} {{Berlin, Heidelberg}},\ \bibinfo {year} {2001})\ pp.\ \bibinfo {pages} {161--190}\BibitemShut {NoStop}%
\bibitem [{\citenamefont {Cheong}\ and\ \citenamefont {Xu}(2022)}]{cheong2022}%
  \BibitemOpen
  \bibfield  {author} {\bibinfo {author} {\bibfnamefont {S.-W.}\ \bibnamefont {Cheong}}\ and\ \bibinfo {author} {\bibfnamefont {X.}~\bibnamefont {Xu}},\ }\bibfield  {title} {\bibinfo {title} {{Magnetic chirality}},\ }\href {https://doi.org/10.1038/s41535-022-00447-5} {\bibfield  {journal} {\bibinfo  {journal} {npj Quantum Mater.}\ }\textbf {\bibinfo {volume} {7}},\ \bibinfo {pages} {40} (\bibinfo {year} {2022})}\BibitemShut {NoStop}%
\bibitem [{\citenamefont {Wen}\ \emph {et~al.}(1989)\citenamefont {Wen}, \citenamefont {Wilczek},\ and\ \citenamefont {Zee}}]{wen1989}%
  \BibitemOpen
  \bibfield  {author} {\bibinfo {author} {\bibfnamefont {X.~G.}\ \bibnamefont {Wen}}, \bibinfo {author} {\bibfnamefont {F.}~\bibnamefont {Wilczek}},\ and\ \bibinfo {author} {\bibfnamefont {A.}~\bibnamefont {Zee}},\ }\bibfield  {title} {\bibinfo {title} {{Chiral spin states and superconductivity}},\ }\href {https://doi.org/10.1103/PhysRevB.39.11413} {\bibfield  {journal} {\bibinfo  {journal} {Phys. Rev. B}\ }\textbf {\bibinfo {volume} {39}},\ \bibinfo {pages} {11413} (\bibinfo {year} {1989})}\BibitemShut {NoStop}%
\bibitem [{\citenamefont {Taguchi}\ \emph {et~al.}(2001)\citenamefont {Taguchi}, \citenamefont {Oohara}, \citenamefont {Yoshizawa}, \citenamefont {Nagaosa},\ and\ \citenamefont {Tokura}}]{taguchi2001}%
  \BibitemOpen
  \bibfield  {author} {\bibinfo {author} {\bibfnamefont {Y.}~\bibnamefont {Taguchi}}, \bibinfo {author} {\bibfnamefont {Y.}~\bibnamefont {Oohara}}, \bibinfo {author} {\bibfnamefont {H.}~\bibnamefont {Yoshizawa}}, \bibinfo {author} {\bibfnamefont {N.}~\bibnamefont {Nagaosa}},\ and\ \bibinfo {author} {\bibfnamefont {Y.}~\bibnamefont {Tokura}},\ }\bibfield  {title} {\bibinfo {title} {{Spin Chirality, Berry Phase, and Anomalous Hall Effect in a Frustrated Ferromagnet}},\ }\href {https://doi.org/10.1126/science.1058161} {\bibfield  {journal} {\bibinfo  {journal} {Science}\ }\textbf {\bibinfo {volume} {291}},\ \bibinfo {pages} {2573} (\bibinfo {year} {2001})}\BibitemShut {NoStop}%
\bibitem [{\citenamefont {Kishine}\ \emph {et~al.}(2020)\citenamefont {Kishine}, \citenamefont {Ovchinnikov},\ and\ \citenamefont {Tereshchenko}}]{kishine2020}%
  \BibitemOpen
  \bibfield  {author} {\bibinfo {author} {\bibfnamefont {J.}~\bibnamefont {Kishine}}, \bibinfo {author} {\bibfnamefont {A.~S.}\ \bibnamefont {Ovchinnikov}},\ and\ \bibinfo {author} {\bibfnamefont {A.~A.}\ \bibnamefont {Tereshchenko}},\ }\bibfield  {title} {\bibinfo {title} {{Chirality-Induced Phonon Dispersion in a Noncentrosymmetric Micropolar Crystal}},\ }\href {https://doi.org/10.1103/PhysRevLett.125.245302} {\bibfield  {journal} {\bibinfo  {journal} {Phys. Rev. Lett.}\ }\textbf {\bibinfo {volume} {125}},\ \bibinfo {pages} {245302} (\bibinfo {year} {2020})}\BibitemShut {NoStop}%
\bibitem [{\citenamefont {Tsunetsugu}\ and\ \citenamefont {Kusunose}(2023)}]{tsunetsugu2023}%
  \BibitemOpen
  \bibfield  {author} {\bibinfo {author} {\bibfnamefont {H.}~\bibnamefont {Tsunetsugu}}\ and\ \bibinfo {author} {\bibfnamefont {H.}~\bibnamefont {Kusunose}},\ }\bibfield  {title} {\bibinfo {title} {{Theory of Energy Dispersion of Chiral Phonons}},\ }\href {https://doi.org/10.7566/JPSJ.92.023601} {\bibfield  {journal} {\bibinfo  {journal} {J. Phys. Soc. Jpn.}\ }\textbf {\bibinfo {volume} {92}},\ \bibinfo {pages} {023601} (\bibinfo {year} {2023})}\BibitemShut {NoStop}%
\bibitem [{\citenamefont {Ishito}\ \emph {et~al.}(2023)\citenamefont {Ishito}, \citenamefont {Mao}, \citenamefont {Kousaka}, \citenamefont {Togawa}, \citenamefont {Iwasaki}, \citenamefont {Zhang}, \citenamefont {Murakami}, \citenamefont {i.~Kishine},\ and\ \citenamefont {Satoh}}]{ishito2023}%
  \BibitemOpen
  \bibfield  {author} {\bibinfo {author} {\bibfnamefont {K.}~\bibnamefont {Ishito}}, \bibinfo {author} {\bibfnamefont {H.}~\bibnamefont {Mao}}, \bibinfo {author} {\bibfnamefont {Y.}~\bibnamefont {Kousaka}}, \bibinfo {author} {\bibfnamefont {Y.}~\bibnamefont {Togawa}}, \bibinfo {author} {\bibfnamefont {S.}~\bibnamefont {Iwasaki}}, \bibinfo {author} {\bibfnamefont {T.}~\bibnamefont {Zhang}}, \bibinfo {author} {\bibfnamefont {S.}~\bibnamefont {Murakami}}, \bibinfo {author} {\bibfnamefont {J.}~\bibnamefont {i.~Kishine}},\ and\ \bibinfo {author} {\bibfnamefont {T.}~\bibnamefont {Satoh}},\ }\bibfield  {title} {\bibinfo {title} {{Truly chiral phonons in $\alpha$-HgS}},\ }\href {https://doi.org/10.1038/s41567-022-01790-x} {\bibfield  {journal} {\bibinfo  {journal} {Nat. Phys.}\ }\textbf {\bibinfo {volume} {19}},\ \bibinfo {pages} {35} (\bibinfo {year} {2023})}\BibitemShut {NoStop}%
\bibitem [{\citenamefont {Ueda}\ \emph {et~al.}(2023)\citenamefont {Ueda}, \citenamefont {Garc^^c3^^ada-Fern^^c3^^a1ndez}, \citenamefont {Agrestini}, \citenamefont {Romao}, \citenamefont {van~den Brink}, \citenamefont {Spaldin}, \citenamefont {Zhou},\ and\ \citenamefont {Staub}}]{ueda2023}%
  \BibitemOpen
  \bibfield  {author} {\bibinfo {author} {\bibfnamefont {H.}~\bibnamefont {Ueda}}, \bibinfo {author} {\bibfnamefont {M.}~\bibnamefont {Garc^^c3^^ada-Fern^^c3^^a1ndez}}, \bibinfo {author} {\bibfnamefont {S.}~\bibnamefont {Agrestini}}, \bibinfo {author} {\bibfnamefont {C.}~\bibnamefont {Romao}}, \bibinfo {author} {\bibfnamefont {J.}~\bibnamefont {van~den Brink}}, \bibinfo {author} {\bibfnamefont {N.}~\bibnamefont {Spaldin}}, \bibinfo {author} {\bibfnamefont {K.-J.}\ \bibnamefont {Zhou}},\ and\ \bibinfo {author} {\bibfnamefont {U.}~\bibnamefont {Staub}},\ }\bibfield  {title} {\bibinfo {title} {{Chiral phonons in quartz probed by X-rays}},\ }\href {https://doi.org/10.1038/s41586-023-06016-5} {\bibfield  {journal} {\bibinfo  {journal} {Nature}\ }\textbf {\bibinfo {volume} {618}},\ \bibinfo {pages} {946} (\bibinfo {year} {2023})}\BibitemShut {NoStop}%
\bibitem [{\citenamefont {Oishi}\ \emph {et~al.}(2024)\citenamefont {Oishi}, \citenamefont {Fujii},\ and\ \citenamefont {Koreeda}}]{oishi2024}%
  \BibitemOpen
  \bibfield  {author} {\bibinfo {author} {\bibfnamefont {E.}~\bibnamefont {Oishi}}, \bibinfo {author} {\bibfnamefont {Y.}~\bibnamefont {Fujii}},\ and\ \bibinfo {author} {\bibfnamefont {A.}~\bibnamefont {Koreeda}},\ }\bibfield  {title} {\bibinfo {title} {{Selective observation of enantiomeric chiral phonons in $\ensuremath{\alpha}$-quartz}},\ }\href {https://doi.org/10.1103/PhysRevB.109.104306} {\bibfield  {journal} {\bibinfo  {journal} {Phys. Rev. B}\ }\textbf {\bibinfo {volume} {109}},\ \bibinfo {pages} {104306} (\bibinfo {year} {2024})}\BibitemShut {NoStop}%
\bibitem [{\citenamefont {Zhang}\ \emph {et~al.}(2025)\citenamefont {Zhang}, \citenamefont {Peshcherenko}, \citenamefont {Yang}, \citenamefont {Ward}, \citenamefont {Raghuvanshi}, \citenamefont {Lindsay}, \citenamefont {Felser}, \citenamefont {Zhang}, \citenamefont {Yan},\ and\ \citenamefont {Miao}}]{zhang2025_nature}%
  \BibitemOpen
  \bibfield  {author} {\bibinfo {author} {\bibfnamefont {H.}~\bibnamefont {Zhang}}, \bibinfo {author} {\bibfnamefont {N.}~\bibnamefont {Peshcherenko}}, \bibinfo {author} {\bibfnamefont {F.}~\bibnamefont {Yang}}, \bibinfo {author} {\bibfnamefont {T.}~\bibnamefont {Ward}}, \bibinfo {author} {\bibfnamefont {P.}~\bibnamefont {Raghuvanshi}}, \bibinfo {author} {\bibfnamefont {L.}~\bibnamefont {Lindsay}}, \bibinfo {author} {\bibfnamefont {C.}~\bibnamefont {Felser}}, \bibinfo {author} {\bibfnamefont {Y.}~\bibnamefont {Zhang}}, \bibinfo {author} {\bibfnamefont {J.}~\bibnamefont {Yan}},\ and\ \bibinfo {author} {\bibfnamefont {H.}~\bibnamefont {Miao}},\ }\bibfield  {title} {\bibinfo {title} {{Measurement of phonon angular momentum}},\ }\href {https://doi.org/10.1038/s41567-025-02952-3} {\bibfield  {journal} {\bibinfo  {journal} {Nat. Phys.}\ }\textbf {\bibinfo {volume} {21}},\ \bibinfo {pages} {1387} (\bibinfo {year} {2025})}\BibitemShut {NoStop}%
\bibitem [{\citenamefont {Sachdev}\ and\ \citenamefont {Bhatt}(1990)}]{sachdev1990}%
  \BibitemOpen
  \bibfield  {author} {\bibinfo {author} {\bibfnamefont {S.}~\bibnamefont {Sachdev}}\ and\ \bibinfo {author} {\bibfnamefont {R.~N.}\ \bibnamefont {Bhatt}},\ }\bibfield  {title} {\bibinfo {title} {{Bond-operator representation of quantum spins: Mean-field theory of frustrated quantum Heisenberg antiferromagnets}},\ }\href {https://doi.org/10.1103/PhysRevB.41.9323} {\bibfield  {journal} {\bibinfo  {journal} {Phys. Rev. B}\ }\textbf {\bibinfo {volume} {41}},\ \bibinfo {pages} {9323} (\bibinfo {year} {1990})}\BibitemShut {NoStop}%
\bibitem [{SM()}]{SM}%
  \BibitemOpen
  \href@noop {} {}\bibinfo {note} {See Supplemental Material at [URL] for details of the displacement modes, symmetry-allowed exchange and DM interactions, cluster mean-field and linear flavor-wave theories, the internal spin polarization of triplons and the spin they carry, the thermal Edelstein effect, and the dynamical spin structure factor.}\BibitemShut {Stop}%
\bibitem [{\citenamefont {Bergman}\ \emph {et~al.}(2007)\citenamefont {Bergman}, \citenamefont {Alicea}, \citenamefont {Gull}, \citenamefont {Trebst},\ and\ \citenamefont {Balents}}]{bergman2007}%
  \BibitemOpen
  \bibfield  {author} {\bibinfo {author} {\bibfnamefont {D.}~\bibnamefont {Bergman}}, \bibinfo {author} {\bibfnamefont {J.}~\bibnamefont {Alicea}}, \bibinfo {author} {\bibfnamefont {E.}~\bibnamefont {Gull}}, \bibinfo {author} {\bibfnamefont {S.}~\bibnamefont {Trebst}},\ and\ \bibinfo {author} {\bibfnamefont {L.}~\bibnamefont {Balents}},\ }\bibfield  {title} {\bibinfo {title} {{Order-by-disorder and spiral spin-liquid in frustrated diamond-lattice antiferromagnets}},\ }\href {https://doi.org/10.1038/nphys622} {\bibfield  {journal} {\bibinfo  {journal} {Nat. Phys.}\ }\textbf {\bibinfo {volume} {3}},\ \bibinfo {pages} {487} (\bibinfo {year} {2007})}\BibitemShut {NoStop}%
\bibitem [{\citenamefont {B\"arwolf}\ \emph {et~al.}(2025)\citenamefont {B\"arwolf}, \citenamefont {Sushchyev}, \citenamefont {Parisen~Toldin},\ and\ \citenamefont {Wessel}}]{barwolf2025}%
  \BibitemOpen
  \bibfield  {author} {\bibinfo {author} {\bibfnamefont {R.}~\bibnamefont {B\"arwolf}}, \bibinfo {author} {\bibfnamefont {A.}~\bibnamefont {Sushchyev}}, \bibinfo {author} {\bibfnamefont {F.}~\bibnamefont {Parisen~Toldin}},\ and\ \bibinfo {author} {\bibfnamefont {S.}~\bibnamefont {Wessel}},\ }\bibfield  {title} {\bibinfo {title} {{Phase transitions in the spin-$\frac{1}{2}$ Heisenberg antiferromagnet on the dimerized diamond lattice}},\ }\href {https://doi.org/10.1103/PhysRevB.111.085136} {\bibfield  {journal} {\bibinfo  {journal} {Phys. Rev. B}\ }\textbf {\bibinfo {volume} {111}},\ \bibinfo {pages} {085136} (\bibinfo {year} {2025})}\BibitemShut {NoStop}%
\bibitem [{\citenamefont {Koyama}\ and\ \citenamefont {Nasu}(2025)}]{koyama2025}%
  \BibitemOpen
  \bibfield  {author} {\bibinfo {author} {\bibfnamefont {S.}~\bibnamefont {Koyama}}\ and\ \bibinfo {author} {\bibfnamefont {J.}~\bibnamefont {Nasu}},\ }\bibfield  {title} {\bibinfo {title} {{Formulation of the spin Nernst effect for spin-nonconserving insulating magnets}},\ }\href {https://doi.org/10.1103/8xgc-d9ch} {\bibfield  {journal} {\bibinfo  {journal} {Phys. Rev. B}\ }\textbf {\bibinfo {volume} {112}},\ \bibinfo {pages} {014447} (\bibinfo {year} {2025})}\BibitemShut {NoStop}%
\bibitem [{\citenamefont {Bergman}\ \emph {et~al.}(2006{\natexlab{a}})\citenamefont {Bergman}, \citenamefont {Shindou}, \citenamefont {Fiete},\ and\ \citenamefont {Balents}}]{bergman2006_prl}%
  \BibitemOpen
  \bibfield  {author} {\bibinfo {author} {\bibfnamefont {D.~L.}\ \bibnamefont {Bergman}}, \bibinfo {author} {\bibfnamefont {R.}~\bibnamefont {Shindou}}, \bibinfo {author} {\bibfnamefont {G.~A.}\ \bibnamefont {Fiete}},\ and\ \bibinfo {author} {\bibfnamefont {L.}~\bibnamefont {Balents}},\ }\bibfield  {title} {\bibinfo {title} {{Quantum Effects in a Half-Polarized Pyrochlore Antiferromagnet}},\ }\href {https://doi.org/10.1103/PhysRevLett.96.097207} {\bibfield  {journal} {\bibinfo  {journal} {Phys. Rev. Lett.}\ }\textbf {\bibinfo {volume} {96}},\ \bibinfo {pages} {097207} (\bibinfo {year} {2006}{\natexlab{a}})}\BibitemShut {NoStop}%
\bibitem [{\citenamefont {Bergman}\ \emph {et~al.}(2006{\natexlab{b}})\citenamefont {Bergman}, \citenamefont {Fiete},\ and\ \citenamefont {Balents}}]{bergman2006_prb1}%
  \BibitemOpen
  \bibfield  {author} {\bibinfo {author} {\bibfnamefont {D.~L.}\ \bibnamefont {Bergman}}, \bibinfo {author} {\bibfnamefont {G.~A.}\ \bibnamefont {Fiete}},\ and\ \bibinfo {author} {\bibfnamefont {L.}~\bibnamefont {Balents}},\ }\bibfield  {title} {\bibinfo {title} {{Ordering in a frustrated pyrochlore antiferromagnet proximate to a spin liquid}},\ }\href {https://doi.org/10.1103/PhysRevB.73.134402} {\bibfield  {journal} {\bibinfo  {journal} {Phys. Rev. B}\ }\textbf {\bibinfo {volume} {73}},\ \bibinfo {pages} {134402} (\bibinfo {year} {2006}{\natexlab{b}})}\BibitemShut {NoStop}%
\bibitem [{\citenamefont {Bergman}\ \emph {et~al.}(2006{\natexlab{c}})\citenamefont {Bergman}, \citenamefont {Shindou}, \citenamefont {Fiete},\ and\ \citenamefont {Balents}}]{bergman2006_prb2}%
  \BibitemOpen
  \bibfield  {author} {\bibinfo {author} {\bibfnamefont {D.~L.}\ \bibnamefont {Bergman}}, \bibinfo {author} {\bibfnamefont {R.}~\bibnamefont {Shindou}}, \bibinfo {author} {\bibfnamefont {G.~A.}\ \bibnamefont {Fiete}},\ and\ \bibinfo {author} {\bibfnamefont {L.}~\bibnamefont {Balents}},\ }\bibfield  {title} {\bibinfo {title} {{Models of degeneracy breaking in pyrochlore antiferromagnets}},\ }\href {https://doi.org/10.1103/PhysRevB.74.134409} {\bibfield  {journal} {\bibinfo  {journal} {Phys. Rev. B}\ }\textbf {\bibinfo {volume} {74}},\ \bibinfo {pages} {134409} (\bibinfo {year} {2006}{\natexlab{c}})}\BibitemShut {NoStop}%
\bibitem [{\citenamefont {Sikora}\ \emph {et~al.}(2009)\citenamefont {Sikora}, \citenamefont {Pollmann}, \citenamefont {Shannon}, \citenamefont {Penc},\ and\ \citenamefont {Fulde}}]{sikora2009}%
  \BibitemOpen
  \bibfield  {author} {\bibinfo {author} {\bibfnamefont {O.}~\bibnamefont {Sikora}}, \bibinfo {author} {\bibfnamefont {F.}~\bibnamefont {Pollmann}}, \bibinfo {author} {\bibfnamefont {N.}~\bibnamefont {Shannon}}, \bibinfo {author} {\bibfnamefont {K.}~\bibnamefont {Penc}},\ and\ \bibinfo {author} {\bibfnamefont {P.}~\bibnamefont {Fulde}},\ }\bibfield  {title} {\bibinfo {title} {{Quantum Liquid with Deconfined Fractional Excitations in Three Dimensions}},\ }\href {https://doi.org/10.1103/PhysRevLett.103.247001} {\bibfield  {journal} {\bibinfo  {journal} {Phys. Rev. Lett.}\ }\textbf {\bibinfo {volume} {103}},\ \bibinfo {pages} {247001} (\bibinfo {year} {2009})}\BibitemShut {NoStop}%
\bibitem [{\citenamefont {Sikora}\ \emph {et~al.}(2011)\citenamefont {Sikora}, \citenamefont {Shannon}, \citenamefont {Pollmann}, \citenamefont {Penc},\ and\ \citenamefont {Fulde}}]{sikora2011}%
  \BibitemOpen
  \bibfield  {author} {\bibinfo {author} {\bibfnamefont {O.}~\bibnamefont {Sikora}}, \bibinfo {author} {\bibfnamefont {N.}~\bibnamefont {Shannon}}, \bibinfo {author} {\bibfnamefont {F.}~\bibnamefont {Pollmann}}, \bibinfo {author} {\bibfnamefont {K.}~\bibnamefont {Penc}},\ and\ \bibinfo {author} {\bibfnamefont {P.}~\bibnamefont {Fulde}},\ }\bibfield  {title} {\bibinfo {title} {{Extended quantum $U(1)$-liquid phase in a three-dimensional quantum dimer model}},\ }\href {https://doi.org/10.1103/PhysRevB.84.115129} {\bibfield  {journal} {\bibinfo  {journal} {Phys. Rev. B}\ }\textbf {\bibinfo {volume} {84}},\ \bibinfo {pages} {115129} (\bibinfo {year} {2011})}\BibitemShut {NoStop}%
\bibitem [{\citenamefont {Dzyaloshinsky}(1958)}]{dzyaloshinskii1958}%
  \BibitemOpen
  \bibfield  {author} {\bibinfo {author} {\bibfnamefont {I.}~\bibnamefont {Dzyaloshinsky}},\ }\bibfield  {title} {\bibinfo {title} {{A thermodynamic theory of “weak” ferromagnetism of antiferromagnetics}},\ }\href {https://doi.org/https://doi.org/10.1016/0022-3697(58)90076-3} {\bibfield  {journal} {\bibinfo  {journal} {J. Phys. Chem. Solids}\ }\textbf {\bibinfo {volume} {4}},\ \bibinfo {pages} {241} (\bibinfo {year} {1958})}\BibitemShut {NoStop}%
\bibitem [{\citenamefont {Moriya}(1960)}]{moriya1960}%
  \BibitemOpen
  \bibfield  {author} {\bibinfo {author} {\bibfnamefont {T.}~\bibnamefont {Moriya}},\ }\bibfield  {title} {\bibinfo {title} {{Anisotropic Superexchange Interaction and Weak Ferromagnetism}},\ }\href {https://doi.org/10.1103/PhysRev.120.91} {\bibfield  {journal} {\bibinfo  {journal} {Phys. Rev.}\ }\textbf {\bibinfo {volume} {120}},\ \bibinfo {pages} {91} (\bibinfo {year} {1960})}\BibitemShut {NoStop}%
\bibitem [{\citenamefont {Sergienko}\ and\ \citenamefont {Dagotto}(2006)}]{sergienko2006}%
  \BibitemOpen
  \bibfield  {author} {\bibinfo {author} {\bibfnamefont {I.~A.}\ \bibnamefont {Sergienko}}\ and\ \bibinfo {author} {\bibfnamefont {E.}~\bibnamefont {Dagotto}},\ }\bibfield  {title} {\bibinfo {title} {{Role of the Dzyaloshinskii-Moriya interaction in multiferroic perovskites}},\ }\href {https://doi.org/10.1103/PhysRevB.73.094434} {\bibfield  {journal} {\bibinfo  {journal} {Phys. Rev. B}\ }\textbf {\bibinfo {volume} {73}},\ \bibinfo {pages} {094434} (\bibinfo {year} {2006})}\BibitemShut {NoStop}%
\bibitem [{\citenamefont {Joshi}\ \emph {et~al.}(1999)\citenamefont {Joshi}, \citenamefont {Ma}, \citenamefont {Mila}, \citenamefont {Shi},\ and\ \citenamefont {Zhang}}]{joshi1999}%
  \BibitemOpen
  \bibfield  {author} {\bibinfo {author} {\bibfnamefont {A.}~\bibnamefont {Joshi}}, \bibinfo {author} {\bibfnamefont {M.}~\bibnamefont {Ma}}, \bibinfo {author} {\bibfnamefont {F.}~\bibnamefont {Mila}}, \bibinfo {author} {\bibfnamefont {D.~N.}\ \bibnamefont {Shi}},\ and\ \bibinfo {author} {\bibfnamefont {F.~C.}\ \bibnamefont {Zhang}},\ }\bibfield  {title} {\bibinfo {title} {{Elementary excitations in magnetically ordered systems with orbital degeneracy}},\ }\href {https://doi.org/10.1103/PhysRevB.60.6584} {\bibfield  {journal} {\bibinfo  {journal} {Phys. Rev. B}\ }\textbf {\bibinfo {volume} {60}},\ \bibinfo {pages} {6584} (\bibinfo {year} {1999})}\BibitemShut {NoStop}%
\bibitem [{\citenamefont {Shiina}\ \emph {et~al.}(2003)\citenamefont {Shiina}, \citenamefont {Shiba}, \citenamefont {Thalmeier}, \citenamefont {Takahashi},\ and\ \citenamefont {Sakai}}]{shiina2003}%
  \BibitemOpen
  \bibfield  {author} {\bibinfo {author} {\bibfnamefont {R.}~\bibnamefont {Shiina}}, \bibinfo {author} {\bibfnamefont {H.}~\bibnamefont {Shiba}}, \bibinfo {author} {\bibfnamefont {P.}~\bibnamefont {Thalmeier}}, \bibinfo {author} {\bibfnamefont {A.}~\bibnamefont {Takahashi}},\ and\ \bibinfo {author} {\bibfnamefont {O.}~\bibnamefont {Sakai}},\ }\bibfield  {title} {\bibinfo {title} {{Dynamics of Multipoles and Neutron Scattering Spectra in Quadrupolar Ordering Phase of CeB$_6$}},\ }\href {https://doi.org/10.1143/JPSJ.72.1216} {\bibfield  {journal} {\bibinfo  {journal} {J. Phys. Soc. Jpn.}\ }\textbf {\bibinfo {volume} {72}},\ \bibinfo {pages} {1216} (\bibinfo {year} {2003})}\BibitemShut {NoStop}%
\bibitem [{\citenamefont {Nasu}\ and\ \citenamefont {Naka}(2021)}]{nasu2021}%
  \BibitemOpen
  \bibfield  {author} {\bibinfo {author} {\bibfnamefont {J.}~\bibnamefont {Nasu}}\ and\ \bibinfo {author} {\bibfnamefont {M.}~\bibnamefont {Naka}},\ }\bibfield  {title} {\bibinfo {title} {{Spin Seebeck effect in nonmagnetic excitonic insulators}},\ }\href {https://doi.org/10.1103/PhysRevB.103.L121104} {\bibfield  {journal} {\bibinfo  {journal} {Phys. Rev. B}\ }\textbf {\bibinfo {volume} {103}},\ \bibinfo {pages} {L121104} (\bibinfo {year} {2021})}\BibitemShut {NoStop}%
\bibitem [{\citenamefont {Colpa}(1978)}]{colpa}%
  \BibitemOpen
  \bibfield  {author} {\bibinfo {author} {\bibfnamefont {J.}~\bibnamefont {Colpa}},\ }\bibfield  {title} {\bibinfo {title} {{Diagonalization of the quadratic boson hamiltonian}},\ }\href {https://doi.org/https://doi.org/10.1016/0378-4371(78)90160-7} {\bibfield  {journal} {\bibinfo  {journal} {Physica}\ }\textbf {\bibinfo {volume} {93A}},\ \bibinfo {pages} {327} (\bibinfo {year} {1978})}\BibitemShut {NoStop}%
\bibitem [{\citenamefont {Zhang}\ and\ \citenamefont {Niu}(2014)}]{zhang2014}%
  \BibitemOpen
  \bibfield  {author} {\bibinfo {author} {\bibfnamefont {L.}~\bibnamefont {Zhang}}\ and\ \bibinfo {author} {\bibfnamefont {Q.}~\bibnamefont {Niu}},\ }\bibfield  {title} {\bibinfo {title} {{Angular Momentum of Phonons and the Einstein--de Haas Effect}},\ }\href {https://doi.org/10.1103/PhysRevLett.112.085503} {\bibfield  {journal} {\bibinfo  {journal} {Phys. Rev. Lett.}\ }\textbf {\bibinfo {volume} {112}},\ \bibinfo {pages} {085503} (\bibinfo {year} {2014})}\BibitemShut {NoStop}%
\bibitem [{\citenamefont {Zhang}\ and\ \citenamefont {Niu}(2015)}]{zhang2015}%
  \BibitemOpen
  \bibfield  {author} {\bibinfo {author} {\bibfnamefont {L.}~\bibnamefont {Zhang}}\ and\ \bibinfo {author} {\bibfnamefont {Q.}~\bibnamefont {Niu}},\ }\bibfield  {title} {\bibinfo {title} {{Chiral Phonons at High-Symmetry Points in Monolayer Hexagonal Lattices}},\ }\href {https://doi.org/10.1103/PhysRevLett.115.115502} {\bibfield  {journal} {\bibinfo  {journal} {Phys. Rev. Lett.}\ }\textbf {\bibinfo {volume} {115}},\ \bibinfo {pages} {115502} (\bibinfo {year} {2015})}\BibitemShut {NoStop}%
\bibitem [{\citenamefont {Zhang}\ and\ \citenamefont {Murakami}(2022)}]{zhang2022}%
  \BibitemOpen
  \bibfield  {author} {\bibinfo {author} {\bibfnamefont {T.}~\bibnamefont {Zhang}}\ and\ \bibinfo {author} {\bibfnamefont {S.}~\bibnamefont {Murakami}},\ }\bibfield  {title} {\bibinfo {title} {{Chiral phonons and pseudoangular momentum in nonsymmorphic systems}},\ }\href {https://doi.org/10.1103/PhysRevResearch.4.L012024} {\bibfield  {journal} {\bibinfo  {journal} {Phys. Rev. Res.}\ }\textbf {\bibinfo {volume} {4}},\ \bibinfo {pages} {L012024} (\bibinfo {year} {2022})}\BibitemShut {NoStop}%
\bibitem [{\citenamefont {Thomasen}\ \emph {et~al.}(2021)\citenamefont {Thomasen}, \citenamefont {Penc}, \citenamefont {Shannon},\ and\ \citenamefont {Romh\'anyi}}]{thomasen2021}%
  \BibitemOpen
  \bibfield  {author} {\bibinfo {author} {\bibfnamefont {A.}~\bibnamefont {Thomasen}}, \bibinfo {author} {\bibfnamefont {K.}~\bibnamefont {Penc}}, \bibinfo {author} {\bibfnamefont {N.}~\bibnamefont {Shannon}},\ and\ \bibinfo {author} {\bibfnamefont {J.}~\bibnamefont {Romh\'anyi}},\ }\bibfield  {title} {\bibinfo {title} {{Fragility of ${\mathcal{Z}}_{2}$ topological invariant characterizing triplet excitations in a bilayer kagome magnet}},\ }\href {https://doi.org/10.1103/PhysRevB.104.104412} {\bibfield  {journal} {\bibinfo  {journal} {Phys. Rev. B}\ }\textbf {\bibinfo {volume} {104}},\ \bibinfo {pages} {104412} (\bibinfo {year} {2021})}\BibitemShut {NoStop}%
\bibitem [{\citenamefont {Esaki}\ \emph {et~al.}(2025)\citenamefont {Esaki}, \citenamefont {Akagi}, \citenamefont {Penc},\ and\ \citenamefont {Katsura}}]{esaki2025}%
  \BibitemOpen
  \bibfield  {author} {\bibinfo {author} {\bibfnamefont {N.}~\bibnamefont {Esaki}}, \bibinfo {author} {\bibfnamefont {Y.}~\bibnamefont {Akagi}}, \bibinfo {author} {\bibfnamefont {K.}~\bibnamefont {Penc}},\ and\ \bibinfo {author} {\bibfnamefont {H.}~\bibnamefont {Katsura}},\ }\bibfield  {title} {\bibinfo {title} {{Spin Nernst and thermal Hall effects of topological triplons in quantum dimer magnets on the maple-leaf and star lattices}},\ }\href {https://doi.org/10.1103/1yv1-wtx8} {\bibfield  {journal} {\bibinfo  {journal} {Phys. Rev. B}\ }\textbf {\bibinfo {volume} {112}},\ \bibinfo {pages} {134435} (\bibinfo {year} {2025})}\BibitemShut {NoStop}%
\bibitem [{\citenamefont {Hirayama}\ \emph {et~al.}(2015)\citenamefont {Hirayama}, \citenamefont {Okugawa}, \citenamefont {Ishibashi}, \citenamefont {Murakami},\ and\ \citenamefont {Miyake}}]{hirayama2015}%
  \BibitemOpen
  \bibfield  {author} {\bibinfo {author} {\bibfnamefont {M.}~\bibnamefont {Hirayama}}, \bibinfo {author} {\bibfnamefont {R.}~\bibnamefont {Okugawa}}, \bibinfo {author} {\bibfnamefont {S.}~\bibnamefont {Ishibashi}}, \bibinfo {author} {\bibfnamefont {S.}~\bibnamefont {Murakami}},\ and\ \bibinfo {author} {\bibfnamefont {T.}~\bibnamefont {Miyake}},\ }\bibfield  {title} {\bibinfo {title} {{Weyl Node and Spin Texture in Trigonal Tellurium and Selenium}},\ }\href {https://doi.org/10.1103/PhysRevLett.114.206401} {\bibfield  {journal} {\bibinfo  {journal} {Phys. Rev. Lett.}\ }\textbf {\bibinfo {volume} {114}},\ \bibinfo {pages} {206401} (\bibinfo {year} {2015})}\BibitemShut {NoStop}%
\bibitem [{\citenamefont {Sakano}\ \emph {et~al.}(2020)\citenamefont {Sakano}, \citenamefont {Hirayama}, \citenamefont {Takahashi}, \citenamefont {Akebi}, \citenamefont {Nakayama}, \citenamefont {Kuroda}, \citenamefont {Taguchi}, \citenamefont {Yoshikawa}, \citenamefont {Miyamoto}, \citenamefont {Okuda}, \citenamefont {Ono}, \citenamefont {Kumigashira}, \citenamefont {Ideue}, \citenamefont {Iwasa}, \citenamefont {Mitsuishi}, \citenamefont {Ishizaka}, \citenamefont {Shin}, \citenamefont {Miyake}, \citenamefont {Murakami}, \citenamefont {Sasagawa},\ and\ \citenamefont {Kondo}}]{sakano2020}%
  \BibitemOpen
  \bibfield  {author} {\bibinfo {author} {\bibfnamefont {M.}~\bibnamefont {Sakano}}, \bibinfo {author} {\bibfnamefont {M.}~\bibnamefont {Hirayama}}, \bibinfo {author} {\bibfnamefont {T.}~\bibnamefont {Takahashi}}, \bibinfo {author} {\bibfnamefont {S.}~\bibnamefont {Akebi}}, \bibinfo {author} {\bibfnamefont {M.}~\bibnamefont {Nakayama}}, \bibinfo {author} {\bibfnamefont {K.}~\bibnamefont {Kuroda}}, \bibinfo {author} {\bibfnamefont {K.}~\bibnamefont {Taguchi}}, \bibinfo {author} {\bibfnamefont {T.}~\bibnamefont {Yoshikawa}}, \bibinfo {author} {\bibfnamefont {K.}~\bibnamefont {Miyamoto}}, \bibinfo {author} {\bibfnamefont {T.}~\bibnamefont {Okuda}}, \bibinfo {author} {\bibfnamefont {K.}~\bibnamefont {Ono}}, \bibinfo {author} {\bibfnamefont {H.}~\bibnamefont {Kumigashira}}, \bibinfo {author} {\bibfnamefont {T.}~\bibnamefont {Ideue}}, \bibinfo {author} {\bibfnamefont {Y.}~\bibnamefont {Iwasa}}, \bibinfo {author} {\bibfnamefont {N.}~\bibnamefont {Mitsuishi}}, \bibinfo {author} {\bibfnamefont {K.}~\bibnamefont {Ishizaka}}, \bibinfo {author} {\bibfnamefont {S.}~\bibnamefont {Shin}}, \bibinfo {author} {\bibfnamefont {T.}~\bibnamefont {Miyake}}, \bibinfo {author} {\bibfnamefont {S.}~\bibnamefont {Murakami}}, \bibinfo {author} {\bibfnamefont {T.}~\bibnamefont {Sasagawa}},\ and\ \bibinfo {author} {\bibfnamefont {T.}~\bibnamefont {Kondo}},\ }\bibfield  {title} {\bibinfo {title} {{Radial Spin Texture in Elemental Tellurium with Chiral Crystal Structure}},\ }\href {https://doi.org/10.1103/PhysRevLett.124.136404} {\bibfield  {journal} {\bibinfo  {journal} {Phys. Rev. Lett.}\ }\textbf {\bibinfo {volume} {124}},\ \bibinfo {pages} {136404} (\bibinfo {year} {2020})}\BibitemShut {NoStop}%
\bibitem [{\citenamefont {Tsunetsugu}\ and\ \citenamefont {Kusunose}(2026)}]{tsunetsugu2026}%
  \BibitemOpen
  \bibfield  {author} {\bibinfo {author} {\bibfnamefont {H.}~\bibnamefont {Tsunetsugu}}\ and\ \bibinfo {author} {\bibfnamefont {H.}~\bibnamefont {Kusunose}},\ }\bibfield  {title} {\bibinfo {title} {{Chiral Phonons in a Cubic Lattice}},\ }\href {https://doi.org/10.7566/JPSJ.95.013601} {\bibfield  {journal} {\bibinfo  {journal} {J. Phys. Soc. Jpn.}\ }\textbf {\bibinfo {volume} {95}},\ \bibinfo {pages} {013601} (\bibinfo {year} {2026})}\BibitemShut {NoStop}%
\bibitem [{\citenamefont {Hamada}\ \emph {et~al.}(2018)\citenamefont {Hamada}, \citenamefont {Minamitani}, \citenamefont {Hirayama},\ and\ \citenamefont {Murakami}}]{hamada2018}%
  \BibitemOpen
  \bibfield  {author} {\bibinfo {author} {\bibfnamefont {M.}~\bibnamefont {Hamada}}, \bibinfo {author} {\bibfnamefont {E.}~\bibnamefont {Minamitani}}, \bibinfo {author} {\bibfnamefont {M.}~\bibnamefont {Hirayama}},\ and\ \bibinfo {author} {\bibfnamefont {S.}~\bibnamefont {Murakami}},\ }\bibfield  {title} {\bibinfo {title} {{Phonon Angular Momentum Induced by the Temperature Gradient}},\ }\href {https://doi.org/10.1103/PhysRevLett.121.175301} {\bibfield  {journal} {\bibinfo  {journal} {Phys. Rev. Lett.}\ }\textbf {\bibinfo {volume} {121}},\ \bibinfo {pages} {175301} (\bibinfo {year} {2018})}\BibitemShut {NoStop}%
\bibitem [{\citenamefont {Li}\ \emph {et~al.}(2020)\citenamefont {Li}, \citenamefont {Mook}, \citenamefont {Raeliarijaona},\ and\ \citenamefont {Kovalev}}]{li2020}%
  \BibitemOpen
  \bibfield  {author} {\bibinfo {author} {\bibfnamefont {B.}~\bibnamefont {Li}}, \bibinfo {author} {\bibfnamefont {A.}~\bibnamefont {Mook}}, \bibinfo {author} {\bibfnamefont {A.}~\bibnamefont {Raeliarijaona}},\ and\ \bibinfo {author} {\bibfnamefont {A.~A.}\ \bibnamefont {Kovalev}},\ }\bibfield  {title} {\bibinfo {title} {{Magnonic analog of the Edelstein effect in antiferromagnetic insulators}},\ }\href {https://doi.org/10.1103/PhysRevB.101.024427} {\bibfield  {journal} {\bibinfo  {journal} {Phys. Rev. B}\ }\textbf {\bibinfo {volume} {101}},\ \bibinfo {pages} {024427} (\bibinfo {year} {2020})}\BibitemShut {NoStop}%
\bibitem [{\citenamefont {Yao}\ and\ \citenamefont {Yokoyama}(2025)}]{yao2025}%
  \BibitemOpen
  \bibfield  {author} {\bibinfo {author} {\bibfnamefont {D.}~\bibnamefont {Yao}}\ and\ \bibinfo {author} {\bibfnamefont {T.}~\bibnamefont {Yokoyama}},\ }\bibfield  {title} {\bibinfo {title} {{Chiral magnon in ferromagnetic chiral crystals}},\ }\href {https://doi.org/10.1103/PhysRevB.111.L060404} {\bibfield  {journal} {\bibinfo  {journal} {Phys. Rev. B}\ }\textbf {\bibinfo {volume} {111}},\ \bibinfo {pages} {L060404} (\bibinfo {year} {2025})}\BibitemShut {NoStop}%
\bibitem [{\citenamefont {Neumann}\ \emph {et~al.}(shed)\citenamefont {Neumann}, \citenamefont {Jaeschke-Ubiergo}, \citenamefont {Zarzuela}, \citenamefont {^^c5^^a0mejkal}, \citenamefont {Sinova},\ and\ \citenamefont {Mook}}]{neumann2026_arxiv}%
  \BibitemOpen
  \bibfield  {author} {\bibinfo {author} {\bibfnamefont {R.~R.}\ \bibnamefont {Neumann}}, \bibinfo {author} {\bibfnamefont {R.}~\bibnamefont {Jaeschke-Ubiergo}}, \bibinfo {author} {\bibfnamefont {R.}~\bibnamefont {Zarzuela}}, \bibinfo {author} {\bibfnamefont {L.}~\bibnamefont {^^c5^^a0mejkal}}, \bibinfo {author} {\bibfnamefont {J.}~\bibnamefont {Sinova}},\ and\ \bibinfo {author} {\bibfnamefont {A.}~\bibnamefont {Mook}},\ }\bibfield  {title} {\bibinfo {title} {{Odd-Parity-Wave Magnons and Nonrelativistic Thermal Edelstein Effect}},\ }\href {https://arxiv.org/abs/2603.05415} {\bibfield  {journal} {\bibinfo  {journal} {arXiv:2603.05415}\ } (\bibinfo {year} {unpublished})}\BibitemShut {NoStop}%
\bibitem [{\citenamefont {Kubo}\ \emph {et~al.}(1957)\citenamefont {Kubo}, \citenamefont {Yokota},\ and\ \citenamefont {Nakajima}}]{kubo1957}%
  \BibitemOpen
  \bibfield  {author} {\bibinfo {author} {\bibfnamefont {R.}~\bibnamefont {Kubo}}, \bibinfo {author} {\bibfnamefont {M.}~\bibnamefont {Yokota}},\ and\ \bibinfo {author} {\bibfnamefont {S.}~\bibnamefont {Nakajima}},\ }\bibfield  {title} {\bibinfo {title} {{Statistical-Mechanical Theory of Irreversible Processes. II. Response to Thermal Disturbance}},\ }\href {https://doi.org/10.1143/JPSJ.12.1203} {\bibfield  {journal} {\bibinfo  {journal} {J. Phys. Soc. Jpn.}\ }\textbf {\bibinfo {volume} {12}},\ \bibinfo {pages} {1203} (\bibinfo {year} {1957})}\BibitemShut {NoStop}%
\bibitem [{\citenamefont {Luttinger}(1964)}]{luttinger1964}%
  \BibitemOpen
  \bibfield  {author} {\bibinfo {author} {\bibfnamefont {J.~M.}\ \bibnamefont {Luttinger}},\ }\bibfield  {title} {\bibinfo {title} {{Theory of Thermal Transport Coefficients}},\ }\href {https://doi.org/10.1103/PhysRev.135.A1505} {\bibfield  {journal} {\bibinfo  {journal} {Phys. Rev.}\ }\textbf {\bibinfo {volume} {135}},\ \bibinfo {pages} {A1505} (\bibinfo {year} {1964})}\BibitemShut {NoStop}%
\bibitem [{\citenamefont {Roessli}\ \emph {et~al.}(2002)\citenamefont {Roessli}, \citenamefont {B\"oni}, \citenamefont {Fischer},\ and\ \citenamefont {Endoh}}]{roessli2002}%
  \BibitemOpen
  \bibfield  {author} {\bibinfo {author} {\bibfnamefont {B.}~\bibnamefont {Roessli}}, \bibinfo {author} {\bibfnamefont {P.}~\bibnamefont {B\"oni}}, \bibinfo {author} {\bibfnamefont {W.~E.}\ \bibnamefont {Fischer}},\ and\ \bibinfo {author} {\bibfnamefont {Y.}~\bibnamefont {Endoh}},\ }\bibfield  {title} {\bibinfo {title} {{Chiral Fluctuations in MnSi above the Curie Temperature}},\ }\href {https://doi.org/10.1103/PhysRevLett.88.237204} {\bibfield  {journal} {\bibinfo  {journal} {Phys. Rev. Lett.}\ }\textbf {\bibinfo {volume} {88}},\ \bibinfo {pages} {237204} (\bibinfo {year} {2002})}\BibitemShut {NoStop}%
\bibitem [{\citenamefont {Lorenzo}\ \emph {et~al.}(2007)\citenamefont {Lorenzo}, \citenamefont {Boullier}, \citenamefont {Regnault}, \citenamefont {Ammerahl},\ and\ \citenamefont {Revcolevschi}}]{lorenzo2007}%
  \BibitemOpen
  \bibfield  {author} {\bibinfo {author} {\bibfnamefont {J.~E.}\ \bibnamefont {Lorenzo}}, \bibinfo {author} {\bibfnamefont {C.}~\bibnamefont {Boullier}}, \bibinfo {author} {\bibfnamefont {L.~P.}\ \bibnamefont {Regnault}}, \bibinfo {author} {\bibfnamefont {U.}~\bibnamefont {Ammerahl}},\ and\ \bibinfo {author} {\bibfnamefont {A.}~\bibnamefont {Revcolevschi}},\ }\bibfield  {title} {\bibinfo {title} {{Dynamical spin chirality and spin anisotropy in ${\mathrm{Sr}}_{14}{\mathrm{Cu}}_{24}{\mathrm{O}}_{41}$: A neutron polarization analysis study}},\ }\href {https://doi.org/10.1103/PhysRevB.75.054418} {\bibfield  {journal} {\bibinfo  {journal} {Phys. Rev. B}\ }\textbf {\bibinfo {volume} {75}},\ \bibinfo {pages} {054418} (\bibinfo {year} {2007})}\BibitemShut {NoStop}%
\bibitem [{\citenamefont {Nambu}\ \emph {et~al.}(2020)\citenamefont {Nambu}, \citenamefont {Barker}, \citenamefont {Okino}, \citenamefont {Kikkawa}, \citenamefont {Shiomi}, \citenamefont {Enderle}, \citenamefont {Weber}, \citenamefont {Winn}, \citenamefont {Graves-Brook}, \citenamefont {Tranquada}, \citenamefont {Ziman}, \citenamefont {Fujita}, \citenamefont {Bauer}, \citenamefont {Saitoh},\ and\ \citenamefont {Kakurai}}]{nambu2020}%
  \BibitemOpen
  \bibfield  {author} {\bibinfo {author} {\bibfnamefont {Y.}~\bibnamefont {Nambu}}, \bibinfo {author} {\bibfnamefont {J.}~\bibnamefont {Barker}}, \bibinfo {author} {\bibfnamefont {Y.}~\bibnamefont {Okino}}, \bibinfo {author} {\bibfnamefont {T.}~\bibnamefont {Kikkawa}}, \bibinfo {author} {\bibfnamefont {Y.}~\bibnamefont {Shiomi}}, \bibinfo {author} {\bibfnamefont {M.}~\bibnamefont {Enderle}}, \bibinfo {author} {\bibfnamefont {T.}~\bibnamefont {Weber}}, \bibinfo {author} {\bibfnamefont {B.}~\bibnamefont {Winn}}, \bibinfo {author} {\bibfnamefont {M.}~\bibnamefont {Graves-Brook}}, \bibinfo {author} {\bibfnamefont {J.~M.}\ \bibnamefont {Tranquada}}, \bibinfo {author} {\bibfnamefont {T.}~\bibnamefont {Ziman}}, \bibinfo {author} {\bibfnamefont {M.}~\bibnamefont {Fujita}}, \bibinfo {author} {\bibfnamefont {G.~E.~W.}\ \bibnamefont {Bauer}}, \bibinfo {author} {\bibfnamefont {E.}~\bibnamefont {Saitoh}},\ and\ \bibinfo {author} {\bibfnamefont {K.}~\bibnamefont {Kakurai}},\ }\bibfield  {title} {\bibinfo {title} {{Observation of Magnon Polarization}},\ }\href {https://doi.org/10.1103/PhysRevLett.125.027201} {\bibfield  {journal} {\bibinfo  {journal} {Phys. Rev. Lett.}\ }\textbf {\bibinfo {volume} {125}},\ \bibinfo {pages} {027201} (\bibinfo {year} {2020})}\BibitemShut {NoStop}%
\bibitem [{\citenamefont {Knetter}\ and\ \citenamefont {Uhrig}(2004)}]{knetter2004}%
  \BibitemOpen
  \bibfield  {author} {\bibinfo {author} {\bibfnamefont {C.}~\bibnamefont {Knetter}}\ and\ \bibinfo {author} {\bibfnamefont {G.~S.}\ \bibnamefont {Uhrig}},\ }\bibfield  {title} {\bibinfo {title} {{Dynamic Structure Factor of the Two-Dimensional Shastry-Sutherland Model}},\ }\href {https://doi.org/10.1103/PhysRevLett.92.027204} {\bibfield  {journal} {\bibinfo  {journal} {Phys. Rev. Lett.}\ }\textbf {\bibinfo {volume} {92}},\ \bibinfo {pages} {027204} (\bibinfo {year} {2004})}\BibitemShut {NoStop}%
\bibitem [{\citenamefont {Loh\"ofer}\ \emph {et~al.}(2015)\citenamefont {Loh\"ofer}, \citenamefont {Coletta}, \citenamefont {Joshi}, \citenamefont {Assaad}, \citenamefont {Vojta}, \citenamefont {Wessel},\ and\ \citenamefont {Mila}}]{lohofer2015}%
  \BibitemOpen
  \bibfield  {author} {\bibinfo {author} {\bibfnamefont {M.}~\bibnamefont {Loh\"ofer}}, \bibinfo {author} {\bibfnamefont {T.}~\bibnamefont {Coletta}}, \bibinfo {author} {\bibfnamefont {D.~G.}\ \bibnamefont {Joshi}}, \bibinfo {author} {\bibfnamefont {F.~F.}\ \bibnamefont {Assaad}}, \bibinfo {author} {\bibfnamefont {M.}~\bibnamefont {Vojta}}, \bibinfo {author} {\bibfnamefont {S.}~\bibnamefont {Wessel}},\ and\ \bibinfo {author} {\bibfnamefont {F.}~\bibnamefont {Mila}},\ }\bibfield  {title} {\bibinfo {title} {{Dynamical structure factors and excitation modes of the bilayer Heisenberg model}},\ }\href {https://doi.org/10.1103/PhysRevB.92.245137} {\bibfield  {journal} {\bibinfo  {journal} {Phys. Rev. B}\ }\textbf {\bibinfo {volume} {92}},\ \bibinfo {pages} {245137} (\bibinfo {year} {2015})}\BibitemShut {NoStop}%
\bibitem [{\citenamefont {Calonge-Mart^^c3^^adnez}\ \emph {et~al.}(shed)\citenamefont {Calonge-Mart^^c3^^adnez}, \citenamefont {Rao}, \citenamefont {Mila},\ and\ \citenamefont {Knolle}}]{martinez2026_arxiv}%
  \BibitemOpen
  \bibfield  {author} {\bibinfo {author} {\bibfnamefont {L.}~\bibnamefont {Calonge-Mart^^c3^^adnez}}, \bibinfo {author} {\bibfnamefont {P.}~\bibnamefont {Rao}}, \bibinfo {author} {\bibfnamefont {F.}~\bibnamefont {Mila}},\ and\ \bibinfo {author} {\bibfnamefont {J.}~\bibnamefont {Knolle}},\ }\bibfield  {title} {\bibinfo {title} {{Topological Triplons in the Pinwheel Valence Bond Solid on the Kagome Lattice}},\ }\href {https://arxiv.org/abs/2606.09823} {\bibfield  {journal} {\bibinfo  {journal} {arXiv:2606.09823}\ } (\bibinfo {year} {unpublished})}\BibitemShut {NoStop}%
\bibitem [{\citenamefont {Di~Matteo}\ \emph {et~al.}(2005)\citenamefont {Di~Matteo}, \citenamefont {Jackeli},\ and\ \citenamefont {Perkins}}]{matteo2005}%
  \BibitemOpen
  \bibfield  {author} {\bibinfo {author} {\bibfnamefont {S.}~\bibnamefont {Di~Matteo}}, \bibinfo {author} {\bibfnamefont {G.}~\bibnamefont {Jackeli}},\ and\ \bibinfo {author} {\bibfnamefont {N.~B.}\ \bibnamefont {Perkins}},\ }\bibfield  {title} {\bibinfo {title} {{Valence-bond crystal and lattice distortions in a pyrochlore antiferromagnet with orbital degeneracy}},\ }\href {https://doi.org/10.1103/PhysRevB.72.024431} {\bibfield  {journal} {\bibinfo  {journal} {Phys. Rev. B}\ }\textbf {\bibinfo {volume} {72}},\ \bibinfo {pages} {024431} (\bibinfo {year} {2005})}\BibitemShut {NoStop}%
\bibitem [{\citenamefont {Garc{\'{\i}}a}\ \emph {et~al.}(2005)\citenamefont {Garc{\'{\i}}a}, \citenamefont {Dahaoui}, \citenamefont {Fertey}, \citenamefont {Wenger},\ and\ \citenamefont {Lecomte}}]{garcia2005}%
  \BibitemOpen
  \bibfield  {author} {\bibinfo {author} {\bibfnamefont {P.}~\bibnamefont {Garc{\'{\i}}a}}, \bibinfo {author} {\bibfnamefont {S.}~\bibnamefont {Dahaoui}}, \bibinfo {author} {\bibfnamefont {P.}~\bibnamefont {Fertey}}, \bibinfo {author} {\bibfnamefont {E.}~\bibnamefont {Wenger}},\ and\ \bibinfo {author} {\bibfnamefont {C.}~\bibnamefont {Lecomte}},\ }\bibfield  {title} {\bibinfo {title} {{Crystallographic investigation of temperature-induced phase transition of the tetrathiafulvalene-$p$-bromanil, TTF-BA charge transfer complex}},\ }\href {https://doi.org/10.1103/PhysRevB.72.104115} {\bibfield  {journal} {\bibinfo  {journal} {Phys. Rev. B}\ }\textbf {\bibinfo {volume} {72}},\ \bibinfo {pages} {104115} (\bibinfo {year} {2005})}\BibitemShut {NoStop}%
\bibitem [{\citenamefont {Geirhos}\ \emph {et~al.}(2021)\citenamefont {Geirhos}, \citenamefont {Langmann}, \citenamefont {Prodan}, \citenamefont {Tsirlin}, \citenamefont {Missiul}, \citenamefont {Eickerling}, \citenamefont {Jesche}, \citenamefont {Tsurkan}, \citenamefont {Lunkenheimer}, \citenamefont {Scherer},\ and\ \citenamefont {K{\'e}zsm{\'a}rki}}]{geirhos2021}%
  \BibitemOpen
  \bibfield  {author} {\bibinfo {author} {\bibfnamefont {K.}~\bibnamefont {Geirhos}}, \bibinfo {author} {\bibfnamefont {J.}~\bibnamefont {Langmann}}, \bibinfo {author} {\bibfnamefont {L.}~\bibnamefont {Prodan}}, \bibinfo {author} {\bibfnamefont {A.~A.}\ \bibnamefont {Tsirlin}}, \bibinfo {author} {\bibfnamefont {A.}~\bibnamefont {Missiul}}, \bibinfo {author} {\bibfnamefont {G.}~\bibnamefont {Eickerling}}, \bibinfo {author} {\bibfnamefont {A.}~\bibnamefont {Jesche}}, \bibinfo {author} {\bibfnamefont {V.}~\bibnamefont {Tsurkan}}, \bibinfo {author} {\bibfnamefont {P.}~\bibnamefont {Lunkenheimer}}, \bibinfo {author} {\bibfnamefont {W.}~\bibnamefont {Scherer}},\ and\ \bibinfo {author} {\bibfnamefont {I.}~\bibnamefont {K{\'e}zsm{\'a}rki}},\ }\bibfield  {title} {\bibinfo {title} {{Cooperative Cluster Jahn-Teller Effect as a Possible Route to Antiferroelectricity}},\ }\href {https://doi.org/10.1103/PhysRevLett.126.187601} {\bibfield  {journal} {\bibinfo  {journal} {Phys. Rev. Lett.}\ }\textbf {\bibinfo {volume} {126}},\ \bibinfo {pages} {187601} (\bibinfo {year} {2021})}\BibitemShut {NoStop}%
\bibitem [{\citenamefont {Langmann}\ \emph {et~al.}(2024)\citenamefont {Langmann}, \citenamefont {Eickerling}, \citenamefont {Prodan}, \citenamefont {Tsirlin}, \citenamefont {Winkler}, \citenamefont {Bord{\'a}cs}, \citenamefont {Tsurkan},\ and\ \citenamefont {K{\'e}zsm{\'a}rki}}]{langmann2024}%
  \BibitemOpen
  \bibfield  {author} {\bibinfo {author} {\bibfnamefont {J.}~\bibnamefont {Langmann}}, \bibinfo {author} {\bibfnamefont {G.}~\bibnamefont {Eickerling}}, \bibinfo {author} {\bibfnamefont {L.}~\bibnamefont {Prodan}}, \bibinfo {author} {\bibfnamefont {A.~A.}\ \bibnamefont {Tsirlin}}, \bibinfo {author} {\bibfnamefont {M.}~\bibnamefont {Winkler}}, \bibinfo {author} {\bibfnamefont {S.}~\bibnamefont {Bord{\'a}cs}}, \bibinfo {author} {\bibfnamefont {V.}~\bibnamefont {Tsurkan}},\ and\ \bibinfo {author} {\bibfnamefont {I.}~\bibnamefont {K{\'e}zsm{\'a}rki}},\ }\bibfield  {title} {\bibinfo {title} {{Atomic-Scale Polar Helix in Inorganic Crystals}},\ }\href {https://doi.org/10.1021/acs.chemmater.4c02060} {\bibfield  {journal} {\bibinfo  {journal} {Chem. Mater.}\ }\textbf {\bibinfo {volume} {36}},\ \bibinfo {pages} {11180} (\bibinfo {year} {2024})}\BibitemShut {NoStop}%
\bibitem [{\citenamefont {Waki}\ \emph {et~al.}(2010)\citenamefont {Waki}, \citenamefont {Kajinami}, \citenamefont {Tabata}, \citenamefont {Nakamura}, \citenamefont {Yoshida}, \citenamefont {Takigawa},\ and\ \citenamefont {Watanabe}}]{waki2010}%
  \BibitemOpen
  \bibfield  {author} {\bibinfo {author} {\bibfnamefont {T.}~\bibnamefont {Waki}}, \bibinfo {author} {\bibfnamefont {Y.}~\bibnamefont {Kajinami}}, \bibinfo {author} {\bibfnamefont {Y.}~\bibnamefont {Tabata}}, \bibinfo {author} {\bibfnamefont {H.}~\bibnamefont {Nakamura}}, \bibinfo {author} {\bibfnamefont {M.}~\bibnamefont {Yoshida}}, \bibinfo {author} {\bibfnamefont {M.}~\bibnamefont {Takigawa}},\ and\ \bibinfo {author} {\bibfnamefont {I.}~\bibnamefont {Watanabe}},\ }\bibfield  {title} {\bibinfo {title} {{Spin-singlet state formation in the cluster Mott insulator ${\text{GaNb}}_{4}{\text{S}}_{8}$ studied by $\ensuremath{\mu}\text{SR}$ and NMR spectroscopy}},\ }\href {https://doi.org/10.1103/PhysRevB.81.020401} {\bibfield  {journal} {\bibinfo  {journal} {Phys. Rev. B}\ }\textbf {\bibinfo {volume} {81}},\ \bibinfo {pages} {020401(R)} (\bibinfo {year} {2010})}\BibitemShut {NoStop}%
\bibitem [{\citenamefont {Ishikawa}\ \emph {et~al.}(2020)\citenamefont {Ishikawa}, \citenamefont {Yajima}, \citenamefont {Matsuo}, \citenamefont {Ihara},\ and\ \citenamefont {Kindo}}]{ishikawa2020}%
  \BibitemOpen
  \bibfield  {author} {\bibinfo {author} {\bibfnamefont {H.}~\bibnamefont {Ishikawa}}, \bibinfo {author} {\bibfnamefont {T.}~\bibnamefont {Yajima}}, \bibinfo {author} {\bibfnamefont {A.}~\bibnamefont {Matsuo}}, \bibinfo {author} {\bibfnamefont {Y.}~\bibnamefont {Ihara}},\ and\ \bibinfo {author} {\bibfnamefont {K.}~\bibnamefont {Kindo}},\ }\bibfield  {title} {\bibinfo {title} {{Nonmagnetic Ground States and a Possible Quadrupolar Phase in $4d$ and $5d$ Lacunar Spinel Selenides $\mathrm{Ga}{M}_{4}{\mathrm{Se}}_{8}$ ($M=\text{Nb}$, Ta)}},\ }\href {https://doi.org/10.1103/PhysRevLett.124.227202} {\bibfield  {journal} {\bibinfo  {journal} {Phys. Rev. Lett.}\ }\textbf {\bibinfo {volume} {124}},\ \bibinfo {pages} {227202} (\bibinfo {year} {2020})}\BibitemShut {NoStop}%
\bibitem [{\citenamefont {Yang}\ \emph {et~al.}(2022)\citenamefont {Yang}, \citenamefont {Kawamoto}, \citenamefont {Higo}, \citenamefont {Wang}, \citenamefont {Stone}, \citenamefont {Neuefeind}, \citenamefont {Ruff}, \citenamefont {Abeykoon}, \citenamefont {Chen}, \citenamefont {Nakatsuji},\ and\ \citenamefont {Plumb}}]{yang2022}%
  \BibitemOpen
  \bibfield  {author} {\bibinfo {author} {\bibfnamefont {T.-H.}\ \bibnamefont {Yang}}, \bibinfo {author} {\bibfnamefont {S.}~\bibnamefont {Kawamoto}}, \bibinfo {author} {\bibfnamefont {T.}~\bibnamefont {Higo}}, \bibinfo {author} {\bibfnamefont {S.~G.}\ \bibnamefont {Wang}}, \bibinfo {author} {\bibfnamefont {M.~B.}\ \bibnamefont {Stone}}, \bibinfo {author} {\bibfnamefont {J.}~\bibnamefont {Neuefeind}}, \bibinfo {author} {\bibfnamefont {J.~P.~C.}\ \bibnamefont {Ruff}}, \bibinfo {author} {\bibfnamefont {A.~M.~M.}\ \bibnamefont {Abeykoon}}, \bibinfo {author} {\bibfnamefont {Y.-S.}\ \bibnamefont {Chen}}, \bibinfo {author} {\bibfnamefont {S.}~\bibnamefont {Nakatsuji}},\ and\ \bibinfo {author} {\bibfnamefont {K.~W.}\ \bibnamefont {Plumb}},\ }\bibfield  {title} {\bibinfo {title} {{Bond ordering and molecular spin-orbital fluctuations in the cluster Mott insulator ${\mathrm{GaTa}}_{4}{\mathrm{Se}}_{8}$}},\ }\href {https://doi.org/10.1103/PhysRevResearch.4.033123} {\bibfield  {journal} {\bibinfo  {journal} {Phys. Rev. Res.}\ }\textbf {\bibinfo {volume} {4}},\ \bibinfo {pages} {033123} (\bibinfo {year} {2022})}\BibitemShut {NoStop}%
\bibitem [{\citenamefont {Sunami}\ \emph {et~al.}(2020)\citenamefont {Sunami}, \citenamefont {Sakai}, \citenamefont {Takehara}, \citenamefont {Adachi}, \citenamefont {Miyagawa}, \citenamefont {Horiuchi},\ and\ \citenamefont {Kanoda}}]{sunami2020}%
  \BibitemOpen
  \bibfield  {author} {\bibinfo {author} {\bibfnamefont {K.}~\bibnamefont {Sunami}}, \bibinfo {author} {\bibfnamefont {Y.}~\bibnamefont {Sakai}}, \bibinfo {author} {\bibfnamefont {R.}~\bibnamefont {Takehara}}, \bibinfo {author} {\bibfnamefont {H.}~\bibnamefont {Adachi}}, \bibinfo {author} {\bibfnamefont {K.}~\bibnamefont {Miyagawa}}, \bibinfo {author} {\bibfnamefont {S.}~\bibnamefont {Horiuchi}},\ and\ \bibinfo {author} {\bibfnamefont {K.}~\bibnamefont {Kanoda}},\ }\bibfield  {title} {\bibinfo {title} {{Magnetic excitations in an ionic spin-chain system with a nonmagnetic ferroelectric instability}},\ }\href {https://doi.org/10.1103/PhysRevResearch.2.043333} {\bibfield  {journal} {\bibinfo  {journal} {Phys. Rev. Res.}\ }\textbf {\bibinfo {volume} {2}},\ \bibinfo {pages} {043333} (\bibinfo {year} {2020})}\BibitemShut {NoStop}%
\bibitem [{\citenamefont {Sawa}\ \emph {et~al.}(2002)\citenamefont {Sawa}, \citenamefont {Ninomiya}, \citenamefont {Ohama}, \citenamefont {Nakao}, \citenamefont {Ohwada}, \citenamefont {Murakami}, \citenamefont {Fujii}, \citenamefont {Noda}, \citenamefont {Isobe},\ and\ \citenamefont {Ueda}}]{sawa2002}%
  \BibitemOpen
  \bibfield  {author} {\bibinfo {author} {\bibfnamefont {H.}~\bibnamefont {Sawa}}, \bibinfo {author} {\bibfnamefont {E.}~\bibnamefont {Ninomiya}}, \bibinfo {author} {\bibfnamefont {T.}~\bibnamefont {Ohama}}, \bibinfo {author} {\bibfnamefont {H.}~\bibnamefont {Nakao}}, \bibinfo {author} {\bibfnamefont {K.}~\bibnamefont {Ohwada}}, \bibinfo {author} {\bibfnamefont {Y.}~\bibnamefont {Murakami}}, \bibinfo {author} {\bibfnamefont {Y.}~\bibnamefont {Fujii}}, \bibinfo {author} {\bibfnamefont {Y.}~\bibnamefont {Noda}}, \bibinfo {author} {\bibfnamefont {M.}~\bibnamefont {Isobe}},\ and\ \bibinfo {author} {\bibfnamefont {Y.}~\bibnamefont {Ueda}},\ }\bibfield  {title} {\bibinfo {title} {{Low-Temperature Structure of the Quarter-Filled Ladder Compound $\alpha'$-NaV$_2$O$_5$}},\ }\href {https://doi.org/10.1143/JPSJ.71.385} {\bibfield  {journal} {\bibinfo  {journal} {J. Phys. Soc. Jpn.}\ }\textbf {\bibinfo {volume} {71}},\ \bibinfo {pages} {385} (\bibinfo {year} {2002})}\BibitemShut {NoStop}%
\bibitem [{\citenamefont {Ohwada}\ \emph {et~al.}(2005)\citenamefont {Ohwada}, \citenamefont {Fujii}, \citenamefont {Katsuki}, \citenamefont {Muraoka}, \citenamefont {Nakao}, \citenamefont {Murakami}, \citenamefont {Sawa}, \citenamefont {Ninomiya}, \citenamefont {Isobe},\ and\ \citenamefont {Ueda}}]{ohwada2005}%
  \BibitemOpen
  \bibfield  {author} {\bibinfo {author} {\bibfnamefont {K.}~\bibnamefont {Ohwada}}, \bibinfo {author} {\bibfnamefont {Y.}~\bibnamefont {Fujii}}, \bibinfo {author} {\bibfnamefont {Y.}~\bibnamefont {Katsuki}}, \bibinfo {author} {\bibfnamefont {J.}~\bibnamefont {Muraoka}}, \bibinfo {author} {\bibfnamefont {H.}~\bibnamefont {Nakao}}, \bibinfo {author} {\bibfnamefont {Y.}~\bibnamefont {Murakami}}, \bibinfo {author} {\bibfnamefont {H.}~\bibnamefont {Sawa}}, \bibinfo {author} {\bibfnamefont {E.}~\bibnamefont {Ninomiya}}, \bibinfo {author} {\bibfnamefont {M.}~\bibnamefont {Isobe}},\ and\ \bibinfo {author} {\bibfnamefont {Y.}~\bibnamefont {Ueda}},\ }\bibfield  {title} {\bibinfo {title} {{Charge-Order Pattern of the Low-Temperature Phase from a Monoclinic Single Domain of ${\mathrm{NaV}}_{2}{\mathrm{O}}_{5}$ Uniquely Determined by Resonant X-Ray Scattering}},\ }\href {https://doi.org/10.1103/PhysRevLett.94.106401} {\bibfield  {journal} {\bibinfo  {journal} {Phys. Rev. Lett.}\ }\textbf {\bibinfo {volume} {94}},\ \bibinfo {pages} {106401} (\bibinfo {year} {2005})}\BibitemShut {NoStop}%
\bibitem [{\citenamefont {Seo}\ and\ \citenamefont {Fukuyama}(1998)}]{seo1998}%
  \BibitemOpen
  \bibfield  {author} {\bibinfo {author} {\bibfnamefont {H.}~\bibnamefont {Seo}}\ and\ \bibinfo {author} {\bibfnamefont {H.}~\bibnamefont {Fukuyama}},\ }\bibfield  {title} {\bibinfo {title} {{Charge Ordering and Spin Gap in NaV$_2$O$_5$}},\ }\href {https://doi.org/10.1143/jpsj.67.2602} {\bibfield  {journal} {\bibinfo  {journal} {J. Phys. Soc. Jpn.}\ }\textbf {\bibinfo {volume} {67}},\ \bibinfo {pages} {2602} (\bibinfo {year} {1998})}\BibitemShut {NoStop}%
\bibitem [{\citenamefont {Mostovoy}\ and\ \citenamefont {Khomskii}(1999)}]{mostovoy1999}%
  \BibitemOpen
  \bibfield  {author} {\bibinfo {author} {\bibfnamefont {M.}~\bibnamefont {Mostovoy}}\ and\ \bibinfo {author} {\bibfnamefont {D.}~\bibnamefont {Khomskii}},\ }\bibfield  {title} {\bibinfo {title} {{Charge ordering and opening of spin gap in NaV$_2$O$_5$}},\ }\href {https://doi.org/https://doi.org/10.1016/S0038-1098(99)00453-6} {\bibfield  {journal} {\bibinfo  {journal} {Solid State Commun.}\ }\textbf {\bibinfo {volume} {113}},\ \bibinfo {pages} {159} (\bibinfo {year} {1999})}\BibitemShut {NoStop}%
\bibitem [{\citenamefont {Yang}\ \emph {et~al.}(2024)\citenamefont {Yang}, \citenamefont {Chang}, \citenamefont {Chen},\ and\ \citenamefont {Plumb}}]{yang2024}%
  \BibitemOpen
  \bibfield  {author} {\bibinfo {author} {\bibfnamefont {T.-H.}\ \bibnamefont {Yang}}, \bibinfo {author} {\bibfnamefont {T.}~\bibnamefont {Chang}}, \bibinfo {author} {\bibfnamefont {Y.-S.}\ \bibnamefont {Chen}},\ and\ \bibinfo {author} {\bibfnamefont {K.~W.}\ \bibnamefont {Plumb}},\ }\bibfield  {title} {\bibinfo {title} {{Jahn-Teller driven quadrupolar ordering and spin-orbital dimer formation in ${\mathrm{GaNb}}_{4}{\mathrm{Se}}_{8}$}},\ }\href {https://doi.org/10.1103/PhysRevB.109.144101} {\bibfield  {journal} {\bibinfo  {journal} {Phys. Rev. B}\ }\textbf {\bibinfo {volume} {109}},\ \bibinfo {pages} {144101} (\bibinfo {year} {2024})}\BibitemShut {NoStop}%
\bibitem [{\citenamefont {Kitou}\ \emph {et~al.}(2017)\citenamefont {Kitou}, \citenamefont {Fujii}, \citenamefont {Kawamoto}, \citenamefont {Katayama}, \citenamefont {Maki}, \citenamefont {Nishibori}, \citenamefont {Sugimoto}, \citenamefont {Takata}, \citenamefont {Nakamura},\ and\ \citenamefont {Sawa}}]{kitou2017}%
  \BibitemOpen
  \bibfield  {author} {\bibinfo {author} {\bibfnamefont {S.}~\bibnamefont {Kitou}}, \bibinfo {author} {\bibfnamefont {T.}~\bibnamefont {Fujii}}, \bibinfo {author} {\bibfnamefont {T.}~\bibnamefont {Kawamoto}}, \bibinfo {author} {\bibfnamefont {N.}~\bibnamefont {Katayama}}, \bibinfo {author} {\bibfnamefont {S.}~\bibnamefont {Maki}}, \bibinfo {author} {\bibfnamefont {E.}~\bibnamefont {Nishibori}}, \bibinfo {author} {\bibfnamefont {K.}~\bibnamefont {Sugimoto}}, \bibinfo {author} {\bibfnamefont {M.}~\bibnamefont {Takata}}, \bibinfo {author} {\bibfnamefont {T.}~\bibnamefont {Nakamura}},\ and\ \bibinfo {author} {\bibfnamefont {H.}~\bibnamefont {Sawa}},\ }\bibfield  {title} {\bibinfo {title} {{Successive Dimensional Transition in ${(\mathrm{TMTTF})}_{2}{\mathrm{PF}}_{6}$ Revealed by Synchrotron X-ray Diffraction}},\ }\href {https://doi.org/10.1103/PhysRevLett.119.065701} {\bibfield  {journal} {\bibinfo  {journal} {Phys. Rev. Lett.}\ }\textbf {\bibinfo {volume} {119}},\ \bibinfo {pages} {065701} (\bibinfo {year} {2017})}\BibitemShut {NoStop}%
\bibitem [{\citenamefont {Luther}\ \emph {et~al.}(1998)\citenamefont {Luther}, \citenamefont {Nojiri}, \citenamefont {Motokawa}, \citenamefont {Isobe},\ and\ \citenamefont {Ueda}}]{luther1998}%
  \BibitemOpen
  \bibfield  {author} {\bibinfo {author} {\bibfnamefont {S.}~\bibnamefont {Luther}}, \bibinfo {author} {\bibfnamefont {H.}~\bibnamefont {Nojiri}}, \bibinfo {author} {\bibfnamefont {M.}~\bibnamefont {Motokawa}}, \bibinfo {author} {\bibfnamefont {M.}~\bibnamefont {Isobe}},\ and\ \bibinfo {author} {\bibfnamefont {Y.}~\bibnamefont {Ueda}},\ }\bibfield  {title} {\bibinfo {title} {{Direct Observation of the Spin Gap in NaV$_2$O$_5$ by Submillimeter Wave ESR}},\ }\href {https://doi.org/10.1143/JPSJ.67.3715} {\bibfield  {journal} {\bibinfo  {journal} {J. Phys. Soc. Jpn.}\ }\textbf {\bibinfo {volume} {67}},\ \bibinfo {pages} {3715} (\bibinfo {year} {1998})}\BibitemShut {NoStop}%
\bibitem [{\citenamefont {Yosihama}\ \emph {et~al.}(1998)\citenamefont {Yosihama}, \citenamefont {Nishi}, \citenamefont {Nakajima}, \citenamefont {Kakurai}, \citenamefont {Fujii}, \citenamefont {Isobe}, \citenamefont {Kagami},\ and\ \citenamefont {Ueda}}]{yoshihama1998}%
  \BibitemOpen
  \bibfield  {author} {\bibinfo {author} {\bibfnamefont {T.}~\bibnamefont {Yosihama}}, \bibinfo {author} {\bibfnamefont {M.}~\bibnamefont {Nishi}}, \bibinfo {author} {\bibfnamefont {K.}~\bibnamefont {Nakajima}}, \bibinfo {author} {\bibfnamefont {K.}~\bibnamefont {Kakurai}}, \bibinfo {author} {\bibfnamefont {Y.}~\bibnamefont {Fujii}}, \bibinfo {author} {\bibfnamefont {M.}~\bibnamefont {Isobe}}, \bibinfo {author} {\bibfnamefont {C.}~\bibnamefont {Kagami}},\ and\ \bibinfo {author} {\bibfnamefont {Y.}~\bibnamefont {Ueda}},\ }\bibfield  {title} {\bibinfo {title} {Spin dynamics in nav$_2$o$_5$ ^^e2^^80^^93 inelastic neutron scattering},\ }\href {https://doi.org/10.1143/JPSJ.67.744} {\bibfield  {journal} {\bibinfo  {journal} {J. Phys. Soc. Jpn.}\ }\textbf {\bibinfo {volume} {67}},\ \bibinfo {pages} {744} (\bibinfo {year} {1998})}\BibitemShut {NoStop}%
\bibitem [{\citenamefont {Grenier}\ \emph {et~al.}(2001)\citenamefont {Grenier}, \citenamefont {Cepas}, \citenamefont {Regnault}, \citenamefont {Lorenzo}, \citenamefont {Ziman}, \citenamefont {Boucher}, \citenamefont {Hiess}, \citenamefont {Chatterji}, \citenamefont {Jegoudez},\ and\ \citenamefont {Revcolevschi}}]{grenier2001}%
  \BibitemOpen
  \bibfield  {author} {\bibinfo {author} {\bibfnamefont {B.}~\bibnamefont {Grenier}}, \bibinfo {author} {\bibfnamefont {O.}~\bibnamefont {Cepas}}, \bibinfo {author} {\bibfnamefont {L.~P.}\ \bibnamefont {Regnault}}, \bibinfo {author} {\bibfnamefont {J.~E.}\ \bibnamefont {Lorenzo}}, \bibinfo {author} {\bibfnamefont {T.}~\bibnamefont {Ziman}}, \bibinfo {author} {\bibfnamefont {J.~P.}\ \bibnamefont {Boucher}}, \bibinfo {author} {\bibfnamefont {A.}~\bibnamefont {Hiess}}, \bibinfo {author} {\bibfnamefont {T.}~\bibnamefont {Chatterji}}, \bibinfo {author} {\bibfnamefont {J.}~\bibnamefont {Jegoudez}},\ and\ \bibinfo {author} {\bibfnamefont {A.}~\bibnamefont {Revcolevschi}},\ }\bibfield  {title} {\bibinfo {title} {{Charge Ordering and Spin Dynamics in ${\mathrm{NaV}}_{2}{O}_{5}$}},\ }\href {https://doi.org/10.1103/PhysRevLett.86.5966} {\bibfield  {journal} {\bibinfo  {journal} {Phys. Rev. Lett.}\ }\textbf {\bibinfo {volume} {86}},\ \bibinfo {pages} {5966} (\bibinfo {year} {2001})}\BibitemShut {NoStop}%
\bibitem [{\citenamefont {R{\~o}{\~o}m}\ \emph {et~al.}(2004)\citenamefont {R{\~o}{\~o}m}, \citenamefont {H{\"u}vonen}, \citenamefont {Nagel}, \citenamefont {Wang},\ and\ \citenamefont {Kremer}}]{room2004}%
  \BibitemOpen
  \bibfield  {author} {\bibinfo {author} {\bibfnamefont {T.}~\bibnamefont {R{\~o}{\~o}m}}, \bibinfo {author} {\bibfnamefont {D.}~\bibnamefont {H{\"u}vonen}}, \bibinfo {author} {\bibfnamefont {U.}~\bibnamefont {Nagel}}, \bibinfo {author} {\bibfnamefont {Y.-J.}\ \bibnamefont {Wang}},\ and\ \bibinfo {author} {\bibfnamefont {R.~K.}\ \bibnamefont {Kremer}},\ }\bibfield  {title} {\bibinfo {title} {{Low-energy excitations and dynamic Dzyaloshinskii-Moriya interaction in $\alpha'$-${\mathrm{NaV}}_{2}{\mathrm{O}}_{5}$ studied by far-infrared spectroscopy}},\ }\href {https://doi.org/10.1103/PhysRevB.69.144410} {\bibfield  {journal} {\bibinfo  {journal} {Phys. Rev. B}\ }\textbf {\bibinfo {volume} {69}},\ \bibinfo {pages} {144410} (\bibinfo {year} {2004})}\BibitemShut {NoStop}%
\bibitem [{\citenamefont {Dumm}\ \emph {et~al.}(2000)\citenamefont {Dumm}, \citenamefont {Loidl}, \citenamefont {Fravel}, \citenamefont {Starkey}, \citenamefont {Montgomery},\ and\ \citenamefont {Dressel}}]{dumm2000}%
  \BibitemOpen
  \bibfield  {author} {\bibinfo {author} {\bibfnamefont {M.}~\bibnamefont {Dumm}}, \bibinfo {author} {\bibfnamefont {A.}~\bibnamefont {Loidl}}, \bibinfo {author} {\bibfnamefont {B.~W.}\ \bibnamefont {Fravel}}, \bibinfo {author} {\bibfnamefont {K.~P.}\ \bibnamefont {Starkey}}, \bibinfo {author} {\bibfnamefont {L.~K.}\ \bibnamefont {Montgomery}},\ and\ \bibinfo {author} {\bibfnamefont {M.}~\bibnamefont {Dressel}},\ }\bibfield  {title} {\bibinfo {title} {{Electron spin resonance studies on the organic linear-chain compounds $(\mathrm{TMT}C\mathrm{F}{)}_{2}X (C=\mathrm{S},\mathrm{Se};$ $X={\mathrm{PF}}_{6},{\mathrm{AsF}}_{6},{\mathrm{ClO}}_{4},\mathrm{Br})$}},\ }\href {https://doi.org/10.1103/PhysRevB.61.511} {\bibfield  {journal} {\bibinfo  {journal} {Phys. Rev. B}\ }\textbf {\bibinfo {volume} {61}},\ \bibinfo {pages} {511} (\bibinfo {year} {2000})}\BibitemShut {NoStop}%
\bibitem [{\citenamefont {Foury-Leylekian}\ \emph {et~al.}(2009)\citenamefont {Foury-Leylekian}, \citenamefont {Petit}, \citenamefont {Coulon}, \citenamefont {Hennion}, \citenamefont {Moradpour},\ and\ \citenamefont {Pouget}}]{leylekian2009}%
  \BibitemOpen
  \bibfield  {author} {\bibinfo {author} {\bibfnamefont {P.}~\bibnamefont {Foury-Leylekian}}, \bibinfo {author} {\bibfnamefont {S.}~\bibnamefont {Petit}}, \bibinfo {author} {\bibfnamefont {C.}~\bibnamefont {Coulon}}, \bibinfo {author} {\bibfnamefont {B.}~\bibnamefont {Hennion}}, \bibinfo {author} {\bibfnamefont {A.}~\bibnamefont {Moradpour}},\ and\ \bibinfo {author} {\bibfnamefont {J.-P.}\ \bibnamefont {Pouget}},\ }\bibfield  {title} {\bibinfo {title} {{Inelastic neutron scattering investigation of magnetic excitations in the spin-Peierls ground state of (TMTTF)$_2$PF$_6$}},\ }\href {https://doi.org/https://doi.org/10.1016/j.physb.2008.11.035} {\bibfield  {journal} {\bibinfo  {journal} {Phys. B: Condens. Matter}\ }\textbf {\bibinfo {volume} {404}},\ \bibinfo {pages} {537} (\bibinfo {year} {2009})}\BibitemShut {NoStop}%
\bibitem [{\citenamefont {Pouget}\ \emph {et~al.}(2017)\citenamefont {Pouget}, \citenamefont {Foury-Leylekian}, \citenamefont {Petit}, \citenamefont {Hennion}, \citenamefont {Coulon},\ and\ \citenamefont {Bourbonnais}}]{pouget2017}%
  \BibitemOpen
  \bibfield  {author} {\bibinfo {author} {\bibfnamefont {J.~P.}\ \bibnamefont {Pouget}}, \bibinfo {author} {\bibfnamefont {P.}~\bibnamefont {Foury-Leylekian}}, \bibinfo {author} {\bibfnamefont {S.}~\bibnamefont {Petit}}, \bibinfo {author} {\bibfnamefont {B.}~\bibnamefont {Hennion}}, \bibinfo {author} {\bibfnamefont {C.}~\bibnamefont {Coulon}},\ and\ \bibinfo {author} {\bibfnamefont {C.}~\bibnamefont {Bourbonnais}},\ }\bibfield  {title} {\bibinfo {title} {{Inelastic neutron scattering investigation of magnetostructural excitations in the spin-Peierls organic system ${(\mathrm{TMTTF})}_{2}\mathrm{P}{\mathrm{F}}_{6}$}},\ }\href {https://doi.org/10.1103/PhysRevB.96.035127} {\bibfield  {journal} {\bibinfo  {journal} {Phys. Rev. B}\ }\textbf {\bibinfo {volume} {96}},\ \bibinfo {pages} {035127} (\bibinfo {year} {2017})}\BibitemShut {NoStop}%
\bibitem [{\citenamefont {Furukawa}\ \emph {et~al.}(2009)\citenamefont {Furukawa}, \citenamefont {Hara},\ and\ \citenamefont {Nakamura}}]{furukawa2009}%
  \BibitemOpen
  \bibfield  {author} {\bibinfo {author} {\bibfnamefont {K.}~\bibnamefont {Furukawa}}, \bibinfo {author} {\bibfnamefont {T.}~\bibnamefont {Hara}},\ and\ \bibinfo {author} {\bibfnamefont {T.}~\bibnamefont {Nakamura}},\ }\bibfield  {title} {\bibinfo {title} {{Anomalous Temperature Dependence of g-Tensor in Organic Conductor, (TMTTF)$_2$X (X=Br, PF$_6$, and SbF$_6$)}},\ }\href {https://doi.org/10.1143/JPSJ.78.104713} {\bibfield  {journal} {\bibinfo  {journal} {J. Phys. Soc. Jpn.}\ }\textbf {\bibinfo {volume} {78}},\ \bibinfo {pages} {104713} (\bibinfo {year} {2009})}\BibitemShut {NoStop}%
\bibitem [{\citenamefont {Yasin}\ \emph {et~al.}(2012)\citenamefont {Yasin}, \citenamefont {Salameh}, \citenamefont {Rose}, \citenamefont {Dumm}, \citenamefont {Krug~von Nidda}, \citenamefont {Loidl}, \citenamefont {Ozerov}, \citenamefont {Untereiner}, \citenamefont {Montgomery},\ and\ \citenamefont {Dressel}}]{yasin2012}%
  \BibitemOpen
  \bibfield  {author} {\bibinfo {author} {\bibfnamefont {S.}~\bibnamefont {Yasin}}, \bibinfo {author} {\bibfnamefont {B.}~\bibnamefont {Salameh}}, \bibinfo {author} {\bibfnamefont {E.}~\bibnamefont {Rose}}, \bibinfo {author} {\bibfnamefont {M.}~\bibnamefont {Dumm}}, \bibinfo {author} {\bibfnamefont {H.-A.}\ \bibnamefont {Krug~von Nidda}}, \bibinfo {author} {\bibfnamefont {A.}~\bibnamefont {Loidl}}, \bibinfo {author} {\bibfnamefont {M.}~\bibnamefont {Ozerov}}, \bibinfo {author} {\bibfnamefont {G.}~\bibnamefont {Untereiner}}, \bibinfo {author} {\bibfnamefont {L.}~\bibnamefont {Montgomery}},\ and\ \bibinfo {author} {\bibfnamefont {M.}~\bibnamefont {Dressel}},\ }\bibfield  {title} {\bibinfo {title} {{Broken magnetic symmetry due to charge-order ferroelectricity discovered in (TMTTF)$_{2}X$ salts by multifrequency ESR}},\ }\href {https://doi.org/10.1103/PhysRevB.85.144428} {\bibfield  {journal} {\bibinfo  {journal} {Phys. Rev. B}\ }\textbf {\bibinfo {volume} {85}},\ \bibinfo {pages} {144428} (\bibinfo {year} {2012})}\BibitemShut {NoStop}%
\bibitem [{\citenamefont {Eremin}\ \emph {et~al.}(2006)\citenamefont {Eremin}, \citenamefont {Zakharov}, \citenamefont {Eremina}, \citenamefont {Deisenhofer}, \citenamefont {Krug~von Nidda}, \citenamefont {Obermeier}, \citenamefont {Horn},\ and\ \citenamefont {Loidl}}]{eremin2006}%
  \BibitemOpen
  \bibfield  {author} {\bibinfo {author} {\bibfnamefont {M.~V.}\ \bibnamefont {Eremin}}, \bibinfo {author} {\bibfnamefont {D.~V.}\ \bibnamefont {Zakharov}}, \bibinfo {author} {\bibfnamefont {R.~M.}\ \bibnamefont {Eremina}}, \bibinfo {author} {\bibfnamefont {J.}~\bibnamefont {Deisenhofer}}, \bibinfo {author} {\bibfnamefont {H.-A.}\ \bibnamefont {Krug~von Nidda}}, \bibinfo {author} {\bibfnamefont {G.}~\bibnamefont {Obermeier}}, \bibinfo {author} {\bibfnamefont {S.}~\bibnamefont {Horn}},\ and\ \bibinfo {author} {\bibfnamefont {A.}~\bibnamefont {Loidl}},\ }\bibfield  {title} {\bibinfo {title} {{Unconventional Anisotropic Superexchange in ${\ensuremath{\alpha}}^{\ensuremath{'}}\mathrm{\text{\ensuremath{-}}}{\mathrm{NaV}}_{2}{\mathrm{O}}_{5}$}},\ }\href {https://doi.org/10.1103/PhysRevLett.96.027209} {\bibfield  {journal} {\bibinfo  {journal} {Phys. Rev. Lett.}\ }\textbf {\bibinfo {volume} {96}},\ \bibinfo {pages} {027209} (\bibinfo {year} {2006})}\BibitemShut {NoStop}%
\end{thebibliography}%


\begin{thebibliography}{14}%
\makeatletter
\providecommand \@ifxundefined [1]{%
 \@ifx{#1\undefined}
}%
\providecommand \@ifnum [1]{%
 \ifnum #1\expandafter \@firstoftwo
 \else \expandafter \@secondoftwo
 \fi
}%
\providecommand \@ifx [1]{%
 \ifx #1\expandafter \@firstoftwo
 \else \expandafter \@secondoftwo
 \fi
}%
\providecommand \natexlab [1]{#1}%
\providecommand \enquote  [1]{``#1''}%
\providecommand \bibnamefont  [1]{#1}%
\providecommand \bibfnamefont [1]{#1}%
\providecommand \citenamefont [1]{#1}%
\providecommand \href@noop [0]{\@secondoftwo}%
\providecommand \href [0]{\begingroup \@sanitize@url \@href}%
\providecommand \@href[1]{\@@startlink{#1}\@@href}%
\providecommand \@@href[1]{\endgroup#1\@@endlink}%
\providecommand \@sanitize@url [0]{\catcode `\\12\catcode `\$12\catcode `\&12\catcode `\#12\catcode `\^12\catcode `\_12\catcode `\%12\relax}%
\providecommand \@@startlink[1]{}%
\providecommand \@@endlink[0]{}%
\providecommand \url  [0]{\begingroup\@sanitize@url \@url }%
\providecommand \@url [1]{\endgroup\@href {#1}{\urlprefix }}%
\providecommand \urlprefix  [0]{URL }%
\providecommand \Eprint [0]{\href }%
\providecommand \doibase [0]{https://doi.org/}%
\providecommand \selectlanguage [0]{\@gobble}%
\providecommand \bibinfo  [0]{\@secondoftwo}%
\providecommand \bibfield  [0]{\@secondoftwo}%
\providecommand \translation [1]{[#1]}%
\providecommand \BibitemOpen [0]{}%
\providecommand \bibitemStop [0]{}%
\providecommand \bibitemNoStop [0]{.\EOS\space}%
\providecommand \EOS [0]{\spacefactor3000\relax}%
\providecommand \BibitemShut  [1]{\csname bibitem#1\endcsname}%
\let\auto@bib@innerbib\@empty
\bibitem [{\citenamefont {Aroyo}(2016)}]{ITC_A}%
  \BibitemOpen
  \bibinfo {editor} {\bibfnamefont {M.~I.}\ \bibnamefont {Aroyo}},\ ed.,\ \href {https://doi.org/10.1107/97809553602060000114} {\emph {\bibinfo {title} {{International Tables for Crystallography, Volume A: Space-Group Symmetry}}}},\ \bibinfo {edition} {6th}\ ed.\ (\bibinfo  {publisher} {Wiley},\ \bibinfo {address} {Chichester},\ \bibinfo {year} {2016})\BibitemShut {NoStop}%
\bibitem [{\citenamefont {Bergman}\ \emph {et~al.}(2007)\citenamefont {Bergman}, \citenamefont {Alicea}, \citenamefont {Gull}, \citenamefont {Trebst},\ and\ \citenamefont {Balents}}]{bergman2007}%
  \BibitemOpen
  \bibfield  {author} {\bibinfo {author} {\bibfnamefont {D.}~\bibnamefont {Bergman}}, \bibinfo {author} {\bibfnamefont {J.}~\bibnamefont {Alicea}}, \bibinfo {author} {\bibfnamefont {E.}~\bibnamefont {Gull}}, \bibinfo {author} {\bibfnamefont {S.}~\bibnamefont {Trebst}},\ and\ \bibinfo {author} {\bibfnamefont {L.}~\bibnamefont {Balents}},\ }\bibfield  {title} {\bibinfo {title} {{Order-by-disorder and spiral spin-liquid in frustrated diamond-lattice antiferromagnets}},\ }\href {https://doi.org/10.1038/nphys622} {\bibfield  {journal} {\bibinfo  {journal} {Nat. Phys.}\ }\textbf {\bibinfo {volume} {3}},\ \bibinfo {pages} {487} (\bibinfo {year} {2007})}\BibitemShut {NoStop}%
\bibitem [{\citenamefont {Joshi}\ \emph {et~al.}(1999)\citenamefont {Joshi}, \citenamefont {Ma}, \citenamefont {Mila}, \citenamefont {Shi},\ and\ \citenamefont {Zhang}}]{joshi1999}%
  \BibitemOpen
  \bibfield  {author} {\bibinfo {author} {\bibfnamefont {A.}~\bibnamefont {Joshi}}, \bibinfo {author} {\bibfnamefont {M.}~\bibnamefont {Ma}}, \bibinfo {author} {\bibfnamefont {F.}~\bibnamefont {Mila}}, \bibinfo {author} {\bibfnamefont {D.~N.}\ \bibnamefont {Shi}},\ and\ \bibinfo {author} {\bibfnamefont {F.~C.}\ \bibnamefont {Zhang}},\ }\bibfield  {title} {\bibinfo {title} {{Elementary excitations in magnetically ordered systems with orbital degeneracy}},\ }\href {https://doi.org/10.1103/PhysRevB.60.6584} {\bibfield  {journal} {\bibinfo  {journal} {Phys. Rev. B}\ }\textbf {\bibinfo {volume} {60}},\ \bibinfo {pages} {6584} (\bibinfo {year} {1999})}\BibitemShut {NoStop}%
\bibitem [{\citenamefont {Shiina}\ \emph {et~al.}(2003)\citenamefont {Shiina}, \citenamefont {Shiba}, \citenamefont {Thalmeier}, \citenamefont {Takahashi},\ and\ \citenamefont {Sakai}}]{shiina2003}%
  \BibitemOpen
  \bibfield  {author} {\bibinfo {author} {\bibfnamefont {R.}~\bibnamefont {Shiina}}, \bibinfo {author} {\bibfnamefont {H.}~\bibnamefont {Shiba}}, \bibinfo {author} {\bibfnamefont {P.}~\bibnamefont {Thalmeier}}, \bibinfo {author} {\bibfnamefont {A.}~\bibnamefont {Takahashi}},\ and\ \bibinfo {author} {\bibfnamefont {O.}~\bibnamefont {Sakai}},\ }\bibfield  {title} {\bibinfo {title} {{Dynamics of Multipoles and Neutron Scattering Spectra in Quadrupolar Ordering Phase of CeB$_6$}},\ }\href {https://doi.org/10.1143/JPSJ.72.1216} {\bibfield  {journal} {\bibinfo  {journal} {J. Phys. Soc. Jpn.}\ }\textbf {\bibinfo {volume} {72}},\ \bibinfo {pages} {1216} (\bibinfo {year} {2003})}\BibitemShut {NoStop}%
\bibitem [{\citenamefont {Colpa}(1978)}]{colpa}%
  \BibitemOpen
  \bibfield  {author} {\bibinfo {author} {\bibfnamefont {J.}~\bibnamefont {Colpa}},\ }\bibfield  {title} {\bibinfo {title} {{Diagonalization of the quadratic boson hamiltonian}},\ }\href {https://doi.org/https://doi.org/10.1016/0378-4371(78)90160-7} {\bibfield  {journal} {\bibinfo  {journal} {Physica}\ }\textbf {\bibinfo {volume} {93A}},\ \bibinfo {pages} {327} (\bibinfo {year} {1978})}\BibitemShut {NoStop}%
\bibitem [{\citenamefont {Rau}\ \emph {et~al.}(2018)\citenamefont {Rau}, \citenamefont {McClarty},\ and\ \citenamefont {Moessner}}]{rau2018}%
  \BibitemOpen
  \bibfield  {author} {\bibinfo {author} {\bibfnamefont {J.~G.}\ \bibnamefont {Rau}}, \bibinfo {author} {\bibfnamefont {P.~A.}\ \bibnamefont {McClarty}},\ and\ \bibinfo {author} {\bibfnamefont {R.}~\bibnamefont {Moessner}},\ }\bibfield  {title} {\bibinfo {title} {{Pseudo-Goldstone Gaps and Order-by-Quantum Disorder in Frustrated Magnets}},\ }\href {https://doi.org/10.1103/PhysRevLett.121.237201} {\bibfield  {journal} {\bibinfo  {journal} {Phys. Rev. Lett.}\ }\textbf {\bibinfo {volume} {121}},\ \bibinfo {pages} {237201} (\bibinfo {year} {2018})}\BibitemShut {NoStop}%
\bibitem [{\citenamefont {Mook}\ \emph {et~al.}(2020)\citenamefont {Mook}, \citenamefont {Klinovaja},\ and\ \citenamefont {Loss}}]{mook2020}%
  \BibitemOpen
  \bibfield  {author} {\bibinfo {author} {\bibfnamefont {A.}~\bibnamefont {Mook}}, \bibinfo {author} {\bibfnamefont {J.}~\bibnamefont {Klinovaja}},\ and\ \bibinfo {author} {\bibfnamefont {D.}~\bibnamefont {Loss}},\ }\bibfield  {title} {\bibinfo {title} {{Quantum damping of skyrmion crystal eigenmodes due to spontaneous quasiparticle decay}},\ }\href {https://doi.org/10.1103/PhysRevResearch.2.033491} {\bibfield  {journal} {\bibinfo  {journal} {Phys. Rev. Res.}\ }\textbf {\bibinfo {volume} {2}},\ \bibinfo {pages} {033491} (\bibinfo {year} {2020})}\BibitemShut {NoStop}%
\bibitem [{\citenamefont {Koyama}\ and\ \citenamefont {Nasu}(2023)}]{koyama2023}%
  \BibitemOpen
  \bibfield  {author} {\bibinfo {author} {\bibfnamefont {S.}~\bibnamefont {Koyama}}\ and\ \bibinfo {author} {\bibfnamefont {J.}~\bibnamefont {Nasu}},\ }\bibfield  {title} {\bibinfo {title} {{Flavor-wave theory with quasiparticle damping at finite temperatures: Application to chiral edge modes in the Kitaev model}},\ }\href {https://doi.org/10.1103/PhysRevB.108.235162} {\bibfield  {journal} {\bibinfo  {journal} {Phys. Rev. B}\ }\textbf {\bibinfo {volume} {108}},\ \bibinfo {pages} {235162} (\bibinfo {year} {2023})}\BibitemShut {NoStop}%
\bibitem [{\citenamefont {Romh\'anyi}\ \emph {et~al.}(2011)\citenamefont {Romh\'anyi}, \citenamefont {Totsuka},\ and\ \citenamefont {Penc}}]{romhanyi2011}%
  \BibitemOpen
  \bibfield  {author} {\bibinfo {author} {\bibfnamefont {J.}~\bibnamefont {Romh\'anyi}}, \bibinfo {author} {\bibfnamefont {K.}~\bibnamefont {Totsuka}},\ and\ \bibinfo {author} {\bibfnamefont {K.}~\bibnamefont {Penc}},\ }\bibfield  {title} {\bibinfo {title} {{Effect of Dzyaloshinskii-Moriya interactions on the phase diagram and magnetic excitations of SrCu${}_{2}$(BO${}_{3}$)${}_{2}$}},\ }\href {https://doi.org/10.1103/PhysRevB.83.024413} {\bibfield  {journal} {\bibinfo  {journal} {Phys. Rev. B}\ }\textbf {\bibinfo {volume} {83}},\ \bibinfo {pages} {024413} (\bibinfo {year} {2011})}\BibitemShut {NoStop}%
\bibitem [{\citenamefont {Hirayama}\ \emph {et~al.}(2015)\citenamefont {Hirayama}, \citenamefont {Okugawa}, \citenamefont {Ishibashi}, \citenamefont {Murakami},\ and\ \citenamefont {Miyake}}]{hirayama2015}%
  \BibitemOpen
  \bibfield  {author} {\bibinfo {author} {\bibfnamefont {M.}~\bibnamefont {Hirayama}}, \bibinfo {author} {\bibfnamefont {R.}~\bibnamefont {Okugawa}}, \bibinfo {author} {\bibfnamefont {S.}~\bibnamefont {Ishibashi}}, \bibinfo {author} {\bibfnamefont {S.}~\bibnamefont {Murakami}},\ and\ \bibinfo {author} {\bibfnamefont {T.}~\bibnamefont {Miyake}},\ }\bibfield  {title} {\bibinfo {title} {{Weyl Node and Spin Texture in Trigonal Tellurium and Selenium}},\ }\href {https://doi.org/10.1103/PhysRevLett.114.206401} {\bibfield  {journal} {\bibinfo  {journal} {Phys. Rev. Lett.}\ }\textbf {\bibinfo {volume} {114}},\ \bibinfo {pages} {206401} (\bibinfo {year} {2015})}\BibitemShut {NoStop}%
\bibitem [{\citenamefont {Kubo}\ \emph {et~al.}(1957)\citenamefont {Kubo}, \citenamefont {Yokota},\ and\ \citenamefont {Nakajima}}]{kubo1957}%
  \BibitemOpen
  \bibfield  {author} {\bibinfo {author} {\bibfnamefont {R.}~\bibnamefont {Kubo}}, \bibinfo {author} {\bibfnamefont {M.}~\bibnamefont {Yokota}},\ and\ \bibinfo {author} {\bibfnamefont {S.}~\bibnamefont {Nakajima}},\ }\bibfield  {title} {\bibinfo {title} {{Statistical-Mechanical Theory of Irreversible Processes. II. Response to Thermal Disturbance}},\ }\href {https://doi.org/10.1143/JPSJ.12.1203} {\bibfield  {journal} {\bibinfo  {journal} {J. Phys. Soc. Jpn.}\ }\textbf {\bibinfo {volume} {12}},\ \bibinfo {pages} {1203} (\bibinfo {year} {1957})}\BibitemShut {NoStop}%
\bibitem [{\citenamefont {Luttinger}(1964)}]{luttinger1964}%
  \BibitemOpen
  \bibfield  {author} {\bibinfo {author} {\bibfnamefont {J.~M.}\ \bibnamefont {Luttinger}},\ }\bibfield  {title} {\bibinfo {title} {{Theory of Thermal Transport Coefficients}},\ }\href {https://doi.org/10.1103/PhysRev.135.A1505} {\bibfield  {journal} {\bibinfo  {journal} {Phys. Rev.}\ }\textbf {\bibinfo {volume} {135}},\ \bibinfo {pages} {A1505} (\bibinfo {year} {1964})}\BibitemShut {NoStop}%
\bibitem [{\citenamefont {Li}\ \emph {et~al.}(2020)\citenamefont {Li}, \citenamefont {Mook}, \citenamefont {Raeliarijaona},\ and\ \citenamefont {Kovalev}}]{li2020}%
  \BibitemOpen
  \bibfield  {author} {\bibinfo {author} {\bibfnamefont {B.}~\bibnamefont {Li}}, \bibinfo {author} {\bibfnamefont {A.}~\bibnamefont {Mook}}, \bibinfo {author} {\bibfnamefont {A.}~\bibnamefont {Raeliarijaona}},\ and\ \bibinfo {author} {\bibfnamefont {A.~A.}\ \bibnamefont {Kovalev}},\ }\bibfield  {title} {\bibinfo {title} {{Magnonic analog of the Edelstein effect in antiferromagnetic insulators}},\ }\href {https://doi.org/10.1103/PhysRevB.101.024427} {\bibfield  {journal} {\bibinfo  {journal} {Phys. Rev. B}\ }\textbf {\bibinfo {volume} {101}},\ \bibinfo {pages} {024427} (\bibinfo {year} {2020})}\BibitemShut {NoStop}%
\bibitem [{\citenamefont {Nasu}\ and\ \citenamefont {Naka}(2021)}]{nasu2021}%
  \BibitemOpen
  \bibfield  {author} {\bibinfo {author} {\bibfnamefont {J.}~\bibnamefont {Nasu}}\ and\ \bibinfo {author} {\bibfnamefont {M.}~\bibnamefont {Naka}},\ }\bibfield  {title} {\bibinfo {title} {{Spin Seebeck effect in nonmagnetic excitonic insulators}},\ }\href {https://doi.org/10.1103/PhysRevB.103.L121104} {\bibfield  {journal} {\bibinfo  {journal} {Phys. Rev. B}\ }\textbf {\bibinfo {volume} {103}},\ \bibinfo {pages} {L121104} (\bibinfo {year} {2021})}\BibitemShut {NoStop}%
\end{thebibliography}%
\end{document}


\let\emph\textit
\renewcommand{\thefigure}{S\arabic{figure}}
\setcounter{figure}{0}
\renewcommand{\thetable}{S\arabic{table}}
\setcounter{table}{0}
\renewcommand{\theequation}{S\arabic{equation}}
\setcounter{equation}{0}

\title{
Supplemental Material for
``Chiral Valence-Bond Solid and Chiral Triplons''
}
\author{Shinnosuke Koyama}
\author{Kazumasa Hattori}
\affiliation{
Department of Physics, Tokyo Metropolitan University,
1-1, Minami-osawa, Hachioji, Tokyo 192-0397, Japan
}

\date{\today}

\maketitle

\section{Construction of Spin-Lattice Hamiltonian}

\subsection{Definition of the X$_4$, Y$_4$, and Z$_4$ displacement modes}

\begin{figure}[t]
\includegraphics[width=0.9\columnwidth]{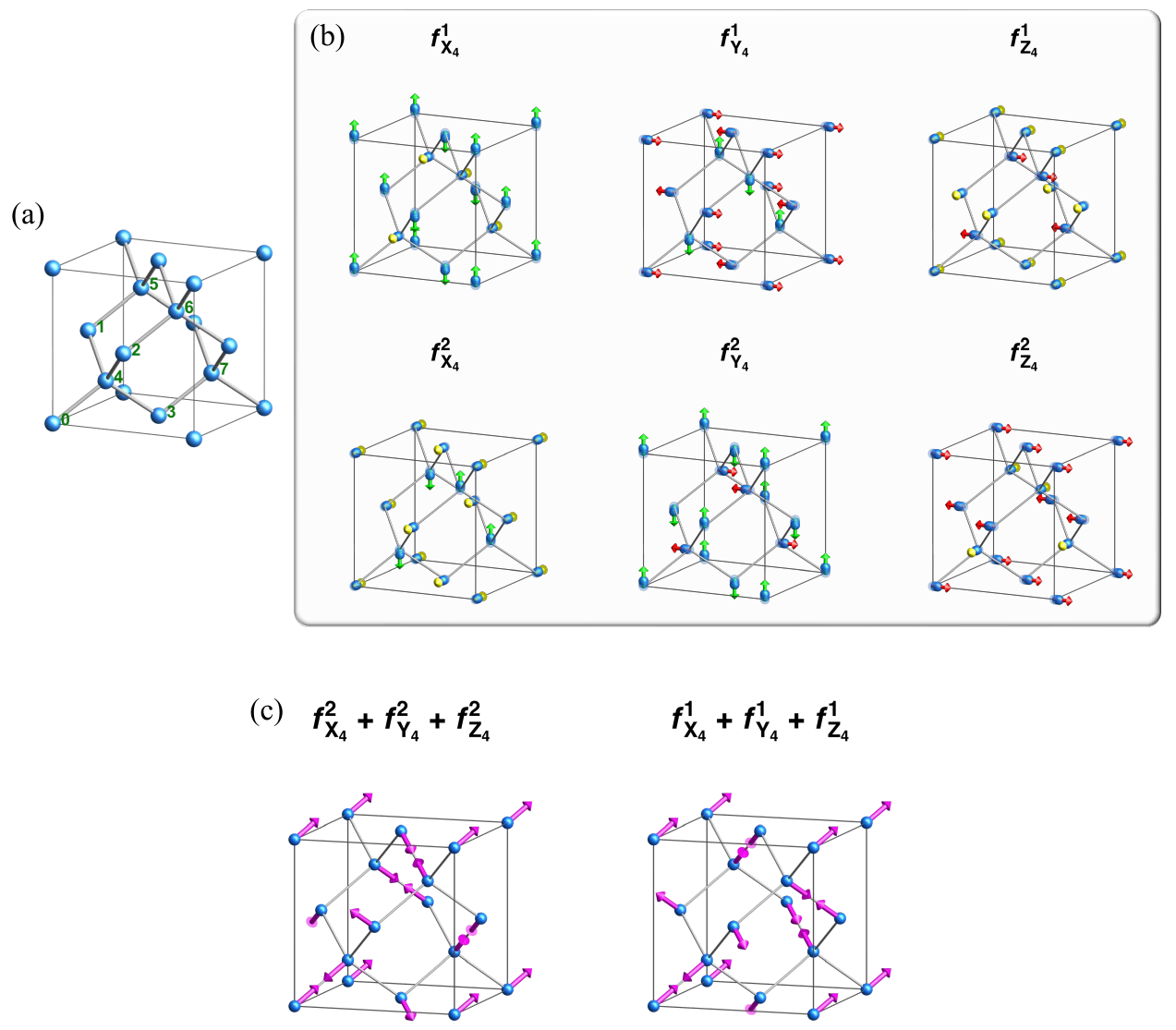}
\caption{
(a) Schematic of the undistorted
diamond lattice.
The numbers
indicate the site labels
in the cubic unit cell.
(b) Six basis vectors of the
X$_4$, Y$_4$, and Z$_4$
displacement modes.
The upper and lower rows show
the $\mu=1$ and $\mu=2$ basis vectors
of each two-dimensional mode, respectively.
The red, yellow, and green arrows at each site
denote displacement components
along the $x$, $y$, and $z$ directions, respectively.
(c) Equal-amplitude triple-$q$ displacement patterns.
The left and right panels show
$\bm{u}_2$ and $\bm{u}_1$, respectively.
The purple arrows represent the resulting site displacements.
}
\label{supfig:cubic_displacement}
\end{figure}

The primitive cell of the diamond lattice
contains two sites.
When the cubic cell containing eight sites is used,
the X, Y, and Z points
in the Brillouin zone~(BZ) associated
with the primitive cell
are folded onto the $\mathrm{\Gamma}$ point.
Therefore, when the displacement modes
at all three points are included,
the smallest real-space unit cell
is enlarged from the primitive cell
to the cubic cell.
Since the cubic cell contains eight sites,
a displacement pattern in this cell
can be represented as a $24$-dimensional vector
consisting of three displacement components
at each site.
We denote the two basis vectors
of the mode at each point by
$\bm{f}_{\mathrm{X}_4}^1$ and
$\bm{f}_{\mathrm{X}_4}^2$ for the X point,
$\bm{f}_{\mathrm{Y}_4}^1$ and
$\bm{f}_{\mathrm{Y}_4}^2$ for the Y point,
and
$\bm{f}_{\mathrm{Z}_4}^1$ and
$\bm{f}_{\mathrm{Z}_4}^2$ for the Z point.
Labeling the sites in the cubic cell
as shown in Fig.~\ref{supfig:cubic_displacement}(a)
and arranging the three displacement components
in ascending order of the site labels
from $0$ to $7$,
we write these basis vectors as
\begin{align}
    \bm{f}_{\mathrm{X}_4}^1
    &= \dfrac{1}{\sqrt{8}}(001,001,00\bar{1},00\bar{1},
    0\bar{1}0,0\bar{1}0,010,010)^T, \quad
    \bm{f}_{\mathrm{X}_4}^2
    = \dfrac{1}{\sqrt{8}}(010,010,0\bar{1}0,0\bar{1}0,
    00\bar{1},00\bar{1},001,001)^T,\\[8pt]
    \bm{f}_{\mathrm{Y}_4}^1
    &= \dfrac{1}{\sqrt{8}}(100,\bar{1}00,100,\bar{1}00,
    00\bar{1},001,00\bar{1},001)^T,\quad
    \bm{f}_{\mathrm{Y}_4}^2
    = \dfrac{1}{\sqrt{8}}(001,00\bar{1},001,00\bar{1},
    \bar{1}00,100,\bar{1}00,100)^T,\\[8pt]
    \bm{f}_{\mathrm{Z}_4}^1
    &= \dfrac{1}{\sqrt{8}}(010,0\bar{1}0,0\bar{1}0,010,
    \bar{1}00,100,100,\bar{1}00)^T,\quad
    \bm{f}_{\mathrm{Z}_4}^2
    = \dfrac{1}{\sqrt{8}}(100,\bar{1}00,\bar{1}00,100,
    0\bar{1}0,010,010,0\bar{1}0)^T,
\end{align}
where, for example,
$001$ denotes a unit displacement
along the $z$ direction,
and $\bar{1}$ denotes $-1$.
These six vectors form an orthonormal basis
for the six-dimensional displacement subspace considered here.
The displacement patterns of the X$_4$, Y$_4$, and Z$_4$ modes
are shown in Fig.~\ref{supfig:cubic_displacement}(b).
Writing the actual site displacements
as a 24-dimensional vector $\bm{u}$,
we define the normal coordinates $Q_{\Lambda}^{\mu}$ by
\begin{align}
    \label{supeq:f_to_u}
    \bm{u}
    =
    \sum_{\Lambda\in\{\mathrm{X}_4,\mathrm{Y}_4,\mathrm{Z}_4\}}
    \sum_{\mu=1}^2
    Q_{\Lambda}^{\mu} \bm{f}_{\Lambda}^{\mu}.
\end{align}
Figure~\ref{supfig:cubic_displacement}(c)
shows the two triple-$q$ displacement patterns
\begin{align}
    \bm{u}_2
    &=
    \bm{f}_{\rm X_4}^{2}
    +
    \bm{f}_{\rm Y_4}^{2}
    +
    \bm{f}_{\rm Z_4}^{2},
    \qquad
    \bm{u}_1
    =
    \bm{f}_{\rm X_4}^{1}
    +
    \bm{f}_{\rm Y_4}^{1}
    +
    \bm{f}_{\rm Z_4}^{1}.
\end{align}
For $\bm{u}_2$,
the two sites forming each of the NN bonds
$(R,0)$--$(R,4)$,
$(R+\bm{a}_1,1)$--$(R,7)$,
$(R+\bm{a}_2,2)$--$(R,5)$,
and
$(R+\bm{a}_3,3)$--$(R,6)$
move in opposite directions along the bond.
For $\bm{u}_1$,
the same motion occurs on the NN bonds
$(R,0)$--$(R,4)$,
$(R+\bm{a}_1,1)$--$(R,6)$,
$(R+\bm{a}_2,2)$--$(R,7)$,
and
$(R+\bm{a}_3,3)$--$(R,5)$.
Here, $R$ labels the cubic unit cell
and the cubic-cell translation vectors are
$\bm{a}_1=a(1,0,0)$,
$\bm{a}_2=a(0,1,0)$,
and
$\bm{a}_3=a(0,0,1)$.
Therefore, in an $S=1/2$ quantum spin model
on the diamond lattice,
the triple-$q$ displacements constructed from
these six basis vectors can promote dimerization
on the corresponding bonds.
This property motivates our choice of the
X$_4$, Y$_4$, and Z$_4$
displacement modes.

Let $g=\{\hat{R}_g|\bm{\tau}_g\}$ be a space-group operation
of the parent $Fd\bar{3}m$ phase,
where $\hat{R}_g$ and $\bm{\tau}_g$
denote its point-operation part
and translation vector, respectively.
The operation $g$ acts on each object
in the appropriate representation.
The $3\times3$ matrix representing $\hat{R}_g$
on polar vectors
is denoted by $R_g$,
whereas its representation on axial vectors is
$\bar{R}_g=\det(R_g)R_g$.
Within the six-dimensional subspace
spanned by the X$_4$, Y$_4$,
and Z$_4$ modes,
the full space-group operation $g$ is represented as
\begin{align}
    \label{supeq:gf}
    g\bm{f}_{\Lambda}^\mu
    =
    \sum_{\Lambda'} \sum_{\mu'}
    U_{\Lambda'\Lambda}^{\mu'\mu}(g)
    \bm{f}_{\Lambda'}^{\mu'},
\end{align}
where 
$U_{\Lambda'\Lambda}^{\mu'\mu}(g)$
is the representation matrix of $g$ in this subspace.
It follows from Eq.~\eqref{supeq:gf}
that the transformed displacement
can be expanded in the same basis as
\begin{align}
    g\bm{u}
    &=
    \sum_{\Lambda}\sum_{\mu}
    Q_{\Lambda}^{\mu} g\bm{f}_{\Lambda}^{\mu} \notag\\
    &=
    \sum_{\Lambda\Lambda'}\sum_{\mu\mu'}
    Q_{\Lambda}^{\mu}
    U_{\Lambda'\Lambda}^{\mu'\mu}(g)
    \bm{f}_{\Lambda'}^{\mu'} \notag\\
    &=
    \sum_{\Lambda'}\sum_{\mu'}
    Q_{\Lambda'}^{\mu'}(g)
    \bm{f}_{\Lambda'}^{\mu'}.
\end{align}
The transformed normal coordinates
are therefore
\begin{align}
    \label{supeq:Q(g)}
    Q_{\Lambda'}^{\mu'}(g)
    =
    \sum_{\Lambda}\sum_{\mu}
    U_{\Lambda'\Lambda}^{\mu'\mu}(g) Q_{\Lambda}^{\mu}.
\end{align}

\subsection{Symmetry-allowed Heisenberg interactions}

We first consider the Heisenberg interactions.
The Heisenberg part
of the spin-lattice Hamiltonian
is written as
\begin{align}
    \mathcal{H}_J(\bm{Q})
    =
    \sum_{\braket{ij}}
    J_{1,ij}(\bm{Q}) \bm{S}_i \cdot \bm{S}_j
    +
    \sum_{\braket{\braket{ij}}}
    J_{2,ij}(\bm{Q}) \bm{S}_i \cdot \bm{S}_j,
    \label{supeq:HJ(Q)}
\end{align}
where $\braket{ij}$ and $\braket{\braket{ij}}$
denote the nearest-neighbor~(NN)
and next-nearest-neighbor~(NNN) pairs,
respectively.
To determine $J_{n,ij}(\bm{Q})$~($n=1,2$)
consistent with the symmetry of
the parent $Fd\bar{3}m$ structure,
we impose
\begin{align}
    \label{supeq:HJ_inv}
    g \mathcal{H}_{J}(\bm{Q}) g^{-1}
    =
    \mathcal{H}_{J}[\bm{Q}(g)].
\end{align}
Since the spin operator
is an axial vector,
it transforms as
\begin{align}
    \label{supeq:gSg-1}
    g S_i^\alpha g^{-1}
    =
    \sum_{\beta}
    (\bar{R}^{-1}_{g})_{\alpha\beta} S_{gi}^{\beta},
\end{align}
where $gi$ denotes
the site to which site $i$
is mapped by $g$.
Transforming the lhs
of Eq.~\eqref{supeq:HJ_inv}
using Eq.~\eqref{supeq:gSg-1}
and comparing the result with
the rhs of Eq.~\eqref{supeq:HJ_inv},
we obtain
\begin{align}
    \label{supeq:J_convert}
    J_{1,ij}(\bm{Q})
    =
    J_{1,gi,gj}[\bm{Q}(g)],
    \qquad
    J_{2,ij}(\bm{Q})
    =
    J_{2,gi,gj}[\bm{Q}(g)].
\end{align}
In addition, the exchange interactions satisfy
$J_{1,ij}(\bm{Q})=J_{1,ji}(\bm{Q})$
and
$J_{2,ij}(\bm{Q})=J_{2,ji}(\bm{Q})$.
Thus, once the exchange modulations
on one representative NN bond
and one representative NNN bond
are determined to first order
in $Q_{\Lambda}^{\mu}$,
the modulations on all remaining
NN and NNN bonds
are generated using
Eq.~\eqref{supeq:J_convert}.

As a representative example,
we consider the NN bond $(R,0)$--$(R,4)$.
The site-symmetry group of its bond center,
whose point group is D$_{3d}$,
is generated by the space-group operations
$\{\hat{C}_{3,[111]}|\bm{0}\}$,
$\{\hat{I}|a(1,1,1)/4\}$,
and
$\{\hat{C}_{2,[1\bar{1}0]}|a(1,1,1)/4\}$,
where
the corresponding point operations act as
$C_{3,[111]}: (x,y,z)\mapsto(z,x,y)$,
$I: (x,y,z)\mapsto(-x,-y,-z)$, and
$C_{2,[1\bar{1}0]}: (x,y,z)\mapsto(-y,-x,-z)$~\cite{ITC_A}.
Under every operation in this bond-center site-symmetry group,
the bond is mapped either to $(R,0)$--$(R,4)$
or to $(R,4)$--$(R,0)$.
For the Heisenberg interaction,
the two orientations are equivalent because
$J_{1,04}(\bm{Q})=J_{1,40}(\bm{Q})$.
Therefore, the exchange interaction satisfies
\begin{align}
    J_{1,04}(\bm{Q})
    =
    J_{1,40}(\bm{Q})
    =
    J_{1,04}[\bm{Q}(g)]
    =
    J_{1,40}[\bm{Q}(g)].
\end{align}
To first order in the normal coordinates,
the form allowed by this D$_{3d}$ symmetry is
\begin{align}
    \lambda_1
    \left(
    Q_{\mathrm{X}_4}^{1}+Q_{\mathrm{X}_4}^{2}
    +Q_{\mathrm{Y}_4}^{1}+Q_{\mathrm{Y}_4}^{2}
    +Q_{\mathrm{Z}_4}^{1}+Q_{\mathrm{Z}_4}^{2}
    \right).
\end{align}

If the Heisenberg exchange
for each bond type
depends only on the distance
between the two sites,
the NN and NNN interactions
can be written as
\begin{align}
    J_{n,ij}(\bm{Q})
    =
    \mathcal{J}_n
    \left(
    \left|
    \bm{r}_j+\bm{u}_j
    -\bm{r}_i-\bm{u}_i
    \right|
    \right),
    \qquad n=1,2,
\end{align}
where $\bm{r}_i$ and $\bm{u}_i$
are the equilibrium position and displacement
of site $i$, respectively,
and $\mathcal{J}_n(r)$ denotes
the distance dependence of the
$n$th-neighbor exchange interactions.
For the bond $(R,0)$--$(R,4)$,
we define
$\bm\tau_{04}=\bm{r}_4-\bm{r}_0$,
$\tau_1=|\bm\tau_{04}|$,
and
$\hat{\bm\tau}_{04}=\bm\tau_{04}/\tau_1$.
Expanding the exchange interaction
to first order in the displacements
gives
\begin{align}
    J_{1,04}(\bm{Q})
    &=
    \mathcal{J}_1
    \left(
    |\bm{\tau}_{04}
    +\bm{u}_4-\bm{u}_0|
    \right)
    \notag\\
    &\simeq
    J_1
    \left[
    1
    -
    \tilde{\lambda}_1
    \hat{\bm{\tau}}_{04}
    \cdot
    (\bm{u}_4-\bm{u}_0)
    \right],
\end{align}
where
$J_1=\mathcal{J}_1(\tau_1)$ and
\begin{align}
    \tilde{\lambda}_1
    =
    -\frac{1}{J_1}
    \left.
    \frac{d\mathcal{J}_1(r)}{dr}
    \right|_{r=\tau_1}.
\end{align}
Using Eq.~\eqref{supeq:f_to_u}, we obtain
\begin{align}
    \hat{\bm{\tau}}_{04}
    \cdot
    (\bm{u}_4-\bm{u}_0)
    =
    -\frac{1}{\sqrt{6}}
    \left(
    Q_{\mathrm{X}_4}^{1}
    +Q_{\mathrm{X}_4}^{2}
    +Q_{\mathrm{Y}_4}^{1}
    +Q_{\mathrm{Y}_4}^{2}
    +Q_{\mathrm{Z}_4}^{1}
    +Q_{\mathrm{Z}_4}^{2}
    \right).
\end{align}
Therefore,
$\lambda_1=J_1\tilde{\lambda}_1/\sqrt{6}$.
In particular, assuming
$\mathcal{J}_n(r)\propto r^{-\eta}$
for both $n=1$ and $2$
gives
$\lambda_1=J_1\eta/(\sqrt{6}\tau_1)$,
as used in the main text.
Applying the same argument
to a representative NNN bond gives
$\lambda_2=J_2\eta/(2\tau_2)$,
where $\tau_2$ is the NNN bond length.
The resulting NN modulations are summarized in
Table~\ref{suptab:NN_heisen_Q},
and the NNN modulations are listed in
Tables~\ref{suptab:NNN_heisen_Q_A}
and~\ref{suptab:NNN_heisen_Q_B}.

\subsection{Symmetry-allowed Dzyaloshinskii--Moriya interactions}

We determine the symmetry-allowed form
of the Dzyaloshinskii--Moriya interaction~(DMI)
in the same manner as for the Heisenberg interactions.
The DMI Hamiltonian satisfies
\begin{align}
    g \mathcal{H}_{\rm DM}(\bm{Q}) g^{-1}
    =
    \mathcal{H}_{\rm DM}[\bm{Q}(g)].
\end{align}
This leads to
\begin{align}
    \label{supeq:HDM_convert}
    \bar{R}_g \bm{D}_{ij}(\bm{Q})
    =
    \bm{D}_{gi,gj}[\bm{Q}(g)].
\end{align}
In deriving this relation,
we have used the identity
\begin{align}
    \label{supeq:RgRg_convert}
    \sum_{\beta\gamma}
    \epsilon_{\alpha\beta\gamma}
    (\bar{R}^{-1}_{g})_{\beta\delta}
    (\bar{R}^{-1}_{g})_{\gamma\epsilon}
    =
    \sum_{\zeta}
    \epsilon_{\delta\epsilon\zeta}
    (\bar{R}^{-1}_{g})_{\alpha\zeta},
\end{align}
which follows from $\det\bar{R}_g=1$.
We also impose the antisymmetry of the DMI vector under bond reversal,
$\bm{D}_{ij}(\bm{Q})=-\bm{D}_{ji}(\bm{Q})$.
Once the DMI on one representative NN bond
is determined to first order in $Q_{\Lambda}^\mu$,
the DMI vectors on all remaining NN bonds are generated
using Eq.~\eqref{supeq:HDM_convert}.

As in the analysis of the Heisenberg interactions,
we consider the NN bond $(R,0)$--$(R,4)$
as the representative bond.
Under every operation of the bond-center site-symmetry group,
this bond is mapped either to $(R,0)$--$(R,4)$
or to $(R,4)$--$(R,0)$.
Imposing this D$_{3d}$ symmetry together with
the antisymmetry under bond reversal,
$\bm{D}_{04}(\bm{Q})=-\bm{D}_{40}(\bm{Q})$,
we obtain the following symmetry-allowed form
\begin{align}
    \bm{D}_{04}(\bm{Q})
    &=
    g_1
    \begin{pmatrix}
        Q_{\mathrm{X}_4}^{-}\\
        Q_{\mathrm{Y}_4}^{-}\\
        Q_{\mathrm{Z}_4}^{-}
    \end{pmatrix}
    +
    g_2
    \begin{pmatrix}
        Q_{\mathrm{Y}_4}^{-}+Q_{\mathrm{Z}_4}^{-}\\
        Q_{\mathrm{X}_4}^{-}+Q_{\mathrm{Z}_4}^{-}\\
        Q_{\mathrm{X}_4}^{-}+Q_{\mathrm{Y}_4}^{-}
    \end{pmatrix},
    \label{supeq:D04}
\end{align}
where $Q_{\Lambda}^{-}=Q_{\Lambda}^{1}-Q_{\Lambda}^{2}$.
The remaining NN DMI vectors listed in
Tables~\ref{suptab:NN_DMI_Q_0}--\ref{suptab:NN_DMI_Q_3}
are generated from
the representative-bond expression
in Eq.~\eqref{supeq:D04}
using the transformation rule
in Eq.~\eqref{supeq:HDM_convert}.

\clearpage
\begingroup
\small
\renewcommand{\arraystretch}{1.25}
\def\qexpr#1{\(\begin{gathered}#1\end{gathered}\)}

\begin{table}[t]
\caption{
NN Heisenberg interaction modulations
linear in $Q_{\Lambda}^\mu$.
The translation vectors are defined as
$\bm a_1=a(100)$, $\bm a_2=a(010)$, and
$\bm a_3=a(001)$.
}
\label{suptab:NN_heisen_Q}
\begin{ruledtabular}
\begin{tabular}{ll}
Bond \(i-j\)
&
$
\displaystyle
\sum_{\Lambda}\sum_{\mu}
A_{1,ij}^{\Lambda\mu} Q_{\Lambda}^\mu
$
\\[8pt]
\hline
$(R,0)-(R,4)$
&
\qexpr{
Q_{\mathrm X_4}^{1}+Q_{\mathrm X_4}^{2}
+Q_{\mathrm Y_4}^{1}+Q_{\mathrm Y_4}^{2}
+Q_{\mathrm Z_4}^{1}+Q_{\mathrm Z_4}^{2}}
\\[8pt]

$(R+\bm a_2+\bm a_3,0)-(R,5)$
&
\qexpr{
-Q_{\mathrm X_4}^{1}-Q_{\mathrm X_4}^{2}
+Q_{\mathrm Y_4}^{1}-Q_{\mathrm Y_4}^{2}
-Q_{\mathrm Z_4}^{1}+Q_{\mathrm Z_4}^{2}}
\\[8pt]

$(R+\bm a_1+\bm a_3,0)-(R,6)$
&
\qexpr{
-Q_{\mathrm X_4}^{1}+Q_{\mathrm X_4}^{2}
-Q_{\mathrm Y_4}^{1}-Q_{\mathrm Y_4}^{2}
+Q_{\mathrm Z_4}^{1}-Q_{\mathrm Z_4}^{2}}
\\[8pt]

$(R+\bm a_1+\bm a_2,0)-(R,7)$
&
\qexpr{
Q_{\mathrm X_4}^{1}-Q_{\mathrm X_4}^{2}
-Q_{\mathrm Y_4}^{1}+Q_{\mathrm Y_4}^{2}
-Q_{\mathrm Z_4}^{1}-Q_{\mathrm Z_4}^{2}}
\\[8pt]

$(R,1)-(R,4)$
&
\qexpr{
-Q_{\mathrm X_4}^{1}-Q_{\mathrm X_4}^{2}
 -Q_{\mathrm Y_4}^{1}+Q_{\mathrm Y_4}^{2}
 +Q_{\mathrm Z_4}^{1}-Q_{\mathrm Z_4}^{2}}
\\[8pt]

$(R,1)-(R,5)$
&
\qexpr{
Q_{\mathrm X_4}^{1}+Q_{\mathrm X_4}^{2}
-Q_{\mathrm Y_4}^{1}-Q_{\mathrm Y_4}^{2}
-Q_{\mathrm Z_4}^{1}-Q_{\mathrm Z_4}^{2}}
\\[8pt]

$(R+\bm a_1,1)-(R,6)$
&
\qexpr{
Q_{\mathrm X_4}^{1}-Q_{\mathrm X_4}^{2}
+Q_{\mathrm Y_4}^{1}-Q_{\mathrm Y_4}^{2}
+Q_{\mathrm Z_4}^{1}+Q_{\mathrm Z_4}^{2}}
\\[8pt]

$(R+\bm a_1,1)-(R,7)$
&
\qexpr{
-Q_{\mathrm X_4}^{1}+Q_{\mathrm X_4}^{2}
+Q_{\mathrm Y_4}^{1}+Q_{\mathrm Y_4}^{2}
-Q_{\mathrm Z_4}^{1}+Q_{\mathrm Z_4}^{2}}
\\[8pt]

$(R,2)-(R,4)$
&
\qexpr{
Q_{\mathrm X_4}^{1}-Q_{\mathrm X_4}^{2}
-Q_{\mathrm Y_4}^{1}-Q_{\mathrm Y_4}^{2}
-Q_{\mathrm Z_4}^{1}+Q_{\mathrm Z_4}^{2}}
\\[8pt]

\((R+\bm a_2,2)-(R,5)\)
&
\qexpr{
-Q_{\mathrm X_4}^{1}+Q_{\mathrm X_4}^{2}
 -Q_{\mathrm Y_4}^{1}+Q_{\mathrm Y_4}^{2}
 +Q_{\mathrm Z_4}^{1}+Q_{\mathrm Z_4}^{2}}
\\[8pt]

$(R,2)-(R,6)$
&
\qexpr{
-Q_{\mathrm X_4}^{1}-Q_{\mathrm X_4}^{2}
+Q_{\mathrm Y_4}^{1}+Q_{\mathrm Y_4}^{2}
-Q_{\mathrm Z_4}^{1}-Q_{\mathrm Z_4}^{2}}
\\[8pt]

$(R+\bm a_2,2)-(R,7)$
&
\qexpr{
Q_{\mathrm X_4}^{1}+Q_{\mathrm X_4}^{2}
+Q_{\mathrm Y_4}^{1}-Q_{\mathrm Y_4}^{2}
+Q_{\mathrm Z_4}^{1}-Q_{\mathrm Z_4}^{2}}
\\[8pt]

$(R,3)-(R,4)$
&
\qexpr{
-Q_{\mathrm X_4}^{1}+Q_{\mathrm X_4}^{2}
+Q_{\mathrm Y_4}^{1}-Q_{\mathrm Y_4}^{2}
-Q_{\mathrm Z_4}^{1}-Q_{\mathrm Z_4}^{2}}
\\[8pt]

$(R+\bm a_3,3)-(R,5)$
&
\qexpr{
Q_{\mathrm X_4}^{1}-Q_{\mathrm X_4}^{2}
+Q_{\mathrm Y_4}^{1}+Q_{\mathrm Y_4}^{2}
+Q_{\mathrm Z_4}^{1}-Q_{\mathrm Z_4}^{2}}
\\[8pt]

$(R+\bm a_3,3)-(R,6)$
&
\qexpr{
Q_{\mathrm X_4}^{1}+Q_{\mathrm X_4}^{2}
-Q_{\mathrm Y_4}^{1}+Q_{\mathrm Y_4}^{2}
-Q_{\mathrm Z_4}^{1}+Q_{\mathrm Z_4}^{2}}
\\[8pt]

$(R,3)-(R,7)$
&
\qexpr{
-Q_{\mathrm X_4}^{1}-Q_{\mathrm X_4}^{2}
-Q_{\mathrm Y_4}^{1}-Q_{\mathrm Y_4}^{2}
+Q_{\mathrm Z_4}^{1}+Q_{\mathrm Z_4}^{2}}
\\[8pt]
\end{tabular}
\end{ruledtabular}
\end{table}
\endgroup

\begingroup
\small
\renewcommand{\arraystretch}{1.25}
\def\qexpr#1{\(\begin{gathered}#1\end{gathered}\)}

\begin{table}[t]
\caption{
NNN Heisenberg interaction modulations
linear in $Q_{\Lambda}^\mu$.
The listed bonds connect sites with labels $0,\ldots,3$.
The site labels are defined in Fig.~\ref{supfig:cubic_displacement}(a).
The translation vectors are defined as
$\bm a_1=a(100)$, $\bm a_2=a(010)$, and
$\bm a_3=a(001)$.
}
\label{suptab:NNN_heisen_Q_A}
\begin{ruledtabular}
\begin{tabular}{ll}
Bond $i-j$
&
$\displaystyle
\sum_{\Lambda}\sum_{\mu}
A_{2,ij}^{\Lambda\mu} Q_{\Lambda}^{\mu}
$
\\[8pt]
\hline
$(R,0)-(R,1)$
&
\qexpr{Q_{\mathrm Y_4}^{2}+Q_{\mathrm Z_4}^{1}}
\\[8pt]

$(R+\bm a_3,0)-(R,1)$
&
\qexpr{-Q_{\mathrm Y_4}^{2}+Q_{\mathrm Z_4}^{1}}
\\[8pt]

$(R+\bm a_2,0)-(R,1)$
&
\qexpr{Q_{\mathrm Y_4}^{2}-Q_{\mathrm Z_4}^{1}}
\\[8pt]

$(R+\bm a_2+\bm a_3,0)-(R,1)$
&
\qexpr{-Q_{\mathrm Y_4}^{2}-Q_{\mathrm Z_4}^{1}}
\\[8pt]

$(R,0)-(R,2)$
&
\qexpr{Q_{\mathrm X_4}^{1}+Q_{\mathrm Z_4}^{2}}
\\[8pt]

$(R+\bm a_3,0)-(R,2)$
&
\qexpr{-Q_{\mathrm X_4}^{1}+Q_{\mathrm Z_4}^{2}}
\\[8pt]

$(R+\bm a_1,0)-(R,2)$
&
\qexpr{Q_{\mathrm X_4}^{1}-Q_{\mathrm Z_4}^{2}}
\\[8pt]

$(R+\bm a_1+\bm a_3,0)-(R,2)$
&
\qexpr{-Q_{\mathrm X_4}^{1}-Q_{\mathrm Z_4}^{2}}
\\[8pt]

$(R,0)-(R,3)$
&
\qexpr{Q_{\mathrm X_4}^{2}+Q_{\mathrm Y_4}^{1}}
\\[8pt]

$(R+\bm a_2,0)-(R,3)$
&
\qexpr{-Q_{\mathrm X_4}^{2}+Q_{\mathrm Y_4}^{1}}
\\[8pt]

$(R+\bm a_1,0)-(R,3)$
&
\qexpr{Q_{\mathrm X_4}^{2}-Q_{\mathrm Y_4}^{1}}
\\[8pt]

$(R+\bm a_1+\bm a_2,0)-(R,3)$
&
\qexpr{-Q_{\mathrm X_4}^{2}-Q_{\mathrm Y_4}^{1}}
\\[8pt]

$(R-\bm a_2,1)-(R,2)$
&
\qexpr{Q_{\mathrm X_4}^{2}-Q_{\mathrm Y_4}^{1}}
\\[8pt]

$(R,1)-(R,2)$
&
\qexpr{-Q_{\mathrm X_4}^{2}-Q_{\mathrm Y_4}^{1}}
\\[8pt]

$(R+\bm a_1-\bm a_2,1)-(R,2)$
&
\qexpr{Q_{\mathrm X_4}^{2}+Q_{\mathrm Y_4}^{1}}
\\[8pt]

$(R+\bm a_1,1)-(R,2)$
&
\qexpr{-Q_{\mathrm X_4}^{2}+Q_{\mathrm Y_4}^{1}}
\\[8pt]

$(R-\bm a_3,1)-(R,3)$
&
\qexpr{Q_{\mathrm X_4}^{1}-Q_{\mathrm Z_4}^{2}}
\\[8pt]

$(R,1)-(R,3)$
&
\qexpr{-Q_{\mathrm X_4}^{1}-Q_{\mathrm Z_4}^{2}}
\\[8pt]

$(R+\bm a_1-\bm a_3,1)-(R,3)$
&
\qexpr{Q_{\mathrm X_4}^{1}+Q_{\mathrm Z_4}^{2}}
\\[8pt]

$(R+\bm a_1,1)-(R,3)$
&
\qexpr{-Q_{\mathrm X_4}^{1}+Q_{\mathrm Z_4}^{2}}
\\[8pt]

$(R-\bm a_3,2)-(R,3)$
&
\qexpr{Q_{\mathrm Y_4}^{2}-Q_{\mathrm Z_4}^{1}}
\\[8pt]

$(R,2)-(R,3)$
&
\qexpr{-Q_{\mathrm Y_4}^{2}-Q_{\mathrm Z_4}^{1}}
\\[8pt]

$(R+\bm a_2-\bm a_3,2)-(R,3)$
&
\qexpr{Q_{\mathrm Y_4}^{2}+Q_{\mathrm Z_4}^{1}}
\\[8pt]

$(R+\bm a_2,2)-(R,3)$
&
\qexpr{-Q_{\mathrm Y_4}^{2}+Q_{\mathrm Z_4}^{1}}
\\[8pt]
\end{tabular}
\end{ruledtabular}
\end{table}
\endgroup

\begingroup
\small
\renewcommand{\arraystretch}{1.25}
\def\qexpr#1{\(\begin{gathered}#1\end{gathered}\)}

\begin{table}[t]
\caption{
NNN Heisenberg interaction modulations
linear in $Q_{\Lambda}^\mu$.
The listed bonds connect sites with labels $4,\ldots,7$.
}
\label{suptab:NNN_heisen_Q_B}
\begin{ruledtabular}
\begin{tabular}{ll}
Bond $i-j$
&
\(\displaystyle
\sum_{\Lambda}\sum_{\mu}
A_{2,ij}^{\Lambda\mu} Q_{\Lambda}^{\mu}
\)
\\[8pt]
\hline
$(R,4)-(R,5)$
&
\qexpr{-Q_{\mathrm Y_4}^{1}-Q_{\mathrm Z_4}^{2}}
\\[8pt]

$(R+\bm a_3,4)-(R,5)$
&
\qexpr{Q_{\mathrm Y_4}^{1}-Q_{\mathrm Z_4}^{2}}
\\[8pt]

$(R+\bm a_2,4)-(R,5)$
&
\qexpr{-Q_{\mathrm Y_4}^{1}+Q_{\mathrm Z_4}^{2}}
\\[8pt]

$(R+\bm a_2+\bm a_3,4)-(R,5)$
&
\qexpr{Q_{\mathrm Y_4}^{1}+Q_{\mathrm Z_4}^{2}}
\\[8pt]

$(R,4)-(R,6)$
&
\qexpr{-Q_{\mathrm X_4}^{2}-Q_{\mathrm Z_4}^{1}}
\\[8pt]

$(R+\bm a_3,4)-(R,6)$
&
\qexpr{Q_{\mathrm X_4}^{2}-Q_{\mathrm Z_4}^{1}}
\\[8pt]

$(R+\bm a_1,4)-(R,6)$
&
\qexpr{-Q_{\mathrm X_4}^{2}+Q_{\mathrm Z_4}^{1}}
\\[8pt]

$(R+\bm a_1+\bm a_3,4)-(R,6)$
&
\qexpr{Q_{\mathrm X_4}^{2}+Q_{\mathrm Z_4}^{1}}
\\[8pt]

$(R,4)-(R,7)$
&
\qexpr{-Q_{\mathrm X_4}^{1}-Q_{\mathrm Y_4}^{2}}
\\[8pt]

$(R+\bm a_2,4)-(R,7)$
&
\qexpr{Q_{\mathrm X_4}^{1}-Q_{\mathrm Y_4}^{2}}
\\[8pt]

$(R+\bm a_1,4)-(R,7)$
&
\qexpr{-Q_{\mathrm X_4}^{1}+Q_{\mathrm Y_4}^{2}}
\\[8pt]

$(R+\bm a_1+\bm a_2,4)-(R,7)$
&
\qexpr{Q_{\mathrm X_4}^{1}+Q_{\mathrm Y_4}^{2}}
\\[8pt]

$(R-\bm a_2,5)-(R,6)$
&
\qexpr{-Q_{\mathrm X_4}^{1}+Q_{\mathrm Y_4}^{2}}
\\[8pt]

$(R,5)-(R,6)$
&
\qexpr{Q_{\mathrm X_4}^{1}+Q_{\mathrm Y_4}^{2}}
\\[8pt]

$(R+\bm a_1-\bm a_2,5)-(R,6)$
&
\qexpr{-Q_{\mathrm X_4}^{1}-Q_{\mathrm Y_4}^{2}}
\\[8pt]

$(R+\bm a_1,5)-(R,6)$
&
\qexpr{Q_{\mathrm X_4}^{1}-Q_{\mathrm Y_4}^{2}}
\\[8pt]

$(R-\bm a_3,5)-(R,7)$
&
\qexpr{-Q_{\mathrm X_4}^{2}+Q_{\mathrm Z_4}^{1}}
\\[8pt]

$(R,5)-(R,7)$
&
\qexpr{Q_{\mathrm X_4}^{2}+Q_{\mathrm Z_4}^{1}}
\\[8pt]

$(R+\bm a_1-\bm a_3,5)-(R,7)$
&
\qexpr{-Q_{\mathrm X_4}^{2}-Q_{\mathrm Z_4}^{1}}
\\[8pt]

$(R+\bm a_1,5)-(R,7)$
&
\qexpr{Q_{\mathrm X_4}^{2}-Q_{\mathrm Z_4}^{1}}
\\[8pt]

$(R-\bm a_3,6)-(R,7)$
&
\qexpr{-Q_{\mathrm Y_4}^{1}+Q_{\mathrm Z_4}^{2}}
\\[8pt]

$(R,6)-(R,7)$
&
\qexpr{Q_{\mathrm Y_4}^{1}+Q_{\mathrm Z_4}^{2}}
\\[8pt]

$(R+\bm a_2-\bm a_3,6)-(R,7)$
&
\qexpr{-Q_{\mathrm Y_4}^{1}-Q_{\mathrm Z_4}^{2}}
\\[8pt]

$(R+\bm a_2,6)-(R,7)$
&
\qexpr{Q_{\mathrm Y_4}^{1}-Q_{\mathrm Z_4}^{2}}
\\[8pt]
\end{tabular}
\end{ruledtabular}
\end{table}
\endgroup


\begingroup
\scriptsize
\renewcommand{\arraystretch}{1.25}
\def\dexpr#1{\(\displaystyle\begin{aligned}#1\end{aligned}\)}

\begin{table}[t]
\caption{
Dzyaloshinskii-Moriya vectors on NN bonds
allowed to first order in $Q_{\Lambda}^\mu$.
The listed bonds are directed from site $0$
to sites $4,\ldots,7$.
Here $Q_{\Lambda}^{\pm} = Q_{\Lambda}^1 \pm Q_{\Lambda}^2$.
The site labels are defined in Fig.~\ref{supfig:cubic_displacement}(a).
}
\label{suptab:NN_DMI_Q_0}
\begin{ruledtabular}
\begin{tabular}{ll}
Directed bond $i\to j$
&
$\bm D_{ij}(\bm Q)$
\\[8pt]
\hline
$(R,0)\to(R,4)$
&
\dexpr{
g_1 \begin{pmatrix}Q_{\mathrm X_4}^{-}\\Q_{\mathrm Y_4}^{-}\\Q_{\mathrm Z_4}^{-}\end{pmatrix}
+g_2 \begin{pmatrix}Q_{\mathrm Y_4}^{-}+Q_{\mathrm Z_4}^{-}\\Q_{\mathrm X_4}^{-}+Q_{\mathrm Z_4}^{-}\\Q_{\mathrm X_4}^{-}+Q_{\mathrm Y_4}^{-}\end{pmatrix}}
\\[8pt]

\((R+\bm a_2+\bm a_3,0)\to(R,5)\)
&
\dexpr{
g_1 \begin{pmatrix}-Q_{\mathrm X_4}^{-}\\-Q_{\mathrm Y_4}^{+}\\Q_{\mathrm Z_4}^{+}\end{pmatrix}
+g_2 \begin{pmatrix}Q_{\mathrm Y_4}^{+}-Q_{\mathrm Z_4}^{+}\\Q_{\mathrm X_4}^{-}+Q_{\mathrm Z_4}^{+}\\Q_{\mathrm X_4}^{-}-Q_{\mathrm Y_4}^{+}\end{pmatrix}}
\\[8pt]

$(R+\bm a_1+\bm a_3,0)\to(R,6)$
&
\dexpr{
g_1 \begin{pmatrix}Q_{\mathrm X_4}^{+}\\-Q_{\mathrm Y_4}^{-}\\-Q_{\mathrm Z_4}^{+}\end{pmatrix}
+g_2 \begin{pmatrix}Q_{\mathrm Y_4}^{-}-Q_{\mathrm Z_4}^{+}\\-Q_{\mathrm X_4}^{+}+Q_{\mathrm Z_4}^{+}\\Q_{\mathrm X_4}^{+}+Q_{\mathrm Y_4}^{-}\end{pmatrix}}
\\[8pt]

$(R+\bm a_1+\bm a_2,0)\to(R,7)$
&
\dexpr{
g_1 \begin{pmatrix}-Q_{\mathrm X_4}^{+}\\Q_{\mathrm Y_4}^{+}\\-Q_{\mathrm Z_4}^{-}\end{pmatrix}
+g_2 \begin{pmatrix}Q_{\mathrm Y_4}^{+}+Q_{\mathrm Z_4}^{-}\\-Q_{\mathrm X_4}^{+}+Q_{\mathrm Z_4}^{-}\\Q_{\mathrm X_4}^{+}-Q_{\mathrm Y_4}^{+}\end{pmatrix}}
\\[8pt]
\end{tabular}
\end{ruledtabular}
\end{table}
\endgroup

\begingroup
\scriptsize
\renewcommand{\arraystretch}{1.25}
\def\dexpr#1{\(\displaystyle\begin{aligned}#1\end{aligned}\)}

\begin{table}[t]
\caption{
Dzyaloshinskii-Moriya vectors on NN bonds
allowed to first order in $Q_{\Lambda}^\mu$.
The listed bonds are directed from site $1$
to sites $4,\ldots,7$.
}
\label{suptab:NN_DMI_Q_1}
\begin{ruledtabular}
\begin{tabular}{ll}
Directed bond $i\to j$
&
$\bm D_{ij}(\bm Q)$
\\[8pt]
\hline
$(R,1)\to(R,4)$
&
\dexpr{
g_1 \begin{pmatrix}-Q_{\mathrm X_4}^{-}\\Q_{\mathrm Y_4}^{+}\\-Q_{\mathrm Z_4}^{+}\end{pmatrix}
+g_2 \begin{pmatrix}-Q_{\mathrm Y_4}^{+}+Q_{\mathrm Z_4}^{+}\\Q_{\mathrm X_4}^{-}-Q_{\mathrm Z_4}^{+}\\Q_{\mathrm X_4}^{-}+Q_{\mathrm Y_4}^{+}\end{pmatrix}}
\\[8pt]

$(R,1)\to(R,5)$
&
\dexpr{
g_1 \begin{pmatrix}Q_{\mathrm X_4}^{-}\\-Q_{\mathrm Y_4}^{-}\\-Q_{\mathrm Z_4}^{-}\end{pmatrix}
+g_2 \begin{pmatrix}-Q_{\mathrm Y_4}^{-}-Q_{\mathrm Z_4}^{-}\\Q_{\mathrm X_4}^{-}-Q_{\mathrm Z_4}^{-}\\Q_{\mathrm X_4}^{-}-Q_{\mathrm Y_4}^{-}\end{pmatrix}}
\\[8pt]

$(R+\bm a_1,1)\to(R,6)$
&
\dexpr{
g_1 \begin{pmatrix}-Q_{\mathrm X_4}^{+}\\-Q_{\mathrm Y_4}^{+}\\Q_{\mathrm Z_4}^{-}\end{pmatrix}
+g_2 \begin{pmatrix}-Q_{\mathrm Y_4}^{+}-Q_{\mathrm Z_4}^{-}\\-Q_{\mathrm X_4}^{+}-Q_{\mathrm Z_4}^{-}\\Q_{\mathrm X_4}^{+}+Q_{\mathrm Y_4}^{+}\end{pmatrix}}
\\[8pt]

$(R+\bm a_1,1)\to(R,7)$
&
\dexpr{
g_1 \begin{pmatrix}Q_{\mathrm X_4}^{+}\\Q_{\mathrm Y_4}^{-}\\Q_{\mathrm Z_4}^{+}\end{pmatrix}
+g_2 \begin{pmatrix}-Q_{\mathrm Y_4}^{-}+Q_{\mathrm Z_4}^{+}\\-Q_{\mathrm X_4}^{+}-Q_{\mathrm Z_4}^{+}\\Q_{\mathrm X_4}^{+}-Q_{\mathrm Y_4}^{-}\end{pmatrix}}
\\[8pt]
\end{tabular}
\end{ruledtabular}
\end{table}
\endgroup

\begingroup
\scriptsize
\renewcommand{\arraystretch}{1.25}
\def\dexpr#1{\(\displaystyle\begin{aligned}#1\end{aligned}\)}

\begin{table}[t]
\caption{
Dzyaloshinskii-Moriya vectors on NN bonds
allowed to first order in $Q_{\Lambda}^\mu$.
The listed bonds are directed from site $2$
to sites $4,\ldots,7$.
}
\label{suptab:NN_DMI_Q_2}
\begin{ruledtabular}
\begin{tabular}{ll}
Directed bond $i\to j$
&
$\bm D_{ij}(\bm Q)$
\\[8pt]
\hline
$(R,2)\to(R,4)$
&
\dexpr{
g_1 \begin{pmatrix}-Q_{\mathrm X_4}^{+}\\-Q_{\mathrm Y_4}^{-}\\Q_{\mathrm Z_4}^{+}\end{pmatrix}
+g_2 \begin{pmatrix}Q_{\mathrm Y_4}^{-}+Q_{\mathrm Z_4}^{+}\\Q_{\mathrm X_4}^{+}-Q_{\mathrm Z_4}^{+}\\-Q_{\mathrm X_4}^{+}+Q_{\mathrm Y_4}^{-}\end{pmatrix}}
\\[8pt]

$(R+\bm a_2,2)\to(R,5)$
&
\dexpr{
g_1 \begin{pmatrix}Q_{\mathrm X_4}^{+}\\Q_{\mathrm Y_4}^{+}\\Q_{\mathrm Z_4}^{-}\end{pmatrix}
+g_2 \begin{pmatrix}Q_{\mathrm Y_4}^{+}-Q_{\mathrm Z_4}^{-}\\Q_{\mathrm X_4}^{+}-Q_{\mathrm Z_4}^{-}\\-Q_{\mathrm X_4}^{+}-Q_{\mathrm Y_4}^{+}\end{pmatrix}}
\\[8pt]

$(R,2)\to(R,6)$
&
\dexpr{
g_1 \begin{pmatrix}-Q_{\mathrm X_4}^{-}\\Q_{\mathrm Y_4}^{-}\\-Q_{\mathrm Z_4}^{-}\end{pmatrix}
+g_2 \begin{pmatrix}Q_{\mathrm Y_4}^{-}-Q_{\mathrm Z_4}^{-}\\-Q_{\mathrm X_4}^{-}-Q_{\mathrm Z_4}^{-}\\-Q_{\mathrm X_4}^{-}+Q_{\mathrm Y_4}^{-}\end{pmatrix}}
\\[8pt]

$(R+\bm a_2,2)\to(R,7)$
&
\dexpr{
g_1 \begin{pmatrix}Q_{\mathrm X_4}^{-}\\-Q_{\mathrm Y_4}^{+}\\-Q_{\mathrm Z_4}^{+}\end{pmatrix}
+g_2 \begin{pmatrix}Q_{\mathrm Y_4}^{+}+Q_{\mathrm Z_4}^{+}\\-Q_{\mathrm X_4}^{-}-Q_{\mathrm Z_4}^{+}\\-Q_{\mathrm X_4}^{-}-Q_{\mathrm Y_4}^{+}\end{pmatrix}}
\\[8pt]
\end{tabular}
\end{ruledtabular}
\end{table}
\endgroup

\begingroup
\scriptsize
\renewcommand{\arraystretch}{1.25}
\def\dexpr#1{\(\displaystyle\begin{aligned}#1\end{aligned}\)}

\begin{table}[t]
\caption{
Dzyaloshinskii-Moriya vectors on NN bonds
allowed to first order in $Q_{\Lambda}^\mu$.
The listed bonds are directed from site $3$
to sites $4,\ldots,7$.
}
\label{suptab:NN_DMI_Q_3}
\begin{ruledtabular}
\begin{tabular}{ll}
Directed bond $i\to j$
&
$\bm D_{ij}(\bm Q)$
\\[8pt]
\hline
$(R,3)\to(R,4)$
&
\dexpr{
g_1 \begin{pmatrix}Q_{\mathrm X_4}^{+}\\-Q_{\mathrm Y_4}^{+}\\-Q_{\mathrm Z_4}^{-}\end{pmatrix}
+g_2 \begin{pmatrix}-Q_{\mathrm Y_4}^{+}+Q_{\mathrm Z_4}^{-}\\Q_{\mathrm X_4}^{+}+Q_{\mathrm Z_4}^{-}\\-Q_{\mathrm X_4}^{+}+Q_{\mathrm Y_4}^{+}\end{pmatrix}}
\\[8pt]

$(R+\bm a_3,3)\to(R,5)$
&
\dexpr{
g_1 \begin{pmatrix}-Q_{\mathrm X_4}^{+}\\Q_{\mathrm Y_4}^{-}\\-Q_{\mathrm Z_4}^{+}\end{pmatrix}
+g_2 \begin{pmatrix}-Q_{\mathrm Y_4}^{-}-Q_{\mathrm Z_4}^{+}\\Q_{\mathrm X_4}^{+}+Q_{\mathrm Z_4}^{+}\\-Q_{\mathrm X_4}^{+}-Q_{\mathrm Y_4}^{-}\end{pmatrix}}
\\[8pt]

$(R+\bm a_3,3)\to(R,6)$
&
\dexpr{
g_1 \begin{pmatrix}Q_{\mathrm X_4}^{-}\\Q_{\mathrm Y_4}^{+}\\Q_{\mathrm Z_4}^{+}\end{pmatrix}
+g_2 \begin{pmatrix}-Q_{\mathrm Y_4}^{+}-Q_{\mathrm Z_4}^{+}\\-Q_{\mathrm X_4}^{-}+Q_{\mathrm Z_4}^{+}\\-Q_{\mathrm X_4}^{-}+Q_{\mathrm Y_4}^{+}\end{pmatrix}}
\\[8pt]

$(R,3)\to(R,7)$
&
\dexpr{
g_1 \begin{pmatrix}-Q_{\mathrm X_4}^{-}\\-Q_{\mathrm Y_4}^{-}\\Q_{\mathrm Z_4}^{-}\end{pmatrix}
+g_2 \begin{pmatrix}-Q_{\mathrm Y_4}^{-}+Q_{\mathrm Z_4}^{-}\\-Q_{\mathrm X_4}^{-}+Q_{\mathrm Z_4}^{-}\\-Q_{\mathrm X_4}^{-}-Q_{\mathrm Y_4}^{-}\end{pmatrix}}
\\[8pt]
\end{tabular}
\end{ruledtabular}
\end{table}
\endgroup

\clearpage

\section{Analysis of spin-lattice Hamiltonian}

\subsection{Cluster mean-field theory with a saddle-point approximation}

We write the spin-lattice Hamiltonian,
including both the Heisenberg interactions and the DMIs,
as
\begin{align}
    \label{supeq:HT(Q)}
    \mathcal{H}_{\rm tot}(\bm{Q})
    =
    \frac{1}{2}
    \sum_{ij}\sum_{\alpha\beta}
    J_{ij}^{\alpha\beta}(\bm{Q})
    S_i^\alpha S_j^\beta
    +
    \frac{N_{\rm cell}K_{\mathrm{X}_4}}{2}\bm{Q}^2.
\end{align}
The factor $1/2$ avoids
double counting in the double sum over
the site indices $i$ and $j$.
The exchange tensor is
\begin{align}
    J_{ij}^{\alpha\beta}(\bm{Q})
    =
    \left[
    J_{1,ij}(\bm{Q})\delta_{\alpha\beta}
    +
    \sum_\gamma
    \epsilon_{\alpha\beta\gamma}D_{ij}^{\gamma}(\bm{Q})
    \right]I_{ij}^{\rm NN}
    +
    J_{2,ij}(\bm{Q})\delta_{\alpha\beta}I_{ij}^{\rm NNN},
\end{align}
where
$I_{ij}^{\rm NN}$ and $I_{ij}^{\rm NNN}$
are indicator functions:
$I_{ij}^{\rm NN}=1$ when sites $i$ and $j$ form an NN pair,
and $I_{ij}^{\rm NN}=0$ otherwise;
$I_{ij}^{\rm NNN}$ is defined
analogously for NNN pairs.

For a given covering of the lattice
by NN two-site clusters,
the Hamiltonian can be
rewritten in terms of
the cluster degrees of freedom as
\begin{align}
    \label{supeq:HT_cluster}
    \mathcal{H}_{\rm tot}(\bm{Q})
    =
    \frac{1}{2}
    \sum_{d\neq d'}\sum_{r,r'=1}^{6}
    J_{dd'}^{rr'}(\bm{Q})
    O_d^r O_{d'}^{r'}
    -
    \sum_{d}\sum_{r=7}^{15}
    h_{d}^{r}(\bm{Q}) O_d^r
    +
    \frac{N_{\rm cell}K_{\mathrm{X}_4}}{2}\bm{Q}^2,
\end{align}
where $d$ labels an NN two-site cluster
and
$N_{\rm cell}$ is the number of cubic unit cells.
The local operators
in the cluster $d$ are defined by
\begin{align}
    \label{supeq:O_definition}
    \begin{array}{lll}
        O^{1}_d = S^x_{d,1}, &
        O^{2}_d = S^y_{d,1}, &
        O^{3}_d = S^z_{d,1},\\[6pt]
        O^{4}_d = S^x_{d,2}, &
        O^{5}_d = S^y_{d,2}, &
        O^{6}_d = S^z_{d,2},\\[6pt]
        O^{7}_d = S^x_{d,1}S^x_{d,2}, &
        O^{8}_d = S^x_{d,1}S^y_{d,2}, &
        O^{9}_d = S^x_{d,1}S^z_{d,2},\\[6pt]
        O^{10}_d = S^y_{d,1}S^x_{d,2}, &
        O^{11}_d = S^y_{d,1}S^y_{d,2}, &
        O^{12}_d = S^y_{d,1}S^z_{d,2},\\[6pt]
        O^{13}_d = S^z_{d,1}S^x_{d,2}, &
        O^{14}_d = S^z_{d,1}S^y_{d,2}, &
        O^{15}_d = S^z_{d,1}S^z_{d,2}.
    \end{array}
\end{align}
The intra-cluster interactions
are incorporated into $h_d^r(\bm{Q})$,
whereas $J_{dd'}^{rr'}(\bm{Q})$
describes the couplings between different clusters.
Accordingly,
the inter-cluster couplings
involve only single-spin operators
$O_d^1,\dots,O_d^6$ on each cluster.

We treat the displacement coordinates
as classical variational parameters.
The saddle-point condition
$\partial\braket{\mathcal{H}_{\rm tot}(\bm{Q})}/\partial Q_\Lambda^\mu=0$
gives
\begin{align}
    \label{supeq:Q_saddle}
    Q_\Lambda^\mu
    &=
    \frac{1}{N_{\rm cell}K_{\mathrm{X}_4}}
    \left[
    \sum_{d}\sum_{r=7}^{15}
    a_{d,\Lambda\mu}^{r}\braket{O_d^r}
    -
    \frac{1}{2}
    \sum_{d\neq d'}\sum_{r,r'=1}^{6}
    a_{dd',\Lambda\mu}^{rr'}
    \braket{O_d^r O_{d'}^{r'}}
    \right],
\end{align}
where $\braket{\cdots}$ denotes
the ground-state expectation value
in the spin Hilbert space
at fixed $\bm{Q}$, and
\begin{align}
    a_{d,\Lambda\mu}^{r}
    =
    \pdiff{h_d^r(\bm{Q})}{Q_\Lambda^\mu},
    \qquad
    a_{dd',\Lambda\mu}^{rr'}
    =
    \pdiff{J_{dd'}^{rr'}(\bm{Q})}{Q_{\Lambda}^\mu}.
\end{align}

We then apply cluster mean-field theory~(CMFT)
by decoupling the inter-cluster part
of Eq.~\eqref{supeq:HT_cluster}.
Introducing the fluctuation
$\delta O_d^r=O_d^r-\braket{O_d^r}$,
we decompose the Hamiltonian as
\begin{align}
    \mathcal{H}_{\rm tot}(\bm{Q})
    =
    \mathcal{H}^{\rm MF}(\bm{Q})
    +
    \mathcal{H}'(\bm{Q}),
\end{align}
where $\mathcal{H}'(\bm{Q})$ contains
terms quadratic in
the fluctuations; these terms are
neglected in the CMFT calculations.
The mean-field Hamiltonian is
\begin{align}
    \mathcal{H}^{\rm MF}(\bm{Q})
    =
    \sum_d \mathcal{H}_{d}^{\rm MF}(\bm{Q})
    -
    \frac{1}{2}
    \sum_{d\neq d'}\sum_{r,r'=1}^{6}
    J_{dd'}^{rr'}(\bm{Q})
    \braket{O_{d}^r}\braket{O^{r'}_{d'}}
    +
    \frac{N_{\rm cell}K_{\mathrm{X}_4}}{2}
    \bm{Q}^2,
\end{align}
where the local mean-field Hamiltonian is
\begin{align}
    \label{supeq:Hd_MF}
    \mathcal{H}_{d}^{\rm MF}(\bm{Q})
    =
    \sum_{r=1}^{6}
    \left[
    \sum_{d'(\neq d)}\sum_{r'=1}^{6}
    J_{dd'}^{rr'}(\bm{Q})\braket{O_{d'}^{r'}}
    \right]
    O_d^r
    -
    \sum_{r=7}^{15}
    h_d^r(\bm{Q})O_d^r.
\end{align}
Within this approximation,
the inter-cluster correlations 
in Eq.~\eqref{supeq:Q_saddle}
are factorized as
\begin{align}
    \braket{O_d^rO_{d'}^{r'}}
    \simeq
    \braket{O_d^r}
    \braket{O_{d'}^{r'}},
\end{align}
and the saddle-point equation becomes
\begin{align}
    \label{supeq:Q_saddle_MF}
    Q_\Lambda^\mu
    &=
    \frac{1}{N_{\rm cell}K_{\mathrm{X}_4}}
    \left[
    \sum_{d}\sum_{r=7}^{15}
    a_{d,\Lambda\mu}^{r}\braket{O_d^r}
    -
    \frac{1}{2}
    \sum_{d\neq d'}\sum_{r,r'=1}^{6}
    a_{dd',\Lambda\mu}^{rr'}
    \braket{O_d^r}\braket{O_{d'}^{r'}}
    \right].
\end{align}
In the numerical calculation,
we iterate the following procedure:
we diagonalize each
$\mathcal H_d^{\rm MF}(\bm{Q})$,
update $\braket{O_d^r}$,
and update $Q_\Lambda^\mu$
using Eq.~\eqref{supeq:Q_saddle_MF}
until self-consistency is reached.
We then compare the energies of the
converged ground-state solutions.

\begin{figure}[b]
\includegraphics[width=0.9\columnwidth]{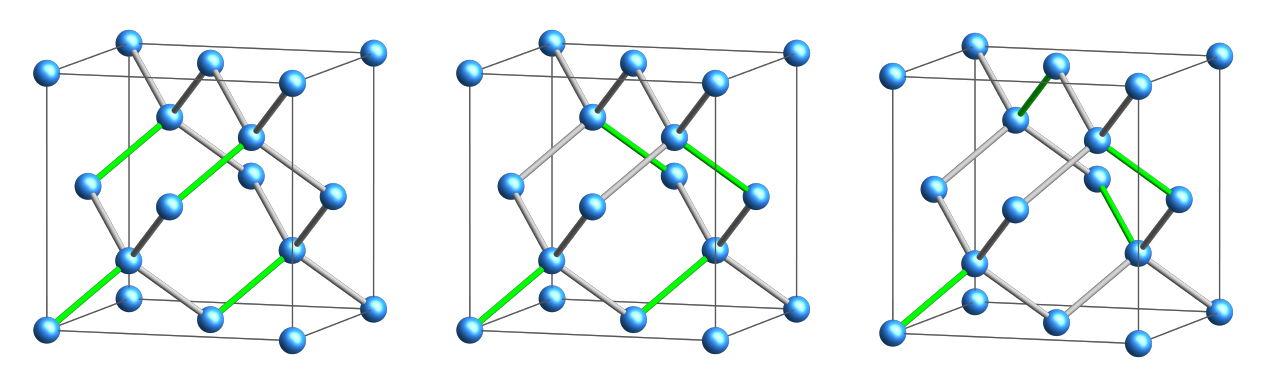}
\caption{
Representative examples of
the three symmetry-inequivalent coverings
of the cubic unit cell
by NN two-site clusters.
Green bonds denote the NN bonds
chosen as two-site clusters.
}
\label{supfig:covering}
\end{figure}

For the two-site cluster covering,
we partition the eight sites
in each cubic unit cell
into four NN two-site clusters
and repeat the same covering pattern
periodically over all cubic cells.
There are $24$ such coverings,
which are classified into
three inequivalent classes under the
symmetry operations of $Fd\bar{3}m$
[Fig.~\ref{supfig:covering}].
We solve the CMFT equations
for a representative covering
from each of the three classes
and select the lowest-energy solution.

We also consider
single-$q$ mean-field solutions.
This is motivated by the fact that
$J_2/J_1=0.10$ is close to
the spiral instability at
$J_2= J_1/8$
in the diamond-lattice $J_1$-$J_2$ model
\cite{bergman2007}.
Although $J_2=0.10J_1$ lies outside
the classical single-$q$
spiral phase,
finite lattice displacements
modulate the exchange interactions.
Single-$q$ solutions
may therefore compete with
the $\bm{q} = \bm{0}$ VBS solutions.
To examine this possibility,
we introduce a single-$q$ ansatz
for the local mean fields
$\braket{O_d^r}$.

For each covering,
we decompose the cluster label as $d=(R,l)$,
where $R$ labels the cubic unit cell
and $l$ distinguishes
the four clusters associated with that cell.
We denote by $\bm{R}$ a representative position
of the cubic unit cell labeled by $R$
and by $\bm{\delta}_l$ the relative position vector
from $\bm{R}$ to the representative position of cluster $l$.
Thus, the representative position
of cluster $d=(R,l)$ is
$\bm{R}+\bm{\delta}_l$.
For $\bm{q}\not\equiv-\bm{q}$
modulo reciprocal lattice vectors of the cubic cell,
the single-$q$ ansatz is
\begin{align}
    \label{supeq:single-q-ansatz}
    \braket{O_d^r}
    =
    \bar{O}_{\bm{q}=\bm{0},l}^r
    +
    2\, {\rm Re}
    \left[
    e^{i\bm{q}\cdot(\bm{R}+\bm{\delta}_l)}
    \bar{O}_{\bm{q},l}^r
    \right].
\end{align}
The corresponding Fourier component is
\begin{align}
    \bar{O}_{\bm{q},l}^{r}
    =
    \frac{1}{N_{\rm cell}}
    \sum_R
    e^{-i\bm{q}\cdot(\bm{R}+\bm{\delta}_l)}
    \braket{O_{R,l}^{r}}.
\end{align}

We approximate general incommensurate wave vectors $\bm{q}$
using a finite set of commensurate wave vectors.
In terms of the reciprocal lattice vectors
$\bm{b}_1,\bm{b}_2$, and
$\bm{b}_3$ of the cubic cell, we discretize
$\bm{q}$ as
\begin{align}
    \bm{q}
    =
    \sum_{\nu=1}^{3}
    \frac{m_\nu}{10}\bm{b}_\nu,
    \qquad
    m_\nu=0,1,\ldots,9.
\end{align}
This amounts to examining
$10^3$ commensurate single-$q$
candidates for each covering.
For each discretized $\bm{q}$,
the phase factor
$e^{i\bm{q}\cdot(\bm{R}+\bm{\delta}_l)}$
is periodic under translations
by the lattice vectors of a finite supercell.
All sampled $\bm{q}$ are commensurate with
a common $10 \times 10 \times 10$ supercell,
although some have smaller periods.
Consequently,
$\mathcal H_d^{\rm MF}$,
$\bar{O}_{\bm{q},l}^r$,
and the energy density at fixed $\bm{q}$
can be evaluated using only
the finite set of phases
$\bm{q}\cdot\bm{R}$,
within the corresponding supercell.
In particular, thermodynamic-limit averages
over $R$ can be replaced by
finite averages over that supercell.
For each $\bm{q}$,
we solve the CMFT and saddle-point equations
self-consistently and select
the lowest-energy solution.
The resulting saddle-point displacement
is denoted by $\bm{Q}_0$.

\subsection{Flavor-wave theory}

We analyze the elementary excitations
in the VBS state using
linear flavor-wave theory~(LFWT).
The VBS state corresponds to
a $\bm{q}=\bm{0}$ solution
with respect to the cubic unit cell.
Therefore,
the local mean-field Hamiltonian
does not depend on the unit-cell label $R$.
For a cluster $d=(R,l)$,
we denote the local mean-field ground state
and excited states by
$\ket{l;0}$ and
$\ket{l;\rho}\ (\rho=1,2,3)$,
respectively.

Since the VBS state has no local magnetization,
$\braket{O_d^r}=0$ for $r=1,\ldots,6$.
Thus, $\delta O_d^r=O_d^r$ for these components.
The quadratic fluctuation term
omitted from the static CMFT decoupling
is therefore
\begin{align}
    \label{supeq:Hprime_Q0}
    \mathcal{H}'(\bm{Q}_0)
    =
    \frac{1}{2}
    \sum_{d\neq d'}\sum_{r,r'=1}^{6}
    J_{dd'}^{rr'}(\bm{Q}_0)
    O_d^r O_{d'}^{r'}.
\end{align}
Both $r$ and $r'$
are restricted to $1,\ldots,6$
because the inter-cluster Hamiltonian
contains only spin-spin couplings.

Expanding the local operator $O_d^r$
in the basis of local states,
we obtain
\begin{align}
    \label{supeq:Odr}
    O_d^r
    &=
    X_d^{00}
    \braket{l;0|O_l^r|l;0}
    +
    \sum_{\rho=1}^{3}
    \left[
    X_d^{\rho0}
    \braket{l;\rho|O_l^r|l;0}
    +
    \mathrm{H.c.}
    \right]
    +
    \sum_{\rho,\rho'=1}^{3}
    X_d^{\rho\rho'}
    \braket{l;\rho|O_l^r|l;\rho'},
\end{align}
where
$X_d^{\rho0}=\ket{l;\rho}\bra{l;0}$,
$X_d^{\rho\rho'}=\ket{l;\rho}\bra{l;\rho'}$,
and $X_d^{00}=\ket{l;0}\bra{l;0}$.
Here, $O_l^r$ denotes
the local operator on the cluster
with internal label $l$.
For the solution with $\bm{q}=\bm{0}$,
the local states and their matrix elements
depend only on $l$;
the unit-cell label $R$ is therefore omitted.

Since $\braket{l;0|O_l^r|l;0}=0$
for $r=1,\ldots,6$,
Eq.~\eqref{supeq:Odr} reduces to
\begin{align}
    O_d^r
    =
    \sum_{\rho=1}^{3}
    \left[
    X_d^{\rho0}
    \braket{l;\rho|O_l^r|l;0}
    +
    \mathrm{H.c.}
    \right]
    +
    \sum_{\rho,\rho'=1}^{3}
    X_d^{\rho\rho'}
    \braket{l;\rho|O_l^r|l;\rho'}.
\end{align}
The operators
$X_d^{00}$, $X_d^{\rho0}$,
and $X_d^{\rho\rho'}$
are represented
in terms of the boson operators
$a_{d,\rho}^\dagger$ and
$a_{d,\rho}$
by the generalized
Holstein--Primakoff transformation~\cite{joshi1999,shiina2003}:
\begin{align}
    X_d^{\rho0}
    &=
    a_{d,\rho}^\dagger
    \left(
    \mathcal{M}
    -
    \sum_{\rho'=1}^{3}
    a_{d,\rho'}^\dagger a_{d,\rho'}
    \right)^{1/2},
    \qquad
    X_d^{0\rho}
    =
    \left(X_d^{\rho0}\right)^\dagger,\\
    X_d^{\rho\rho'}
    &=
    a_{d,\rho}^\dagger a_{d,\rho'},\\
    X_d^{00}
    &=
    \mathcal{M}
    -
    \sum_{\rho=1}^{3}
    a_{d,\rho}^\dagger a_{d,\rho}.
\end{align}
The parameter $\mathcal{M}$
is fixed by the local completeness constraint
\begin{align}
    \mathcal{M}
    =
    X_d^{00}
    +
    \sum_{\rho=1}^{3}
    a_{d,\rho}^\dagger a_{d,\rho}.
\end{align}
Here, the physical Hilbert space corresponds to
$\mathcal{M}=1$.

Assuming that the excitation densities are low,
we retain only terms up to quadratic order
in the boson operators.
This corresponds to the approximation
\begin{align}
    \left(
    \mathcal{M}
    -
    \sum_{\rho'=1}^{3}
    a_{d,\rho'}^\dagger a_{d,\rho'}
    \right)^{1/2}
    \simeq
    \sqrt{\mathcal{M}}.
\end{align}
Setting $\mathcal{M}=1$,
the local operators for $r=1,\ldots,6$
are linearized as
\begin{align}
    \label{supeq:O_expand}
    O_d^r
    \simeq
    \sum_{\rho=1}^{3}
    \left[
    \braket{l;\rho|O_l^r|l;0}
    a_{d,\rho}^\dagger
    +
    \mathrm{H.c.}
    \right].
\end{align}
The terms containing
$X_d^{\rho\rho'}=a_{d,\rho}^\dagger a_{d,\rho'}$
generate only third- and higher-order bosonic terms in
$\mathcal{H}'\sim O_d^rO_{d'}^{r'}$, and are therefore neglected.
We refer to this approximation as LFWT.

Substituting Eq.~\eqref{supeq:O_expand} into
Eq.~\eqref{supeq:Hprime_Q0}, we obtain
\begin{align}
    \mathcal{H}'(\bm{Q}_0)
    \simeq
    \frac{1}{2}
    \sum_{d\neq d'}
    \sum_{r,r'=1}^{6}
    \sum_{\rho,\rho'=1}^{3}
    J_{dd'}^{rr'}(\bm{Q}_0)
    &
    \left[
    \braket{l;\rho|O_l^r|l;0}
    a_{R,l\rho}^\dagger
    +
    \mathrm{H.c.}
    \right]
    \left[
    \braket{l';\rho'|O_{l'}^{r'}|l';0}
    a_{R',l'\rho'}^\dagger
    +
    \mathrm{H.c.}
    \right],
\end{align}
where $d=(R,l)$ and $d'=(R',l')$,
and we have written
$a_{d,\rho}=a_{R,l\rho}$.
There are
$N=4\times3=12$
local excitation degrees of freedom
in each cubic unit cell,
giving $N$ positive-energy
excitation branches.

Since $\ket{l;0}$ and $\ket{l;\rho}$ 
are eigenstates of the local mean-field Hamiltonian,
the mean-field contribution is written,
up to a constant, as
\begin{align}
    \mathcal{H}^{\rm MF}(\bm{Q}_0)
    =
    \frac{1}{2}
    \sum_{R,l}\sum_{\rho=1}^{3}
    \Delta E_{\rho}^{l}(\bm{Q}_0)
    \left(
    a_{R,l\rho}^\dagger a_{R,l\rho}
    +
    a_{R,l\rho} a_{R,l\rho}^\dagger
    \right),
\end{align}
where $\Delta E_{\rho}^{l}(\bm{Q}_0)$
is the energy difference
between $\ket{l;\rho}$ and $\ket{l;0}$.
The LFWT Hamiltonian is therefore
\begin{align}
    \mathcal{H}_{\rm FW}(\bm{Q}_0)
    \equiv
    \mathcal{H}^{\rm MF}(\bm{Q}_0)
    +
    \mathcal{H}'(\bm{Q}_0),
\end{align}
which is quadratic
in the boson operators.
In the following, we omit the argument $\bm{Q}_0$.

To focus on the effects of VBS order,
we use the undistorted lattice positions
in the Fourier phase factors,
while retaining the $\bm{Q}_0$ dependence
of the Hamiltonian.
We define the Fourier transform of the boson operators by
\begin{align}
    \label{supeq:a_fourier}
    a_{R,l\rho}
    =
    \frac{1}{\sqrt{N_{\rm cell}}}
    \sum_{\bm{k}}
    a_{\bm{k},l\rho}
    e^{i\bm{k}\cdot(\bm{R}+\bm{\delta}_l)}.
\end{align}
Substituting this transformation into $\mathcal{H}_{\rm FW}$ gives
\begin{align}
    \label{supeq:HFW}
    \mathcal{H}_{\rm FW}
    =
    \frac{1}{2}
    \sum_{\bm{k}}
    \mathcal{A}_{\bm{k}}^\dagger
    H_{\bm{k}}
    \mathcal{A}_{\bm{k}},
\end{align}
where $H_{\bm{k}}$ is a
$2N\times2N$ Hermitian matrix
and
$\mathcal{A}_{\bm{k}}^\dagger$ is
a $2N$-component vector.
Introducing the composite index
$s=(l,\rho)$ with $s=1,\ldots,N$,
we define
\begin{align}
    \mathcal{A}_{\bm{k},s}^\dagger
    =
    a_{\bm{k},l\rho}^\dagger,
    \qquad
    \mathcal{A}_{\bm{k},s+N}^\dagger
    =
    a_{-\bm{k},l\rho}.
\end{align}

We diagonalize Eq.~\eqref{supeq:HFW} using a Bogoliubov transformation.
To preserve the bosonic commutation relations,
the transformation matrix
$T_{\bm{k}}$ must be paraunitary~\cite{colpa}.
We use the paraunit matrix
\begin{align}
    \sigma_3
    =
    {\rm diag}
    (
    \underbrace{1,\ldots,1}_{N},
    \underbrace{-1,\ldots,-1}_{N}
    ).
\end{align}
The paraunitary matrix satisfies
$
T_{\bm{k}}^\dagger \sigma_3 T_{\bm{k}}
=
T_{\bm{k}}\sigma_3 T_{\bm{k}}^\dagger
=
\sigma_3
$.
We introduce the Bogoliubov bosons through
$
\mathcal{A}_{\bm{k}}^\dagger
=
\mathcal{B}_{\bm{k}}^\dagger T_{\bm{k}}^\dagger,
$
where
\begin{align}
    \mathcal{B}_{\bm{k}}^\dagger
    =
    \left(
    b_{\bm{k},1}^\dagger,\ldots,b_{\bm{k},N}^\dagger,
    b_{-\bm{k},1},\ldots,b_{-\bm{k},N}
    \right).
\end{align}
Equation~\eqref{supeq:HFW} is then diagonalized as
\begin{align}
    \mathcal{H}_{\rm FW}
    =
    \frac{1}{2}
    \sum_{\bm{k}}
    \mathcal{B}_{\bm{k}}^\dagger
    \mathcal{E}_{\bm{k}}
    \mathcal{B}_{\bm{k}},
\end{align}
where
$
\mathcal{E}_{\bm{k}}
=
T_{\bm{k}}^\dagger H_{\bm{k}}T_{\bm{k}}
=
{\rm diag}
(
\varepsilon_{\bm{k},1},\ldots,\varepsilon_{\bm{k},N},
\varepsilon_{-\bm{k},1},\ldots,\varepsilon_{-\bm{k},N}
)
$
is diagonal.
The Bogoliubov vacuum and one-triplon states
are defined by
\begin{align}
    b_{\bm{k},n}\ket{\ket{0}}=0,
    \qquad
    \ket{\ket{\bm{k},n}}
    =
    b_{\bm{k},n}^\dagger\ket{\ket{0}}.
\end{align}

The paraunitary matrix can be written
in block form as
\begin{align}
    T_{\bm{k}}
    =
    \begin{pmatrix}
        U_{\bm{k}} & V_{-\bm{k}}^*\\
        V_{\bm{k}} & U_{-\bm{k}}^*
    \end{pmatrix},
\end{align}
where $U_{\bm{k}}$ and $V_{\bm{k}}$ are $N\times N$ matrices.
With this convention,
the Holstein--Primakoff bosons and
Bogoliubov bosons are related by
Eqs.~\eqref{supeq:atob}
and~\eqref{supeq:adaggertobdagger}
\cite{rau2018,mook2020,koyama2023}:
\begin{align}
    \label{supeq:atob}
    a_{\bm{k},s}
    &=
    \sum_{n=1}^{N}
    \left[
    (U_{\bm{k}})_{sn} b_{\bm{k},n}
    +
    (V_{-\bm{k}}^*)_{sn} b_{-\bm{k},n}^\dagger
    \right],
    \\
    \label{supeq:adaggertobdagger}
    a_{\bm{k},s}^\dagger
    &=
    \sum_{n=1}^{N}
    \left[
    (U_{\bm{k}}^*)_{sn} b_{\bm{k},n}^\dagger
    +
    (V_{-\bm{k}})_{sn} b_{-\bm{k},n}
    \right].
\end{align}

\section{Internal spin polarization of triplons}

We define the Fourier component of the spin operator as
\begin{align}
    S_{\bm{k}}^\alpha
    &\equiv
    \frac{1}{\sqrt{N_s}}
    \sum_{i=1}^{N_s}
    S_i^\alpha
    e^{-i\bm{k}\cdot\bm{r}_i}
    \notag\\
    &=
    \sqrt{\frac{N_{\rm cell}}{N_s}}
    \sum_{l,c}
    S_{\bm{k},lc}^\alpha ,
\end{align}
where $N_s=8N_{\rm cell}$ is the total number of sites.
We decompose the site label as $i=(d,c)=(R,l,c)$,
where $c=1,2$ labels the two sites within cluster $d$.
The $lc$ component of the Fourier-transformed
spin operator is defined by
\begin{align}
    \label{supeq:Slc_fourier}
    S_{\bm{k},lc}^\alpha
    =
    \frac{1}{\sqrt{N_{\rm cell}}}
    \sum_{R}
    S_{R,lc}^\alpha
    e^{-i\bm{k}\cdot
    (\bm{R}+\bm{\delta}_l+\tilde{\bm{\delta}}_{l,c})}.
\end{align}
Here,
$\tilde{\bm{\delta}}_{l,c}$
is the relative position vector
from the representative position
of cluster $l$ to site $c$.
Thus, the undistorted position
of site $i=(R,l,c)$ is
$\bm{r}_i=
\bm{R}+\bm{\delta}_l+\tilde{\bm{\delta}}_{l,c}$.

Within the linearized approximation,
applying the local-operator expansion
in Eq.~\eqref{supeq:O_expand}
to $S_{R,lc}^\alpha$
and using Eq.~\eqref{supeq:a_fourier}, we obtain
\begin{align}
    \label{supeq:Slc_fourier_expand}
    S_{\bm{k},lc}^\alpha
    =
    \sum_{\rho=1}^{3}
    e^{-i\bm{k}\cdot\tilde{\bm{\delta}}_{l,c}}
    \left[
    (S_c^\alpha)_{l;\rho0}^{*}
    a_{\bm{k},l\rho}
    +
    (S_c^\alpha)_{l;\rho0}
    a_{-\bm{k},l\rho}^{\dagger}
    \right],
\end{align}
where
$(S_c^\alpha)_{l;\rho0}=\braket{l;\rho|S_c^\alpha|l;0}$.
Since the local states and
the matrix elements of the local operators
are independent of $R$,
the unit-cell label is omitted in the
matrix elements.

Using Eqs.~\eqref{supeq:atob}
and~\eqref{supeq:adaggertobdagger},
Eq.~\eqref{supeq:Slc_fourier_expand}
can be written as
\begin{align}
    \label{supeq:Slc_F}
    S_{\bm{k},lc}^\alpha
    =
    \sum_{n=1}^{N}
    \left[
    F_{lc,n}^{\alpha}(\bm{k}) b_{\bm{k},n}
    +
    F_{lc,n}^{\alpha *}(-\bm{k})
    b_{-\bm{k},n}^{\dagger}
    \right],
\end{align}
where
\begin{align}
    \label{supeq:Fk}
    F_{lc,n}^{\alpha}(\bm{k})
    =
    e^{-i\bm{k}\cdot\tilde{\bm{\delta}}_{l,c}}
    \sum_{\rho=1}^{3}
    \left[
    (S_c^\alpha)_{l;\rho0}^{*}
    (U_{\bm{k}})_{l\rho,n}
    +
    (S_c^\alpha)_{l;\rho0}
    (V_{\bm{k}})_{l\rho,n}
    \right].
\end{align}
The quantity $F_{lc,n}^{\alpha}(\bm{k})$
is the transition spin amplitude
between the Bogoliubov vacuum and the one-triplon state:
\begin{align}
    \label{supeq:Flc}
    F_{lc,n}^{\alpha}(\bm{k})
    =
    \braket{\braket{0|S_{\bm{k},lc}^{\alpha}|\bm{k},n}}.
\end{align}

To visualize the transition spin amplitude
as a classical spin fluctuation,
we define
\begin{align}
    \label{supeq:deltaS}
    \delta \bm{S}_{lc,n}(\bm{k},t)
    =
    {\rm Re}
    \left[
    \bm{F}_{lc,n}(\bm{k})
    e^{-i\varepsilon_{\bm{k},n}t/\hbar}
    \right],
\end{align}
where
$\bm{F}_{lc,n}(\bm{k})
=
(F_{lc,n}^{x},F_{lc,n}^{y},F_{lc,n}^{z})$.
As a measure of the polarization of
this spin fluctuation, we introduce
\begin{align}
    \label{supeq:chi}
    \bm{\chi}_{lc,n}(\bm{k})
    =
    -\frac{1}{2}
    {\rm Im}
    \left[
    \bm{F}_{lc,n}(\bm{k})
    \times
    \bm{F}_{lc,n}^{*}(\bm{k})
    \right].
\end{align}

To clarify the physical meaning of
$\delta\bm{S}_{lc,n}$ and $\bm{\chi}_{lc,n}$,
we write
$\bm{F}_{lc,n}(\bm{k})=\bm{A}+i\bm{B}$,
where $\bm{A}$ and $\bm{B}$ are real vectors.
The corresponding spin fluctuation is
\begin{align}
    \delta\bm{S}_{lc,n}(\bm{k},t)
    =
    \bm{A}
    \cos\frac{\varepsilon_{\bm{k},n}t}{\hbar}
    +
    \bm{B}
    \sin\frac{\varepsilon_{\bm{k},n}t}{\hbar}.
\end{align}
When $\bm{A}$ and $\bm{B}$ are not parallel,
$\delta \bm{S}_{lc,n}(\bm{k},t)$
traces an ellipse in the plane
spanned by $\bm{A}$ and $\bm{B}$,
representing an internal rotation
in spin space.
The polarization vector becomes
\begin{align}
    \bm\chi_{lc,n}(\bm{k})
    =
    \bm{A}\times\bm{B}.
\end{align}
Thus, $\bm{\chi}_{lc,n}(\bm{k})$ specifies
the axis and sense of the internal rotation.
We use $\bm{\chi}_{lc,n}(\bm{k})$
to characterize the internal polarization
of triplons.

We next discuss the transformation property of
$\bm{\chi}_{lc,n}(\bm{k})$.
Let $g$ be an operation of the parent space group.
For a structure with space group $\mathcal{G}$,
the transformed structure has the space group
$\mathcal{G}_g = g \mathcal{G}g^{-1}$.
We denote the
Bogoliubov vacuum and the one-triplon state by
$\ket{\ket{0;\mathcal{G}}}$ and
$\ket{\ket{\bm{k},n;\mathcal{G}}}$, respectively.
Under $g$, these states transform as
\begin{align}
    g\ket{\ket{0;\mathcal{G}}}
    =
    \ket{\ket{0;\mathcal{G}_g}},
    \qquad
    g\ket{\ket{\bm{k},n;\mathcal{G}}}
    =
    \ket{\ket{R_g\bm{k},n;\mathcal{G}_g}},
\end{align}
up to overall phases, which are suppressed below.
We assume that the relevant branch is nondegenerate
and use the same energy-ordered band index $n$
for the symmetry-related states.

The transition spin amplitude in the structure with the space group
$\mathcal{G}$ is written as
\begin{align}
    \bm{F}_{lc,n}^{(\mathcal{G})}(\bm{k})
    =
    \braket{\braket{
    0;\mathcal{G}|\bm{S}_{\bm{k},lc}|\bm{k},n;\mathcal{G}
    }}.
\end{align}
Using Eq.~\eqref{supeq:gSg-1},
its Cartesian components
transform as
\begin{align}
    F^{\alpha(\mathcal{G})}_{lc,n}(\bm{k})
    &=
    \braket{\braket{
    0;\mathcal{G}|g^{-1}gS^\alpha_{\bm{k},lc}g^{-1}g|\bm{k},n;\mathcal{G}
    }}\notag\\
    &=
    \sum_{\beta}
    \braket{\braket{
    0;\mathcal{G}_g|(\bar{R}_g^{-1})_{\alpha\beta}S^\beta_{R_g\bm{k},glc}|R_g\bm{k},n;\mathcal{G}_g
    }}\notag\\
    &=
    \sum_\beta
    (\bar{R}_g^{-1})_{\alpha\beta}
    F^{\beta(\mathcal{G}_g)}_{glc,n}(R_g\bm{k}),
\end{align}
where $glc$ denotes the image of
the site label $(l,c)$ under $g$.
Equivalently,
\begin{align}
    \label{supeq:F_trans}
    \bm{F}_{glc,n}^{(\mathcal{G}_g)}(R_g\bm{k})
    =
    \bar{R}_g
    \bm{F}_{lc,n}^{(\mathcal{G})}(\bm{k}).
\end{align}
It follows that
\begin{align}
    \label{supeq:chi_trans}
    \bm{\chi}_{glc,n}^{(\mathcal{G}_g)}(R_g\bm{k})
    =
    \bar{R}_g
    \bm{\chi}_{lc,n}^{(\mathcal{G})}(\bm{k}) .
\end{align}
In deriving this relation, we have used Eq.~\eqref{supeq:RgRg_convert}.
Thus, the three spin components of
$\bm{\chi}_{lc,n}(\bm{k})$ transform as an axial vector,
namely according to the T$_{1g}$ representation
of the parent point group.

\begin{figure}[t]
\includegraphics[width=0.9\columnwidth]{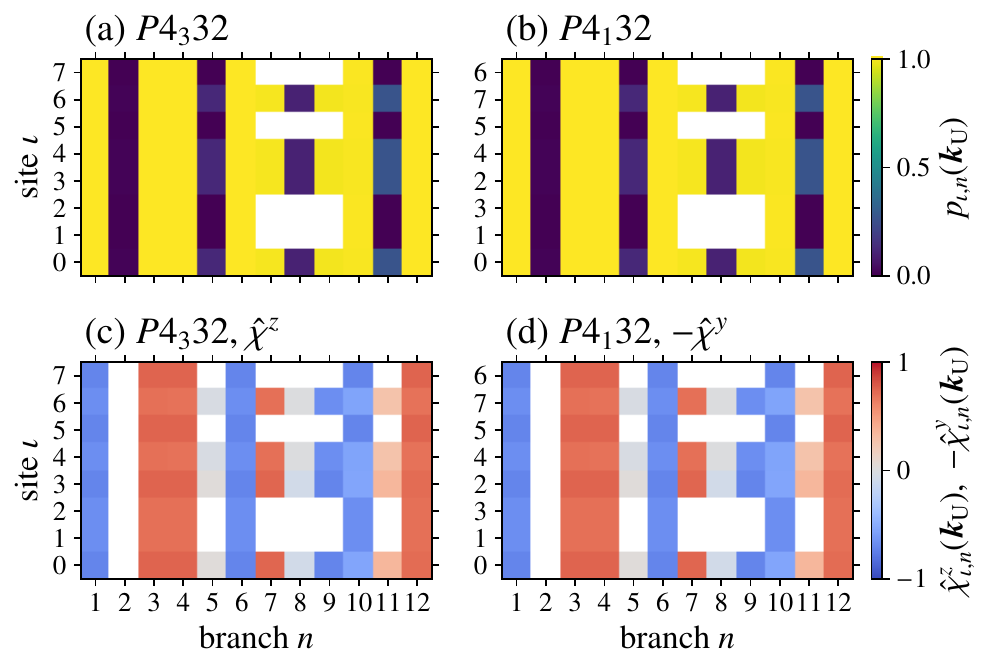}
\caption{
The upper panels show $p_{\iota,n}(\bm{k}_{\rm U})$ for
(a) $P4_332$ and (b) $P4_132$, respectively.
White regions correspond to entries satisfying
$|\bm{F}_{\iota,n}|^2/[S(S+1)]<0.01$.
The lower panels show the normalized internal polarization
$\hat{\chi}_{\iota,n}^\alpha(\bm{k}_{\rm U})
=\chi_{\iota,n}^\alpha(\bm{k}_{\rm U})/|\bm{\chi}_{\iota,n}(\bm{k}_{\rm U})|$:
(c) $\hat{\chi}_{\iota,n}^{z}$
for $P4_332$ and
(d) $-\hat{\chi}_{\iota,n}^{y}$
for $P4_132$.
All panels are calculated for
$J_2=0.1J_1$, $\eta=6$,
$\bar{K}_{\rm X_4}=600$, and
$g_1=g_2=0.2\lambda_1$.
White regions in the lower panels correspond to entries satisfying
$|\bm{F}_{\iota,n}|^2/[S(S+1)]<0.01$ or $p_{\iota,n}(\bm{k}_{\rm U})<0.01$.
}
\label{supfig:pcirc}
\end{figure}

To quantify the degree of circular polarization,
we define
\begin{align}
    p_{\iota,n}(\bm{k})
    =
    2
    \frac{|\bm{\chi}_{\iota,n}(\bm{k})|}
    {|\bm{F}_{\iota,n}(\bm{k})|^2}.
\end{align}
We use the composite index
$\iota=(l,c)$
for the internal site degrees of freedom
within the cubic unit cell.
As shown in Fig.~\ref{supfig:cubic_displacement}(a),
this index can equivalently be identified with
the site label $\iota=0,\ldots,7$.
The polarization is linear when
$p_{\iota,n}(\bm{k})=0$,
circular when
$p_{\iota,n}(\bm{k})=1$,
and elliptical when
$0<p_{\iota,n}(\bm{k})<1$.

Figures~\ref{supfig:pcirc}(a)
and~\ref{supfig:pcirc}(b)
show $p_{\iota,n}(\bm{k}_{\rm U})$ for the two enantiomorphic states
at $\bm{k}_\mathrm{U}=\pi(2,1/2,1/2)/a$,
with $J_2=0.1J_1$, $\eta=6$, $\bar{K}_{\rm X_4}=600$,
and $g_1=g_2=0.2\lambda_1$.
We choose the U point because
all branches at $\bm{k}_{\rm U}$
are nondegenerate,
allowing the polarization of each mode to be compared 
unambiguously.
Since the polarization of a component
with a very small transition
amplitude is not meaningful,
entries satisfying
$|\bm{F}_{\iota,n}|^2/[S(S+1)]<0.01$
are masked in white.

In the parent $Fd\bar{3}m$ structure,
the little group of $\bm{k}_{\rm U}$
contains the mirror operation
$\{\hat{\sigma}_{yz}|\bm{0}\}$,
where 
$\sigma_{yz}: (x,y,z)\mapsto(x,z,y)$.
The mirror operation maps
the $P4_332$ enantiomer onto
the $P4_132$ enantiomer.
Under this mapping, the sites
$(0,1,2,3,4,5,6,7)$ in the $P4_332$ state correspond to
$(0,1,3,2,4,5,7,6)$ in the $P4_132$ state.
Denoting by $\iota'$ the site
in the $P4_132$ state corresponding to
the site $\iota$ in the $P4_332$ state,
we obtain
$p_{\iota,n}^{(P4_332)}(\bm{k}_{\rm U})
=
p_{\iota',n}^{(P4_132)}(\bm{k}_{\rm U})$.
The numerical results in
Figs.~\ref{supfig:pcirc}(a)
and~\ref{supfig:pcirc}(b)
satisfy this relation.

Since $\bm{\chi}_{\iota,n}(\bm{k}_{\rm U})$ transforms as an
axial vector according to Eq.~\eqref{supeq:chi_trans},
we also obtain
\begin{align}
    \bm{\chi}_{\iota',n}^{(P4_132)}(\bm{k}_{\rm U})
    =
    \bar{\sigma}_{yz}
    \bm{\chi}_{\iota,n}^{(P4_332)}(\bm{k}_{\rm U}),
\end{align}
which gives
\begin{align}
    \label{supeq:chiz and chiy}
    \chi_{\iota,n}^{z,(P4_332)}(\bm{k}_{\rm U})
    =
    -\chi_{\iota',n}^{y,(P4_132)}(\bm{k}_{\rm U}).
\end{align}
Because the transformation preserves
$|\bm{\chi}_{\iota,n}(\bm{k}_{\rm U})|$,
the normalized components
$\hat{\chi}_{\iota,n}^\alpha(\bm{k}_{\rm U})
=\chi_{\iota,n}^\alpha(\bm{k}_{\rm U})/|\bm{\chi}_{\iota,n}(\bm{k}_{\rm U})|$
satisfy
\begin{align}
\hat{\chi}_{\iota,n}^{z,(P4_332)}(\bm{k}_{\rm U})
=
-\hat{\chi}_{\iota',n}^{y,(P4_132)}(\bm{k}_{\rm U}).
\end{align}
Figures~\ref{supfig:pcirc}(c)
and~\ref{supfig:pcirc}(d)
show $\hat{\chi}_{\iota,n}^{z}(\bm{k}_{\rm U})$ for $P4_332$ and
$-\hat{\chi}_{\iota,n}^{y}(\bm{k}_{\rm U})$ for $P4_132$, respectively.
Entries satisfying
$p_{\iota,n}(\bm{k}_{\rm U})<0.01$
are nearly linearly polarized
and are therefore also masked in white,
in addition to those satisfying
$|\bm{F}_{\iota,n}|^2 / [S(S+1)] < 0.01$.
The results in Figs.~\ref{supfig:pcirc}(c)
and~\ref{supfig:pcirc}(d) confirm
the relation above.
The real-space spin motion of the $n=1$ mode
is shown in Fig.~3 of the main text.

For the subsequent discussion of
the dynamical spin structure factor,
we define
\begin{align}
    \label{supeq:Fk_tot}
    \bm{F}_n(\bm{k})=\sum_{lc}\bm{F}_{lc,n}(\bm{k})
\end{align}
and introduce
\begin{align}
    \label{supeq:chi_coh}
    \bm{\chi}_{n}(\bm{k})
    =
    -\frac{1}{2}{\rm Im}
    \left[
    \bm{F}_n(\bm{k})\times\bm{F}_n^*(\bm{k})
    \right].
\end{align}
The uniform sum over $l$ and $c$
selects the A$_{1g}$ component
in the internal site space.
Thus, $\bm{F}_n(\bm{k})$ transforms
as
${\rm A}_{1g}\otimes{\rm T}_{1g}
={\rm T}_{1g}$
with respect to the spin indices.
Furthermore, because $\bm{\chi}_n(\bm{k})$
is given by the
antisymmetric product of
two ${\rm T}_{1g}$ vectors,
$\bm{\chi}_n(\bm{k})$ also transforms
according to the T$_{1g}$ representation
with respect to the spin indices.

\section{Spin carried by triplons}

To evaluate the spin carried by a triplon,
we introduce the site spin operator
projected onto the local excited-state subspace,
\begin{align}
    \bm{\mathcal{S}}_{d,c}
    \equiv
    P_d^{\rm ex}\bm{S}_{d,c} P_d^{\rm ex}
    =
    \sum_{\rho,\rho'=1}^{3}
    \braket{l;\rho|\bm{S}_{d,c}|l;\rho'}
    a_{d,\rho}^{\dagger}a_{d,\rho'},
\end{align}
and define its site-resolved expectation value
in the one-triplon state
$|\bm{k},n\rangle\rangle$ as
\begin{align}
    \bm{s}_{lc,n}(\bm{k})
    =
    \braket{\braket{\bm{k},n|\sum_{R}\bm{\mathcal{S}}_{R,lc}|\bm{k},n}}.
\end{align}
Here,
$d=(R,l)$,
$P_d^{\rm ex}$ is the projection
operator onto the local excited-state subspace,
and $\bm{S}_{d,c}$ is the spin operator at site $c$ of dimer $d$.
The six Cartesian components of
$\bm{S}_{d,c}$~($c=1,2$)
correspond to $r=1,\ldots,6$
in Eq.~\eqref{supeq:O_definition}.

We consider a time-reversal~(TR)-symmetric
local Hamiltonian.
Since the local Hamiltonian used in this work
acts on two $S=1/2$ spins~[Eq.~\eqref{supeq:Hd_MF}],
the TR operator $\Theta$ satisfies $\Theta^2=1$.
Its local eigenstates can therefore
be chosen to be individually TR invariant.
A TR-invariant state is mapped to itself
up to an overall phase,
and the phase of each state can be
chosen such that
$\Theta\ket{l;0}=\ket{l;0}$ and
$\Theta\ket{l;\rho}=\ket{l;\rho}$.
In this basis,
$\Theta$ is represented by
the complex conjugation operator $\mathcal{K}$.
Since $\Theta P_d^{\rm ex}\Theta^{-1}=P_d^{\rm ex}$
and the spin operator changes sign
under TR, we have
\begin{align}
    \Theta\bm{\mathcal{S}}_{d,c}\Theta^{-1}
    =
    -\bm{\mathcal{S}}_{d,c}
    =
    \bm{\mathcal{S}}_{d,c}^*.
\end{align}
Therefore, each Cartesian component
$\mathcal{S}_{d,c}^\alpha$
is represented by
a purely imaginary Hermitian matrix
in the basis of local excited states.

For a real linear combination
$\ket{l;m}=\sum_{\rho=1}^{3}C_{m\rho}\ket{l;\rho}$
with $C_{m\rho}\in\mathbb{R}$,
we obtain
\begin{align}
    \braket{l;m|\mathcal{S}^\alpha_{d,c}|l;m}
    &=
    (C\mathcal{S}_{d,c}^\alpha C^T)_{mm}
    =
    -(C \mathcal{S}_{d,c}^\alpha C^T)_{mm}
    =
    0,
\end{align}
where 
$(\mathcal{S}_{d,c}^\alpha)^T=-\mathcal{S}_{d,c}^\alpha$
has been used.
Thus, in the boson-number-conserving case,
a finite
$\bm{s}_{lc,n}(\bm{k})$
requires relative phases
among the expansion coefficients of the one-triplon state.
In this sense,
$\bm{s}_{lc,n}(\bm{k})$
characterizes the site-resolved
internal spin polarization of a triplon.

To clarify the relation between
$\bm{s}_{lc,n}(\bm{k})$
and
$\bm{\chi}_{lc,n}(\bm{k})$,
we consider the limit
in which the local ground state is a pure singlet state
and the three local excited states
are pure triplets.
We further assume that
the one-triplon Hamiltonian
conserves the triplon number,
so that no Bogoliubov transformation
is required.
In this limit,
we denote the vacuum and
one-triplon states by ordinary kets.
As shown below,
an exact proportionality holds
between $\bm{s}_{lc,n}(\bm{k})$
and $\bm{\chi}_{lc,n}(\bm{k})$.
We define the local singlet state
$\ket{l;s}$ and
the local triplet states
$\ket{l;t_x}, \ket{l;t_y}$, and
$\ket{l;t_z}$ as
\begin{align}
    \label{supeq:s}
    \ket{l;s}
    &=
    \frac{1}{\sqrt{2}} \left( \ket{\uparrow\downarrow} - \ket{\downarrow\uparrow} \right),\\
    \ket{l;t_x}
    &=
    \frac{i}{\sqrt{2}} \left( \ket{\uparrow\uparrow} - \ket{\downarrow\downarrow} \right),\\
    \ket{l;t_y}
    &=
    \frac{1}{\sqrt{2}} \left( \ket{\uparrow\uparrow} + \ket{\downarrow\downarrow} \right),\\
    \ket{l;t_z}
    &=
    -\frac{i}{\sqrt{2}} \left( \ket{\uparrow\downarrow} + \ket{\downarrow\uparrow} \right),
    \label{supeq:tz}
\end{align}
respectively~\cite{romhanyi2011}.
We denote the boson creation operators
$a_{R,l\rho}^\dagger$
corresponding to these three local excited states
by
$t_{R,lx}^\dagger, t_{R,ly}^\dagger$, and
$t_{R,lz}^\dagger$, respectively.
The vacuum is given by
$\ket{\rm vac}=\bigotimes_{R,l} \ket{l;s}$,
where the local singlet state $\ket{l;s}$
is independent of the unit-cell label $R$.
An arbitrary normalized one-triplon state
can be written as
\begin{align}
    \ket{\bm{k},n}
    =
    \sum_l\sum_{\alpha=x,y,z} C_{l\alpha,n}(\bm{k}) t_{\bm{k},l\alpha}^\dagger \ket{\rm vac},
\end{align}
where $C_{l\alpha,n}(\bm{k})$
are the expansion coefficients
and
\begin{align}
    t_{\bm{k},l\alpha}^\dagger
    =
    \frac{1}{\sqrt{N_{\rm{cell}}}} \sum_{R} t^\dagger_{R,l\alpha}
    e^{i\bm{k}\cdot(\bm{R}+\bm{\delta}_l)}.
\end{align}
The transition spin amplitude
$\bm{F}_{lc,n}(\bm{k})$
defined in Eq.~\eqref{supeq:Flc}
is
\begin{align}
    F^\alpha_{lc,n}(\bm{k})
    &=
    \braket{{\rm vac}|S^\alpha_{\bm{k},lc}|\bm{k},n}\notag\\
    &=
    \frac{1}{N_{\rm cell}}
    \sum_{RR'}\sum_{l'}\sum_{\beta}
     e^{-i\bm{k}\cdot(\bm{R}+\bm{\delta}_l + \tilde{\bm{\delta}}_{l,c})}
    C_{l'\beta,n}(\bm{k})e^{i\bm{k}\cdot(\bm{R}'+\bm{\delta}_{l'})}
    \braket{{\rm vac}|S_{R,lc}^\alpha t_{R',l'\beta}^\dagger
    |{\rm vac}}.
\end{align}
Introducing the sign factor
$\xi_{c=1}=+1$ and $\xi_{c=2}=-1$,
Eqs.~\eqref{supeq:s}--\eqref{supeq:tz}
give
\begin{align}
    \braket{{\rm vac}|S_{R,lc}^\alpha t_{R',l'\beta}^\dagger
    |{\rm vac}}
    =
    -\frac{i\xi_c}{2}\delta_{R,R'} \delta_{l,l'} \delta_{\alpha\beta}.
\end{align}
We therefore obtain
\begin{align}
    F^\alpha_{lc,n}(\bm{k})
    &=
    -\frac{i\xi_c}{2}
    C_{l\alpha,n}(\bm{k}) e^{-i\bm{k}\cdot\tilde{\bm{\delta}}_{l,c}}.
\end{align}
Substituting this expression into the definition of
$\bm{\chi}_{lc,n}(\bm{k})$ yields
\begin{align}
    \label{supeq:chi_puretriplet}
    \bm{\chi}_{lc,n}(\bm{k})
    =
    -\frac{1}{8} \mathrm{Im}
    \left[ \bm{C}_{l,n}(\bm{k}) \times \bm{C}_{l,n}^*(\bm{k}) \right],
\end{align}
where
$\bm{C}_{l,n}(\bm{k})
=
[C_{lx,n}(\bm{k}),
C_{ly,n}(\bm{k}),
C_{lz,n}(\bm{k})]$.
Similarly, evaluating
$\bm{s}_{lc,n}(\bm{k})$ gives
\begin{align}
    s^\alpha_{lc,n}(\bm{k})
    &=
    \sum_{R}\braket{\bm{k},n|\mathcal{S}_{R,lc}^\alpha|\bm{k},n}\notag\\
    &=
    \frac{1}{N_{\rm cell}}
    \sum_{RR'R''}\sum_{l'l''}\sum_{\beta\gamma}
    C_{l'\beta,n}^*(\bm{k}) C_{l''\gamma,n}(\bm{k})
    e^{-i\bm{k}\cdot(\bm{R}'+\bm{\delta}_{l'})}
    e^{i\bm{k}\cdot(\bm{R}''+\bm{\delta}_{l''})}
    \braket{{\rm vac}|t_{R',l'\beta} \mathcal{S}^\alpha_{R,lc} t_{R'',l''\gamma}^\dagger|{\rm vac}}\notag\\
    &=
    -\frac{i}{2}
    \sum_{\beta\gamma}
    \epsilon_{\alpha\beta\gamma}
    C_{l\beta,n}^*(\bm{k}) C_{l\gamma,n}(\bm{k}).
\end{align}
Here, we have used
\begin{align}
    \braket{{\rm vac}|t_{R',l'\beta} \mathcal{S}^\alpha_{R,lc} t_{R'',l''\gamma}^\dagger|{\rm vac}}
    =
    -\frac{i\epsilon_{\alpha\beta\gamma}}{2}
    \delta_{R,R'}\delta_{R,R''} \delta_{l,l'} \delta_{l,l''}.
\end{align}
It follows that
\begin{align}
    \bm{s}_{lc,n}(\bm{k})
    =
    -\frac{1}{2} \mathrm{Im}
    \left[ \bm{C}_{l,n}(\bm{k}) \times \bm{C}_{l,n}^*(\bm{k}) \right].
\end{align}
Comparison with Eq.~\eqref{supeq:chi_puretriplet}
then yields the exact proportionality
\begin{align}
    \label{supeq:s_chi_relation}
    \bm{s}_{lc,n}(\bm{k})
    =
    4\bm{\chi}_{lc,n}(\bm{k}).
\end{align}
Equation~\eqref{supeq:s_chi_relation}
shows that,
in the pure-singlet--triplet and
boson-number-conserving limit,
$\bm{\chi}_{lc,n}(\bm{k})$,
defined through the transition matrix elements,
and $\bm{s}_{lc,n}(\bm{k})$,
defined through the diagonal matrix elements,
are exactly proportional
and provide equivalent measures
of the internal spin polarization.

In the present system,
the intradimer DMI induces
singlet--triplet mixing in
the local ground and excited states
already at first order~\cite{romhanyi2011}.
Furthermore, the one-triplon excitations are
Bogoliubov quasiparticles containing
both particle and hole components.
Thus, the DMI-induced mixing
and particle--hole structure
modify the quantitative relation
between $\bm{s}_{lc,n}(\bm{k})$
and $\bm{\chi}_{lc,n}(\bm{k})$
from its pure-limit form
$\bm{s}_{lc,n}(\bm{k})=4\bm{\chi}_{lc,n}(\bm{k})$.
In the present Bogoliubov problem,
the two quantities provide closely related measures
of the same internal spin polarization.

\begin{figure}[t]
\includegraphics[width=0.7\columnwidth]{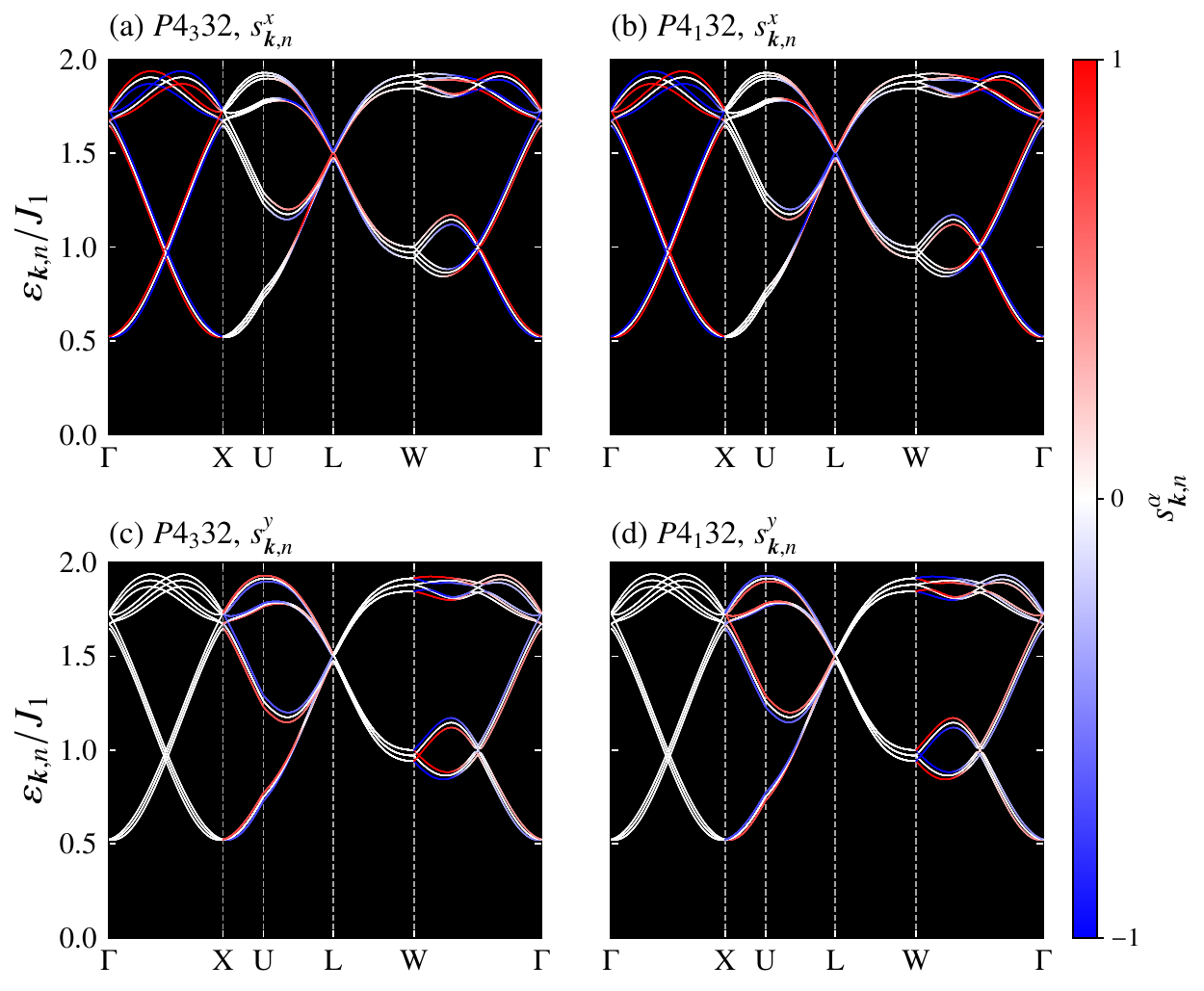}
\caption{
Triplon dispersions colored by the spin polarization.
The upper panels show $s_{\bm{k},n}^{x}$ for
(a) $P4_332$ and (b) $P4_132$,
while the lower panels show $s_{\bm{k},n}^{y}$ for
(c) $P4_332$ and (d) $P4_132$.
The parameters are the same as those in Fig.~\ref{supfig:pcirc}.
At degenerate points, the spin polarization is averaged over the
degenerate subspace.
}
\label{supfig:sx_sy}
\end{figure}

The net spin carried by a one-triplon state
is defined as
\begin{align}
    \bm{s}_{\bm{k},n}
    =
    \braket{\braket{
    \bm{k},n|
    \bm{\mathcal{S}}_{\rm tot}
    |\bm{k},n
    }},
\end{align}
where
$\bm{\mathcal{S}}_{\rm tot}=\sum_{d,c}\bm{\mathcal{S}}_{d,c}$.
Using
Eqs.~\eqref{supeq:atob} and
\eqref{supeq:adaggertobdagger},
this becomes
\begin{align}
    \bm{s}_{\bm{k},n}
    =
    \sum_l\sum_{\rho,\rho'=1}^{3}
    \braket{l;\rho|\bm{S}_{d,1} + \bm{S}_{d,2}|l;\rho'}
    \left[
    (U_{\bm{k}}^*)_{l\rho,n}
    (U_{\bm{k}})_{l\rho',n}
    +
    (V_{\bm{k}})_{l\rho,n}
    (V_{\bm{k}}^*)_{l\rho',n}
    \right].
\end{align}
The composite index $l\rho$
combines the cluster-sublattice index $l$
and the local excited-state index $\rho$.

Figure~\ref{supfig:sx_sy} shows
$s^x_{\bm{k},n}$ and $s^y_{\bm{k},n}$
for the enantiomorphic pair.
Both components reverse sign
between the two enantiomorphic states.
The component $s^x_{\bm{k},n}$ vanishes
along the X--U path.
In the BZ of the cubic unit cell,
the points
$\bm{k}_\mathrm{X}=\pi(2,0,0)/a$ and
$\bm{k}_\mathrm{U}=\pi(2,1/2,1/2)/a$
are equivalent to
$(0,0,0)$ and
$\pi(0,1/2,1/2)/a$, respectively.
Thus, the X--U path can be represented as
a path with $k_x = 0$.
Similarly,
$s^y_{\bm{k},n}$ vanishes
along the
$\mathrm{\Gamma}$--X path,
where $k_y = 0$.
These nodal lines
are consistent with a hedgehog-like spin texture 
$s^\alpha_{\bm{k},n}\propto k_\alpha$~\cite{hirayama2015}.
The reversal of the spin texture
between the two enantiomers,
for which $G_0$ has opposite signs,
suggests that the triplon Hamiltonian
contains a leading-order effective coupling
of the form
$G_0\bm{k}\cdot\bm{\mathcal{S}}_{\rm tot}$.

Figure~\ref{supfig:sx_sy} also shows that
$s^y_{\bm{k},n}$ vanishes along
the L--W path.
This vanishing can be understood from
the little-group symmetry along this path.
Since
$\bm{k}_\mathrm{L}=\pi(1,1,1)/a$ and
$\bm{k}_\mathrm{W}=\pi(2,1,0)/a$,
a wave vector along this path
can be parameterized as
\begin{align}
    \bm{k}_{[{\rm L\text{-}W}]}
    =
    \frac{\pi}{a}(1+t,1,1-t),
    \qquad
    0\leq t\leq 1.
\end{align}
Both space groups $P4_332$ and $P4_132$
contain an operation whose point-operation part is
$C_{2,[10\bar{1}]}:(x,y,z)\mapsto(-z,-y,-x)$
~\cite{ITC_A}.
Since this rotation leaves
$\bm{k}_{[{\rm L\text{-}W}]}$
invariant up to a reciprocal lattice vector,
the corresponding space-group operation belongs
to its little group.
The relevant branches are degenerate
along the L--W path.
As stated in the caption of Fig.~\ref{supfig:sx_sy},
the spin polarization at each degenerate
point is averaged over the degenerate subspace.
With this convention, the averaged spin polarization, denoted again by
$\bm{s}_{\bm{k}_{[{\rm L\text{-}W}]},n}$, must satisfy
\begin{align}
    C_{2,[10\bar{1}]}
    \bm{s}_{\bm{k}_{[{\rm L\text{-}W}]},n}
    =
    \bm{s}_{\bm{k}_{[{\rm L\text{-}W}]},n}.
\end{align}
This condition gives
\begin{align}
    s^x_{\bm{k}_{[{\rm L\text{-}W}]},n}
    =
    -s^z_{\bm{k}_{[{\rm L\text{-}W}]},n},
    \qquad
    s^y_{\bm{k}_{[{\rm L\text{-}W}]},n}
    =
    -s^y_{\bm{k}_{[{\rm L\text{-}W}]},n}.
\end{align}
It therefore follows that
$s^y_{\bm{k}_{[{\rm L\text{-}W}]},n}=0$.
Comparison of $s^z_{\bm{k},n}$ in
Fig.~4(a) and (b) of the main text
with $s^x_{\bm{k},n}$ in Figs.~\ref{supfig:sx_sy}(a) and \ref{supfig:sx_sy}(b)
also confirms that
$s^x_{\bm{k}_{[{\rm L\text{-}W}]},n}
=-s^z_{\bm{k}_{[{\rm L\text{-}W}]},n}$
along the L--W path.

\section{Thermal Edelstein effect}

We define the spin-angular-momentum response
to a temperature gradient by
\begin{align}
    \frac{\hbar\braket{\mathcal{S}_{\rm tot}^{\alpha}}_{\nabla T}}
    {V}
    =
    \kappa_{\alpha\beta}(-\nabla_{\beta}T),
\end{align}
where $V$ is the volume and
$\braket{\cdots}_{\nabla T}$ denotes
the thermal average in the presence of
a temperature gradient.
Using linear-response theory
\cite{kubo1957,luttinger1964,li2020},
we obtain
\begin{align}
    \kappa_{\alpha\beta}
    =
    -\frac{\hbar^2}{\Gamma}
    \frac{1}{VT}
    \sum_{\bm{k}}\sum_{n=1}^{N}
    s_{\bm{k},n}^{\alpha}
    \varepsilon_{\bm{k},n}
    v_{\bm{k},n}^{\beta}
    \frac{\partial f(\varepsilon_{\bm{k},n})}
    {\partial \varepsilon_{\bm{k},n}},
\end{align}
where
$f(x)=(e^{x/k_{\rm B}T}-1)^{-1}$ is the
Bose distribution function,
and
\begin{align}
    \bm{v}_{\bm{k},n}
    =
    \frac{1}{\hbar}
    \bm{\nabla}_{\bm{k}}
    \varepsilon_{\bm{k},n}
\end{align}
is the triplon group velocity.
The parameter $\Gamma$ denotes
the phenomenological relaxation width.
Defining the relaxation time
$\tau=\hbar/\Gamma$,
the response coefficient can be written as
\begin{align}
    \kappa_{\alpha\beta}
    =
    -\frac{\tau\hbar}{VT}
    \sum_{\bm{k}}\sum_{n=1}^{N}
    s_{\bm{k},n}^{\alpha}
    \varepsilon_{\bm{k},n}
    v_{\bm{k},n}^{\beta}
    \frac{\partial f(\varepsilon_{\bm{k},n})}
    {\partial \varepsilon_{\bm{k},n}}.
\end{align}

\section{Dynamical spin structure factor}

We define the dynamical spin structure factor as
\begin{align}
    \label{supeq:dssf}
    S^{\alpha\beta}(\bm{k},\omega)
    =
    \frac{1}{N_s}
    \sum_{ij}
    e^{-i\bm{k}\cdot(\bm{r}_i-\bm{r}_j)}
    \int_{-\infty}^{\infty}
    \frac{dt}{2\pi\hbar}
    e^{i\omega t/\hbar}
    \braket{\braket{
    0|S_i^\alpha(t)S_j^\beta|0
    }},
\end{align}
where $\omega$ has the dimension of energy
and
$O(t)=e^{i\mathcal{H}_{\rm FW}t/\hbar}
Oe^{-i\mathcal{H}_{\rm FW}t/\hbar}$.
Using Eq.~\eqref{supeq:Slc_fourier},
Eq.~\eqref{supeq:dssf} can be rewritten as
\begin{align}
    S^{\alpha\beta}(\bm{k},\omega)
    =
    \frac{N_{\rm cell}}{N_s}
    \sum_{ll'}\sum_{cc'}
    \int_{-\infty}^{\infty}
    \frac{dt}{2\pi\hbar}
    e^{i\omega t/\hbar}
    \braket{\braket{
    0|
    S_{\bm{k},lc}^{\alpha}(t)
    S_{-\bm{k},l'c'}^{\beta}
    |0
    }}.
\end{align}
Within LFWT,
inserting the one-triplon states gives
\cite{nasu2021,koyama2023}
\begin{align}
    S^{\alpha\beta}(\bm{k},\omega)
    =
    \frac{N_{\rm cell}}{N_s}
    \sum_{n=1}^{N}
    F_n^\alpha(\bm{k})
    F_n^{\beta *}(\bm{k})
    \delta(\omega-\varepsilon_{\bm{k},n}),
\end{align}
where
$\bm F_n(\bm{k})=\sum_{lc}\bm F_{lc,n}(\bm{k})$
is defined in Eq.~\eqref{supeq:Fk_tot}.

We next consider the
antisymmetric part of the dynamical spin
structure factor.
Since 
$S^{\beta\alpha}(\bm{k},\omega)
=
[S^{\alpha\beta}(\bm{k},\omega)]^*$,
the antisymmetric off-diagonal part
is equivalent to the imaginary part of
the off-diagonal components.
Explicitly,
\begin{align}
    \mathrm{Im}\,S^{\alpha\beta}(\bm{k},\omega)
    &=
    \frac{1}{2i}
    \left[
    S^{\alpha\beta}(\bm{k},\omega)
    -
    S^{\beta\alpha}(\bm{k},\omega)
    \right] \notag\\
    &=
    \frac{N_{\rm cell}}{N_s}
    \sum_{n=1}^{N}
    \mathrm{Im}
    \left[
    F_n^\alpha(\bm{k})
    F_n^{\beta *}(\bm{k})
    \right]
    \delta(\omega-\varepsilon_{\bm{k},n}) \notag\\
    &=
    -
    \frac{N_{\rm cell}}{N_s}
    \epsilon_{\alpha\beta\gamma}
    \sum_{n=1}^{N}
    \chi_n^\gamma(\bm{k})
    \delta(\omega-\varepsilon_{\bm{k},n}).
\end{align}
The quantity $\chi_n^\gamma(\bm{k})$
is the $\gamma$ component
of the polarization vector defined in
Eq.~\eqref{supeq:chi_coh}.
Thus, the imaginary parts of the off-diagonal components
of the dynamical spin structure factor are directly related to
the internal polarization of triplons.

\bibliography{./refs}